%% file: main.tex
\documentclass[letterpaper,twocolumn,10pt]{article}
\usepackage{usenix}

\usepackage{amsmath}

\usepackage{00_preamble}
\usepackage{00_macros}

\begin{document}

\date{}

\title{\Large \bf Anywhere, Any-Stymie: Remote Activation of Trojan Malware on LiDAR with Modulated Signals}

\author{
{\rm R. Spencer Hallyburton}\\
Duke University\\
spencer.hallyburton@duke.edu
\and
{\rm Miroslav Pajic}\\
Duke University\\
miroslav.pajic@duke.edu
}

\maketitle

\input{0-abstract}

\input{1-introduction}
\input{2-background}
\input{3-vulnerability}
\input{4-threat-model}
\input{5-activation}
\input{6-feasibility}

\input{7-impact}
\input{8-discussion}
\input{9-conclusion}
\input{9-acks}

\appendix
\input{99-appendix-open-science}

\bibliographystyle{plain}
\bibliography{references}

\input{99-appendix-A-experimental-protocol}
\input{99-appendix-B-detector-verification}

\input{figures/figure_appendix_setup_indoor}
\input{figures/figure_appendix_setup_outdoor_mobile}
\input{figures/figure_appendix_triptych_static}
\input{figures/figure_appendix_triptych_mobile}

\end{document}

%% file: 0-abstract.tex
\begin{abstract}
LiDAR sensors are widely deployed in autonomous systems for safety-critical 3D perception. We identify a previously unexplored attack surface in which dormant malware, embedded in the LiDAR sensing pipeline, remains inactive during normal operation and is triggered externally after deployment, without access to sensor hardware or networking at attack time. This malware was developed in a closed research test environment with vendor technical support, rather than by exploiting an inherent production supply-chain vulnerability.
To trigger the malware, we repurpose the sensor's own native dual-return signal processing as a covert channel for a modulated optical signal delivered into the environment. Once triggered, the malware performs false data injection and suppression on static and mobile platforms. Unlike optical spoofing, where the transmitted signal is itself the attack, here it carries only a small key that the sensor decodes with existing signal processing.
We establish feasibility with static triggering at 300~ft and drive-by runs reaching 35~mph, and verify that injected artifacts are taken up as detections by a fine-tuned 3D object detector. On a deployed tactical autonomous vehicle running an unmodified production autonomy stack, we demonstrate both fail-safe and fail-dangerous impact: emergency stop, mission abort, and collision. These results motivate stronger integrity guarantees across LiDAR development and deployment pipelines.
\end{abstract}

%% file: 1-introduction.tex
\section{Introduction}

Light Detection and Ranging (LiDAR) has become a foundational sensor for safety-critical decision-making across domains including autonomous ground vehicles, aerial platforms, intelligent transportation systems, and precision agriculture~\cite{sun2020scalability, kellner2019new, cui2019automatic, wang2021challenges}. As reliance on LiDAR has increased, so too has the impact of adversarial manipulation. Prior work has shown that perturbations to LiDAR inputs can induce emergency braking, unsafe maneuvers, or collisions in autonomous vehicles~\cite{cao2021invisible, sato2025realism}. Although many systems employ multi-sensor fusion, LiDAR remains disproportionately trusted for spatial inference and situational awareness~\cite{hallyburton2023partial}. Consequently, failures or attacks affecting LiDAR perception can have immediate and severe consequences.

Most existing LiDAR security analyses focus on adversaries who remain outside the sensor, whether at the network layer or attacking solely across physical channels. Network-layer attacks, such as packet injection, require the strong assumption of an attacker with network access. This threat model breaks down when networks are cryptographically secured and authenticated. Physical-channel attacks, including LiDAR spoofing, require less assumed access but often require meeting tight spatial constraints, careful timing, and continued, uninterrupted interaction with the victim sensor throughout the attack window~\cite{2019cao-spoofing, hallyburton2022security, jin2023pla}. These constraints limit practicality against mobile platforms.

A different and concerning threat model places malware inside the LiDAR sensing stack itself. Rather than require privileged access at attack time, such an attacker needs only a one-time foothold in the sensor's supply chain or maintenance chain. This is plausible because a LiDAR is a globally sourced, complex cyber-physical system (CPS): its firmware, third-party components, and post-deployment update paths open multiple routes through which an adversary can implant malicious code, whether before the sensor ships or through later maintenance. Such code can be inserted before it is signed, or through surfaces that signing does not cover, so integrity controls like secure boot do not close the threat. We develop this feasibility argument in Section~\ref{sec:vulnerability}.

Trojan malware in CPS is broadly recognized, yet the sensor-internal case has received little attention because sensors are conventionally trusted as measurement sources rather than treated as potential adversaries. How, and to what extent, an adversary can drive a compromised LiDAR to manipulate its own output from within the pipeline remains open.

For sensor-internal LiDAR malware to succeed, two conditions must hold. \underline{First}, it must be feasibly triggered after deployment without access to the vehicle or sensor internals. \underline{Second}, once activated, it must be capable of inducing safety-critical effects on LiDAR output, whether by producing plausible manipulated data or by forcing the sensor into failure states that meaningfully alter downstream autonomy behavior. These conditions make dormant, selectively triggered malware especially dangerous: implantation and activation are decoupled, so it stays unobtrusive through normal operation, integration, and validation, and the adversary triggers it later with a brief optical signal from a distance, requiring no network access, precise aiming, or sustained presence at attack time. These are the constraints that limit prior spoofing against moving platforms, making the sensor-internal vector valuable against mobile, safety-critical targets that matter most.

To address the first condition, we implant malware inside the LiDAR and build a handheld optical emitter to trigger it. Dual-return processing is a native sensor feature that resolves multiple echoes per laser pulse. We repurpose it as a covert channel that needs no added hardware and hides within normal signal processing: the emitter encodes a key into the returns, and the malware decodes it. Unlike prior optical attacks, where the transmitted signal is itself the attack, here it carries only this key, so briefly illuminating the environment activates the malware (Section~\ref{sec:activation}). To address the second condition, the same malware injects synthetic returns or suppresses genuine ones in real time from inside the pipeline, before the point cloud reaches downstream perception; we implement and evaluate the full attack on a real LiDAR sensor.

Across every condition we sampled, the optical trigger activated the malware, and the selected manipulation executed as intended. Indoors, activation succeeded at roughly 150~ft, whether by direct line of sight, through reflections off walls and ceilings, or even across separate floors of a building. Outdoors, we triggered the malware at standoffs out to 300~ft, and against a victim vehicle driving past at 35~mph. For an injected person-like artifact, a fine-tuned 3D object detector reports it as a person-class detection, and its point density matches that of real objects closely enough to defeat a low-level plausibility check. Suppression is the converse: the targeted returns are removed, so a real object vanishes from the point cloud and the stack sees only free space.

The consequences are most acute on a deployed tactical autonomous vehicle (TAV) running an unmodified production autonomy stack. A trigger from roughly 100~ft produces both fail-safe and fail-dangerous outcomes. The fail-safe outcomes disrupt the mission without a crash: cutting off the point cloud halts the vehicle in an emergency stop indistinguishable from a benign fault, and injecting phantom obstacles forces it to maneuver evasively and abort the mission. The fail-dangerous outcome is worse: suppressing returns from an obstacle in its path drives the vehicle into a collision it could otherwise avoid. The platform fuses the compromised LiDAR with benign cameras, yet corrupting the point cloud alone is enough to drive these outcomes, consistent with a stack that weights LiDAR heavily for obstacle geometry. Multi-sensor fusion does not save it.

Finally, we analyze why common perception defenses, such as filtering, redundancy, and anomaly detection, are poorly aligned with this threat model, and we argue for stronger integrity guarantees on the manufacturing and supply chains of sensing components (Section~\ref{sec:discussion}).

\vspace{4pt}
\noindent\textbf{Contributions.} This work makes the following contributions:
\begin{itemize}[leftmargin=16pt,itemsep=3pt,topsep=4pt,parsep=1pt]
    \item We identify sensor-internal LiDAR malware with post-deployment activation as a distinct attack surface, different from prior spoofing and network-oriented attacks.
    \item We design and implement an end-to-end attack that repurposes the LiDAR's native dual-return processing as a covert channel, so the transmitted signal carries only a key rather than the attack itself.
    \item We characterize trigger feasibility across indoor/outdoor, static/drive-by conditions, with standoff to 300~ft, speeds to 35~mph, and delivery via loosely-aimed reflections.
    \item We show that a fine-tuned 3D object detector reports injected person-like artifacts as person-class detections.
    \item We demonstrate attacks on a \emph{real deployed TAV running an unmodified production autonomy stack}, causing both fail-safe and fail-dangerous outcomes, including a collision with an obstacle hidden by suppression.
\end{itemize}

\noindent The remainder of this paper is organized as follows. Sections~\ref{sec:background} and~\ref{sec:vulnerability} provide background and motivate the sensor-internal threat model. Sections~\ref{sec:threat-model} and~\ref{sec:activation} define the threat model and activation mechanism. Sections~\ref{sec:evaluation} and~\ref{sec:impact} present the feasibility and deployed-vehicle evaluation, and Section~\ref{sec:discussion} discusses implications, limitations, and mitigations.

%% file: 2-background.tex
\section{Background: LiDAR as a Critical Sensor}
\label{sec:background}

This section reviews how LiDAR sensors operate, why autonomy stacks trust them for safety-critical decisions, and what prior security work has and has not examined. Across that work, the adversary always operates outside the sensor; we instead develop a sensor-internal threat.

\subsection{LiDAR Sensing in Autonomy}
\label{sec:background-lidar}

LiDAR sensors are CPS that integrate optical, electronic, and software components. They emit coded laser pulses and use photodetectors to measure time-of-flight distance~\cite{nofriandi2024ultra}. A single pulse striking successive surfaces yields multiple echoes; to handle edges and partial occlusions, sensors commonly resolve the two strongest, a feature known as \emph{dual-return} processing. Embedded processors analyze these returns, performing signal conditioning, filtering, and packetization before publishing point clouds. These sensors rely on custom silicon and non-volatile memory that stores firmware.

LiDAR provides the spatial measurements that autonomy stacks use for obstacle detection, localization, and navigation, and it is trusted for safety-critical decisions such as collision avoidance and emergency stopping, where errors cause physical harm~\cite{li2020lidar, chang2012infrastructure, zhang2014loam}. Because this trust is rarely questioned, compromising LiDAR data carries immediate consequences, and its firmware, often signed at manufacture yet updated after deployment, is an exposed avenue for that compromise.

\subsection{Prior LiDAR Security Analyses}

Prior LiDAR security work has focused on adversarial manipulation through physical channels external to the sensor. These attacks disrupt or falsify LiDAR measurements by interacting with the emitted or received optical signals.

\myparagraph{Jamming and blinding attacks.}
Jamming and blinding attacks overwhelm LiDAR receivers with high-intensity or continuous signals, degrading measurements or causing temporary denial~\cite{2017illusiondazzle, 2016contactlessattacks}. They are straightforward to execute but typically detectable through abrupt density changes, and their effects cease once the interference stops.

\myparagraph{Spoofing attacks.}
Spoofing attacks inject timed laser pulses to introduce false points or shift object locations~\cite{2014cyberattacks, 2017illusiondazzle, 2019cao-spoofing, 2020sun-spoofing}. Although stealthier than jamming, they require \emph{precise aiming, tight timing synchronization to the scan, and continuous interaction with the sensor}, making them hard to sustain against mobile platforms~\cite{2021caoNdssDemo, sato2025realism}.

\myparagraph{Adversarial objects.}
Physical adversarial objects are crafted and placed in a scene to deceive LiDAR perception models~\cite{2020tuphysicalatt, zhu2021can}, but exploit weaknesses in learned pipelines and are constrained by placement, limited spatial influence, and dependence on specific model architectures.

All of these attacks share one assumption: the adversary operates \emph{outside} the sensor and must interact continuously over a physical channel, which limits their persistence, scalability, and stealth. We instead consider malware embedded \emph{inside} the sensor, triggered remotely after deployment.

\subsection{Trojans, Backdoors, and Covert Channels}
\label{sec:background-trojans}

The sensor-internal model connects to three lines of work outside LiDAR security, each distinct from ours.

\myparagraph{Hardware and firmware Trojans.}
Hardware and firmware Trojans embed dormant logic that activates on a rare trigger and evades functional test~\cite{yang2016a2, hardwareTrojans2014, subraman2019demonstrating}. Our payload adopts this structure but resides in a LiDAR sensor and falsifies its point-cloud output rather than subverting a host's computation.

\myparagraph{Neural backdoors.}
Neural backdoors implant a trigger-conditioned behavior into a perception model~\cite{gu2019badnets}. They poison the model that consumes sensor data, whereas we corrupt the data upstream, so the manipulation is model-agnostic.

\myparagraph{Signal-injection and transduction attacks.}
Signal-injection and transduction attacks falsify other sensors by delivering an external analog signal over their physical channel, as in acoustic injection into inertial sensors or projected light into cameras~\cite{trippel2017walnut, nassi2021phantom}. They perturb perception during delivery, while in our setting the external signal only \emph{activates} the sensor-internal malware and carries no perception payload.

None of these share the activation channel: we repurpose the LiDAR's native dual-return processing as a decoder for a covert message that gates the embedded malware (Section~\ref{sec:activation}).

%% file: 3-vulnerability.tex
\section{Feasibility of Sensor-Internal Malware}
\label{sec:vulnerability}


Implanting malware inside a LiDAR is feasible: the sensor combines complex, globally sourced hardware with onboard firmware and post-deployment update paths, opening routes for a dormant implant. Fig.~\ref{fig:threat-model} illustrates the attack surface and operating scenario. We examine implantation, firmware integrity, and why LiDAR is a high-value target.

\input{figures/figure_threat_model}

\subsection{Feasibility of Malware Implantation}

\myparagraph{A trusted but unaudited peripheral.}
A LiDAR is a complex CPS: its lasers, photodetectors, optics, and chips come from a global supply chain, and its firmware is proprietary, often written by third-party vendors~\cite{zhao2019recent, hamamatsu2020lasers}. The autonomy integrator that trusts its output rarely inspects or controls firmware, so the code running inside the sensor goes largely unscrutinized.

\myparagraph{Access paths.}
Attacker-controlled code can enter through several established paths: (1)~malicious insertion during fabrication or board integration, by a compromised component supplier or contract manufacturer~\cite{hardwareTrojans2014}; (2)~compromised firmware or over-the-air update channels, which deliver code to the sensor after deployment~\cite{kuppusamy2018uptane}; and (3)~calibration, configuration, and maintenance interfaces, which are often writable in the field and receive less scrutiny than the primary code image. None require the privileged in-vehicle network access that prior attacks assume~\cite{checkoway2011comprehensive}; each targets a single component's supply or maintenance chain. Firmware is a particularly favorable host: malicious logic embedded in existing signal-processing is difficult to distinguish from legitimate sensing.

\myparagraph{Dormancy and test evasion.}
A dormant implant stays inert until an external trigger, so it passes checks that test only nominal performance and compliance, not adversarial activation. At the hardware level, an analog malicious circuit dormant until a rare trigger evades functional, structural, and side-channel testing~\cite{yang2016a2}. We apply the same idea in sensor firmware, with an optical trigger rather than internal state.

\myparagraph{Amortized cost of implantation.}
Network and physical-channel attacks pay \emph{per attack}, each activation demanding renewed network access or sustained optical interaction with the victim~\cite{2019cao-spoofing, hallyburton2022security}. Implantation pays \emph{once}: a single compromised firmware image, replicated across a production run, leaves a whole fleet remotely activatable long after manufacture, with no further access. The attacker needs only a one-time foothold in the supply or maintenance chain.

\subsection{Firmware Integrity in Practice}
\label{sec:vulnerability-integrity}

Secure boot, firmware signing, and attestation are the controls expected to reject unauthorized code. On sensor peripherals, three gaps limit what they cover.

\myparagraph{Uneven enforcement.}
The primary autonomy mainframe is typically protected by a hardware root of trust and a verified boot chain, but that protection does not extend to the sensor tier, which may lack its own secure boot or signature verification. If the sensor does not attest its code state to a downstream consumer, the autonomy stack accepts its point cloud on trust, and a modified image runs unchallenged.

\myparagraph{Insertion before signing.}
A signature protects an image from when it is applied. An adversary positioned earlier, such as a compromised firmware vendor, produces a validly-signed malicious image that passes every check. Signing-key compromise has the same effect. A valid signature proves only that an image is authentic, not that its code is benign.

\myparagraph{Surfaces beyond the signed image.}
Code signing typically covers the primary firmware image and not the calibration tables, configuration parameters, and field-writable maintenance surfaces that also shape sensor output. Manipulation delivered through these channels need not defeat code signing.

Firmware compromise therefore remains reachable: the controls that would block it are unevenly deployed and bypassable in the supply chain. This motivates the mitigation we return to in Section~\ref{sec:discussion}: runtime attestation of sensor code state to a downstream verifier~\cite{nunes2019vrased, surminski2021realswatt}, which would raise the attack's cost but is not yet standard for the sensor tier.

\subsection{LiDAR As a High-Value Target}

\myparagraph{Sensors over actuators.}
Sensors and actuators sit at opposite ends of the platform's protection spectrum. Steering, braking, and powertrain control are the most hardened and monitored: they fall under functional-safety regimes such as ISO~26262, sit behind authenticated control buses increasingly guarded by intrusion detection, and are the classical target of vehicle-compromise research~\cite{checkoway2011comprehensive}. Sensors receive far less scrutiny, yet they supply the inputs that safety logic trusts. Corrupting a trusted sensor subverts safety decisions by falsifying their inputs, without touching the hardened logic.

\myparagraph{Concentration of trust in LiDAR.}
Among sensors, LiDAR concentrates this risk. Cameras, radar, and satellite navigation share the same supply-chain exposure, but LiDAR is often the authoritative source of 3D geometry for obstacle detection, localization, and navigation, even under sensor fusion~\cite{mao2023survey, hallyburton2023partial}. Automotive-security standards such as ISO/SAE~21434 and UL~4600~\cite{costantino2022depth, koopman2022ul} lean on vendor self-certification across international supply chains~\cite{singleton2024laser}, so an embedded LiDAR receives limited independent review once integrated.

\myparagraph{A LiDAR-specific mechanism.}
Our activation mechanism is specific to LiDAR: it repurposes the sensor's native signal processing, detailed in Section~\ref{sec:activation}. The broader principle, that a dormant sensor implant can be triggered through its own signal processing, may extend to other modalities, but the concrete attack here depends on a property of LiDAR itself.

Feasibility, however, is not demonstration: the malware must also be triggerable after deployment and produce safety-critical effects. These two conditions structure what follows.

%% file: figures/figure_threat_model.tex
\begin{figure}[!t]
\centering
\resizebox{\columnwidth}{!}{\input{diagrams/diagram_threat_model}}
\caption{Threat model. Malware is embedded on the LiDAR prior to attack execution and remains dormant. Supply-chain compromise is a motivating path, but other firmware-level implantation routes are possible. Post-deployment activation occurs via a transient modulated optical signal, without network or OS access and without the need for precise aiming.}
\label{fig:threat-model}
\end{figure}

%% file: diagrams/diagram_threat_model.tex
\begin{tikzpicture}[
    node distance=1.7cm,
    font=\small,
    base/.style={draw, rounded corners, align=center, line width=0.8pt, inner sep=6pt},
    stage/.style={base, fill=black!5, draw=black!70},
    sensor/.style={base, fill=red!12, draw=red!60!black},
    victim/.style={base, fill=ACMDarkBlue!8, draw=ACMDarkBlue, minimum width=5cm},
    attacker/.style={base, fill=red!12, draw=red!60!black},
    flow/.style={-{Stealth[length=2.4mm]}, line width=1pt, draw=ACMDarkBlue},
    place/.style={-{Stealth[length=2.4mm]}, line width=1pt, draw=black!75},
    attack/.style={-{Stealth[length=2.4mm]}, line width=1pt, draw=red!70!black, dashed},
    lbl/.style={draw=none, align=center, font=\footnotesize, inner sep=2pt}
]

\node[stage] (implant) {\textbf{Implantation Path}\\{\footnotesize (e.g., supply chain, updates, maintenance)}};
\node[sensor] (lidar) [below of=implant] {\textbf{LiDAR Sensor}\\Firmware $+$ \textbf{Malware}};
\node[victim] (vehicle) [below of=lidar] {\textbf{Autonomous Vehicle}\\Perception \& Control};

\node[attacker] (attacker) [right of=lidar, xshift=2.9cm, yshift=-3mm] {\textbf{External}\\\textbf{Attacker}};

\draw[place] (implant) -- (lidar) node[lbl,midway,left] {place malware};
\draw[flow] (lidar) -- (vehicle) node[lbl,midway,left] {point cloud};

\draw[attack] (attacker) -- (lidar)
    node[lbl,midway,above,text=red!70!black,yshift=1pt] {modulated\\optical signal};

\end{tikzpicture}

%% file: 4-threat-model.tex
\section{Threat Model}
\label{sec:threat-model}


We consider an adversary that has placed dormant malware into LiDAR firmware or drivers. It aims to activate that malware after deployment and, using only a remote optical signal, induce safety-critical failures by altering the sensor's outputs. Fig.~\ref{fig:threat-model} situates this adversary within the operating scenario.

\subsection{System and Trust Model}

\myparagraph{System model.}
The victim is an autonomous platform relying on LiDAR for safety-critical perception; our evaluation targets a tactical autonomous vehicle (TAV) running an unmodified production autonomy stack (Section~\ref{sec:impact}). The stack uses LiDAR data for obstacle detection, localization, motion planning, and emergency stopping, where errors produce harm.

\myparagraph{Trust boundary.}
The adversary compromises only the LiDAR pipeline. The autonomy compute, the cameras and other sensors, and the in-vehicle network remain benign. The stack fuses the compromised LiDAR with these healthy modalities.

\subsection{Attacker Capabilities}

Before the attack, the adversary can place malware into the LiDAR sensing stack through the implantation paths of Section~\ref{sec:vulnerability}. After deployment, the adversary does not have access to the vehicle's network, operating system, or other sensors, and needs no physical access to the vehicle or LiDAR.

At attack time, the adversary can transmit optical signals into the operating environment using a handheld emitter built from a modified commodity optical source (Section~\ref{sec:evaluation}). The adversary interacts with the system only through the LiDAR's physical sensing channel. The embedded malware executes within the normal LiDAR sensing pipeline under typical firmware resource constraints and can read and modify LiDAR data before it is consumed by downstream perception.

\subsection{Attacker Knowledge}

We assume the following knowledge model.

\begin{enumerate}[label=\textbf{K.\arabic*},itemsep=3pt,topsep=4pt,parsep=1pt]
    \item \label{know:trigger} The adversary knows the preconfigured trigger patterns that will activate the malware.
    \item \label{know:model} The adversary knows coarse victim information relevant to trigger delivery, such as the operating wavelength, but \emph{not} device-specific identifiers or internal state.
    \item \label{know:noobserve} The adversary \emph{cannot} observe victim data or malware state during triggering; attack activation is open-loop.
    \item \label{know:environment} The adversary needs no knowledge of the vehicle's pose, planned route, or any scene configuration.
\end{enumerate}

\subsection{Constraints and Non-Goals}

We place several threats out of scope: network compromise, network denial-of-service, and physical tampering at attack time. Implantation precedes the attack and is in scope. The victim may deploy firmware-integrity controls, but Section~\ref{sec:vulnerability-integrity} shows implantation remains reachable despite them; Section~\ref{sec:discussion} examines defenses that detect an implant at runtime.

This threat model motivates a stealthy post-deployment activation mechanism operating solely through the physical sensing channel, which we develop next.

%% file: 5-activation.tex
\section{Triggering via Modulated Signals}
\label{sec:activation}


\input{figures/figure_trigger_flow}

Embedded LiDAR malware stays dormant until triggered, and under the threat model of Section~\ref{sec:threat-model} that trigger must work after deployment without network access, runtime observability, or physical access to the sensor. We survey candidate triggers and adopt modulated physical-channel activation. Fig.~\ref{fig:trigger-flow} summarizes the attacker-to-victim triggering pipeline, and Fig.~\ref{fig:lidar-pipeline} shows where the malware resides in the sensing pipeline, reading processed LiDAR data and modifying the point cloud.

\input{figures/figure_malware_placement}

\subsection{Triggering Trojan Malware}

Our attack triggers through the LiDAR's own physical channel. A dormant trigger can instead fire on a timer, scene content, or a network message, but none fits our threat model: a temporal trigger offers no control over the environment at activation and surfaces during pre-deployment testing; a situational trigger must interpret scene content, adding perception-like logic and a continuous loop that consumes firmware resources and invites detection; and a network trigger requires the secured-channel or platform access we place out of scope.

\myparagraph{Physical-channel triggers.}
A physical-channel trigger avoids all three limitations. Because a LiDAR continuously emits and receives optical signals, the trigger can be delivered transiently and covertly, with no network access or persistent interaction. The malware then detects it from outputs the LiDAR \emph{already computes}, rather than analyzing scene content (Section~\ref{sec:activation-modulated}), so detection costs little. Activation occurs only with an adversary present, invisibly to pre-deployment testing.

\subsection{Modulated Signal Activation}
\label{sec:activation-modulated}

\input{figures/figure_trigger_symbol_detector}

\input{figures/figure_trigger_timing}

We propose \emph{modulated signal activation}, which repurposes the LiDAR's own dual-return sensing as a trigger channel, adding only lightweight aggregation and matching, not a new execution path. End to end, the attacker emits an optical sequence, the sensor's native processing renders it as symbols, and the malware reconstructs them into a message that, on a match, activates the selected manipulation of LiDAR output. Below we trace this pipeline, why the channel is novel, how message length trades against robustness, and how easily the signal is delivered.

The attacker emits a \emph{transmitted optical sequence} into the scene: bright pulses that the LiDAR receives alongside ordinary reflections. For each laser shot, dual-return processing reports the two strongest echoes, and the attacker times its pulses so that the separation between the two peaks encodes one symbol (Fig.~\ref{fig:trigger-symbol-detector}). Other encodings over LiDAR signal-processing outputs would serve.

The malware then assembles these symbols into a \emph{reconstructed message}, potentially across repeated transmissions (Fig.~\ref{fig:trigger-message-reconstruction}), and compares it against \emph{target messages} embedded in the malware. A match satisfies the \emph{trigger condition}, selects the corresponding \emph{attack mode} (the LiDAR manipulation executed after activation, Section~\ref{sec:impact}), and causes \emph{activation}: the malware executes the selected behavior.

\myparagraph{Open-loop activation.}
The novelty is the activation channel: \emph{native LiDAR signal-processing outputs repurposed as a selective, feedback-free trigger that delivers attacker-specified messages after deployment.} Repurposing this path is not mechanical: dual-return handling exists to resolve multiple echoes per pulse, not to carry information, so the malware must reconstruct the message open-loop, under ambient noise. Consistent with Assumptions~\ref{know:trigger} and~\ref{know:noobserve}, the adversary does not infer or observe the encoding scheme; the trigger patterns are embedded into the malware before deployment. The signal only activates the malware, and unlike optical spoofing it does not itself alter perceived geometry at trigger time.

\myparagraph{Message length and robustness.}
In our implementation, activation uses a finite-length trigger of four code symbols. Each symbol is read from a single dual-return measurement, but the malware evaluates one symbol per LiDAR rotation (about 100~ms in our experiments): because the sensor is scanning, it revisits the attacker's bearing only once per revolution, so the signal must hold each symbol steady for a full rotation to be sampled. A four-symbol trigger therefore requires roughly 400~ms of sustained delivery (Fig.~\ref{fig:trigger-timing}). Longer codes and stricter matching reduce accidental activation in noisy environments, but raise activation latency and demand that the aggregator reconstruct more symbols correctly. Section~\ref{sec:evaluation-setups} summarizes the implementation parameters.

\myparagraph{Delivery.}
Delivering the trigger is far simpler than mounting a network or spoofing attack: it needs no precise aiming at the sensor aperture and tolerates interrupted line of sight, since the signal reaches the victim over direct or reflected paths. In our evaluation, the attacker uses the handheld emitter of Section~\ref{sec:threat-model} to deliver it toward the victim platform.

Section~\ref{sec:evaluation} evaluates this trigger across indoor, outdoor, and drive-by conditions, and Section~\ref{sec:impact} demonstrates the safety-critical effects the malware produces once activated.

%% file: figures/figure_trigger_flow.tex
\begin{figure}[t]
\centering
\resizebox{\columnwidth}{!}{\input{diagrams/diagram_trigger_flow}}
\caption{End-to-end triggering pipeline. On the victim, the malware reuses \emph{native} dual-return symbol detection and filtering (gray) and adds only lightweight logic (red): assembling symbols into a message, matching against stored targets, and, on a match, modifying the outgoing point cloud.}
\label{fig:trigger-flow}
\end{figure}

%% file: diagrams/diagram_trigger_flow.tex
\begin{tikzpicture}[
    font=\footnotesize,
    node distance=5mm and 5mm,
    box/.style={draw, rounded corners, align=center, line width=0.7pt, inner sep=4pt, text width=20mm, minimum height=10mm},
    stage/.style={box, fill=black!5, draw=black!62},
    act/.style={box, fill=red!12, draw=red!60!black},
    flow/.style={-{Stealth[length=2mm]}, line width=0.9pt, draw=black!78},
    sig/.style={-{Stealth[length=2.4mm]}, line width=1.1pt, draw=red!70!black, dashed},
    lbl/.style={draw=none, align=center, font=\footnotesize, inner sep=2pt},
    title/.style={font=\bfseries\footnotesize, inner sep=1.5pt}
]

\node[act] (ams) {Attack Mode\\Selection};
\node[act, right=of ams] (tml) {Transmit\\Message Lookup};
\node[act, right=of tml] (ts) {Emit Optical\\Sequence};
\draw[flow] (ams) -- (tml);
\draw[flow] (tml) -- (ts);

\begin{scope}[on background layer]
    \node[draw, rounded corners, line width=0.9pt, draw=red!45!black, fill=red!4,
          fit=(ams)(ts), inner sep=4mm] (abox) {};
\end{scope}
\node[title, fill=red!3, anchor=west] at ([xshift=4mm]abox.north west) {Attacker-side logic};

\node[stage, below=19mm of ams] (sd) {Symbol\\Detector};
\node[stage, right=of sd] (sf) {Symbol\\Filter};
\node[act, right=of sf] (mr) {Message\\Reconstruction};

\node[act, below=6mm of sd] (tmm) {Target Message\\Match};
\node[act, right=of tmm] (mpc) {Modify\\Point Cloud};
\node[lbl, right=of mpc, text=red!60!black] (ll) {\itshape lightweight\\\itshape added logic};

\draw[flow] (sd) -- (sf);
\draw[flow] (sf) -- (mr);
\draw[flow] (mr.south) -- ++(0,-4mm) -| (tmm.north);
\draw[flow] (tmm) -- (mpc);

\begin{scope}[on background layer]
    \node[draw, rounded corners, line width=0.9pt, draw=ACMDarkBlue!70, fill=ACMDarkBlue!5,
          fit=(sd)(mr)(tmm)(mpc)(ll), inner sep=4mm] (vbox) {};
\end{scope}
\node[title, fill=ACMDarkBlue!4, anchor=east] at ([xshift=-4mm]vbox.south east) {Victim malware logic};

\draw[sig] (abox.south) -- (vbox.north);
\node[lbl, text=red!70!black, anchor=west] at ($(abox.south)!0.5!(vbox.north)$) {\;modulated optical signal};

\node[draw, rounded corners, fill=black!5, draw=black!62, minimum width=5mm, minimum height=3mm, inner sep=0pt, anchor=north west] (sw1) at ([yshift=-4mm]vbox.south west) {};
\node[lbl, anchor=west] (l1) at (sw1.east) {\,native LiDAR (reused)};
\node[draw, rounded corners, fill=red!12, draw=red!60!black, minimum width=5mm, minimum height=3mm, inner sep=0pt, anchor=west] (sw2) at ([xshift=6mm]l1.east) {};
\node[lbl, anchor=west] (l2) at (sw2.east) {\,adversary-controlled};

\end{tikzpicture}

%% file: figures/figure_malware_placement.tex
\begin{figure*}[t]
\centering
\input{diagrams/diagram_malware_placement}
\caption{Placement of the malware within the LiDAR sensing pipeline. The malware resides after low-level signal processing and before point cloud publication, where it can read processed LiDAR data and modify the outgoing point cloud consumed by downstream decision algorithms including perception and control.}
\label{fig:lidar-pipeline}
\end{figure*}

%% file: diagrams/diagram_malware_placement.tex
\begin{tikzpicture}[
    font=\footnotesize,
    node distance=6mm,
    base/.style={draw, rounded corners, align=center, line width=0.8pt, inner sep=5pt, text width=22mm, minimum height=11mm},
    stage/.style={base, fill=black!5, draw=black!65},
    malware/.style={base, fill=red!12, draw=red!60!black},
    downstream/.style={base, fill=ACMDarkBlue!8, draw=ACMDarkBlue},
    flow/.style={-{Stealth[length=2.4mm]}, line width=1pt, draw=black!78},
    lbl/.style={draw=none, align=center, font=\footnotesize, inner sep=2pt}
]

\node[stage] (front) {Optical Sensing\\Front End};
\node[stage, right=of front] (sigproc) {Signal\\Processing};
\node[malware, right=of sigproc] (mw) {\textbf{Malware}\\{\footnotesize trigger logic \&}\\{\footnotesize attack execution}};
\node[stage, right=of mw] (pub) {Point Cloud\\Publication};
\node[downstream, right=22mm of pub] (ds) {Perception \&\\Control Stack};

\draw[flow] (front) -- (sigproc);
\draw[flow] (sigproc) -- (mw);
\draw[flow] (mw) -- (pub);
\draw[flow] (pub) -- (ds) node[lbl, pos=0.62, above] {point cloud};

\begin{scope}[on background layer]
    \node[draw, rounded corners, line width=1pt, draw=black!55, fill=black!3,
          fit=(front)(pub), inner sep=6mm] (sensor) {};
\end{scope}
\node[anchor=north west, font=\bfseries] at ([shift={(2.5mm,-1.5mm)}]sensor.north west) {LiDAR Sensor};

\end{tikzpicture}

%% file: figures/figure_trigger_symbol_detector.tex
\begin{figure}[t]
    \centering
    \begin{subfigure}[b]{0.5\linewidth}
        \centering
        \input{diagrams/diagram_symbol_detector}
        \caption{Symbol detection.}
        \label{fig:trigger-symbol-detector}
    \end{subfigure}

    \vspace{6pt}

    \begin{subfigure}[b]{\linewidth}
        \centering
        \input{diagrams/diagram_message_reconstruction}
        \caption{Message reconstruction.}
        \label{fig:trigger-message-reconstruction}
    \end{subfigure}
    \caption{The two-step decode inside the malware. (a) Dual-return timing in a measurement event maps to a symbol; (b) filtered symbols are aggregated into a reconstructed message and compared against stored target messages to satisfy the trigger condition.}
\end{figure}

%% file: diagrams/diagram_symbol_detector.tex
\begin{tikzpicture}[
    x=1mm, y=1mm,
    bin/.style={fill=ACMDarkBlue!14, draw=ACMDarkBlue!35, line width=0.2pt},
    peak/.style={fill=ACMDarkBlue, draw=ACMDarkBlue!80!black, line width=0.2pt},
    meas/.style={{Bar[width=2.6pt]}-{Bar[width=2.6pt]}, line width=0.5pt, draw=black!80},
    lbl/.style={font=\footnotesize, inner sep=1pt}
]
\def\nbins{11}
\def\bw{2.5}
\def\bs{3.1}
\def\bh{4}
\def\ph{7}

\newcommand{\symrow}[4]{%
  \begin{scope}[shift={(0,#1)}]
    \foreach \i in {0,...,\nbins}{ \fill[bin] ({\i*\bs},0) rectangle ({\i*\bs+\bw},\bh); }
    \fill[peak] ({#2*\bs},0) rectangle ({#2*\bs+\bw},\ph);
    \fill[peak] ({#3*\bs},0) rectangle ({#3*\bs+\bw},\ph);
    \draw[meas] ({#2*\bs+\bw/2},{\ph+1.6}) -- ({#3*\bs+\bw/2},{\ph+1.6})
        node[lbl, midway, above] {#4};
  \end{scope}
}

\symrow{26}{3}{5}{Symbol 1}
\symrow{13}{3}{7}{Symbol 2}
\symrow{0}{3}{9}{Symbol 3}

\draw[-{Stealth[length=1.8mm]}, line width=0.6pt] (0,-3) -- ({\nbins*\bs+\bw},-3)
    node[lbl, below left] {time};
\end{tikzpicture}

%% file: diagrams/diagram_message_reconstruction.tex
\begin{tikzpicture}[
    x=1mm, y=1mm,
    bin/.style={fill=ACMDarkBlue!14, draw=ACMDarkBlue!35, line width=0.2pt},
    peak/.style={fill=ACMDarkBlue, draw=ACMDarkBlue!80!black, line width=0.2pt},
    lbl/.style={font=\footnotesize, inner sep=1pt}
]
\def\nbins{9}
\def\bw{2.5}
\def\bs{3.1}
\def\bh{3.2}
\def\ph{5.6}

\newcommand{\mrow}[2]{%
  \begin{scope}[shift={(0,#1)}]
    \foreach \i in {0,...,\nbins}{ \fill[bin] ({\i*\bs},0) rectangle ({\i*\bs+\bw},\bh); }
    \fill[peak] ({2*\bs},0) rectangle ({2*\bs+\bw},\ph);
    \pgfmathsetmacro{\bidx}{2+2*#2}
    \fill[peak] ({\bidx*\bs},0) rectangle ({\bidx*\bs+\bw},\ph);
    \node[lbl, anchor=west] at ({\nbins*\bs+\bw+1.5},{\ph/2}) {\,#2};
  \end{scope}
}

\newcommand{\msgcol}[6]{%
  \begin{scope}[shift={(#1,0)}]
    \node[font=\bfseries\footnotesize, anchor=south west] at (0,42) {#2};
    \mrow{30}{#3}\mrow{20}{#4}\mrow{10}{#5}\mrow{0}{#6}
    \draw[-{Stealth[length=1.6mm]}, line width=0.5pt, draw=black!60]
        (-2.5,{\ph+30}) -- (-2.5,-1) node[lbl, midway, left, rotate=90, anchor=south] {sequence};
  \end{scope}
}

\msgcol{0}{Message A: [1\,2\,1\,3]}{1}{2}{1}{3}
\msgcol{44}{Message B: [3\,2\,3\,2]}{3}{2}{3}{2}
\end{tikzpicture}

%% file: figures/figure_trigger_timing.tex
\begin{figure*}[t]
\centering
\resizebox{0.70\textwidth}{!}{\input{diagrams/diagram_trigger_timing}}
\caption{Trigger timing. The four-symbol trigger is delivered across consecutive per-rotation windows (about 100~ms each), for roughly 400~ms of sustained delivery. Within each window the malware reads one symbol from the dual-return separation.}
\label{fig:trigger-timing}
\end{figure*}

%% file: diagrams/diagram_trigger_timing.tex
\begin{tikzpicture}[
    x=1mm, y=1mm,
    win/.style={draw=ACMDarkBlue!55, fill=ACMDarkBlue!5, rounded corners=1pt, line width=0.7pt},
    bin/.style={fill=ACMDarkBlue!14, draw=ACMDarkBlue!35, line width=0.15pt},
    peak/.style={fill=ACMDarkBlue, draw=ACMDarkBlue!80!black, line width=0.15pt},
    lbl/.style={font=\footnotesize, inner sep=1pt},
    call/.style={font=\footnotesize, inner sep=1pt},
    meas/.style={{Bar[width=2.2pt]}-{Bar[width=2.2pt]}, line width=0.5pt, draw=black!75}
]
\def\nb{8}     
\def\bw{1.7}   
\def\bs{2.4}   
\def\bh{2.6}   
\def\ph{6}     
\def\ww{18}    
\def\wpad{2}   
\def\wg{6}     
\pgfmathsetmacro{\wwidth}{\nb*\bs+\bw+2*\wpad}   
\def\wh{11}    

\foreach \k/\val in {0/1, 1/2, 2/1, 3/3}{  
    \pgfmathsetmacro{\x}{\k*(\wwidth+\wg)}
    \node[win, minimum width=\wwidth mm, minimum height=\wh mm, anchor=south west] at (\x,0) {};
    \begin{scope}[shift={(\x+\wpad,2)}]
        \foreach \i in {0,...,\nb}{ \fill[bin] ({\i*\bs},0) rectangle ({\i*\bs+\bw},\bh); }
        \fill[peak] ({1*\bs},0) rectangle ({1*\bs+\bw},\ph);
        \pgfmathsetmacro{\bidx}{1+2*\val}
        \fill[peak] ({\bidx*\bs},0) rectangle ({\bidx*\bs+\bw},\ph);
    \end{scope}
    \node[lbl, anchor=north] at ({\x+\wwidth/2},-1) {Symbol \val};
}

\draw[meas] (0,-6) -- (\wwidth,-6) node[call, midway, below] {$\approx$100\,ms (1 rotation)};

\pgfmathsetmacro{\total}{4*\wwidth+3*\wg}
\draw[decorate, decoration={brace, amplitude=5pt}, line width=0.6pt] (0,{\wh+2}) -- (\total,{\wh+2});
\node[font=\small, anchor=south] at ({\total/2},{\wh+4}) {4-symbol trigger \;($\approx$400\,ms sustained)};

\draw[-{Stealth[length=1.8mm]}, line width=0.6pt] (0,-15) -- ({\total+3},-15) node[lbl, anchor=north east, yshift=-0.5mm] {time};
\end{tikzpicture}

%% file: 6-feasibility.tex
\section{Attack Feasibility and Detectability}
\label{sec:evaluation}

The attack must satisfy the two conditions of Section~\ref{sec:vulnerability}: the malware must be triggerable after deployment, and its activation must produce safety-critical effects. This section establishes the first, trigger feasibility across sampled indoor, outdoor, and drive-by settings (Section~\ref{sec:evaluation-feasibility}), and takes up the second at the perception level, showing that injected output is taken up by a downstream detector (Section~\ref{sec:evaluation-detectability}). Section~\ref{sec:impact} completes the second on a deployed vehicle.

\subsection{Protocol and Instrumentation}
\label{sec:evaluation-setups}

\input{figures/figure_victim_platform}
\input{figures/figure_modulating_laser}
\input{figures/figure_experiment_setups}

The experiments below share a common setup. The victim is an Ouster OS1-128 LiDAR running custom firmware that contains the Trojan malware, co-mounted with a camera on a shared rigid platform (Fig.~\ref{fig:victim-platform}) so their relative transformation stays fixed. This custom firmware was developed for research purposes with technical support from Ouster engineers within a closed test environment and does not represent inherent vulnerabilities in the standard production supply chain. The camera is unmodified and serves only to record reference views of the scene; it is never a target of attack. Both sensors connect through the LiDAR interface box and an Ethernet switch to a data recorder. The attacker illuminates this platform with a handheld optical emitter and never communicates with the victim over any network.

All manipulation occurs inside the compromised LiDAR sensing pipeline, after signal processing and before point cloud publication, so downstream perception receives the altered point cloud as ordinary sensor output.

\myparagraph{Mobile mounting.}
For the mobile experiments, the shared platform was mounted on the roof rails of a vehicle (Fig.~\ref{fig:appendix-setup-outdoor-mobile}) to approximate a realistic deployed viewpoint, while keeping the recorded sensing stack isolated from any vehicle feedback-control function for driver safety.

\myparagraph{Attack effects.}
We evaluate three attack effects: \emph{error induction}, which forces the sensor into a fault state; \emph{injection}, which adds person-like false returns; and \emph{suppression}, which removes genuine returns. Injection and suppression are characterized here across feasibility conditions; all three are demonstrated on the deployed vehicle in Section~\ref{sec:impact}.

The handheld optical emitter (Fig.~\ref{fig:modulating-laser}) is built from a modified LiDAR-based optical source configured to transmit along a fixed direction rather than rotate; a microcontroller gates emission timing, firmware implements the coded pulse sequence, and a small keyboard extension allows the attacker to select the desired attack mode at runtime.

\myparagraph{Trigger-code implementation parameters.}
Trigger parameters follow Section~\ref{sec:activation}: four code symbols sampled one per LiDAR rotation, for a total duration of roughly 400~ms. Fixing the encoding left open there, a symbol is represented by a dual-return peak separation of approximately 10--20~ns.

\subsection{Static and Mobile Feasibility}
\label{sec:evaluation-feasibility}

We characterize whether suppression and injection can be activated and executed across sampled static and mobile settings. A successful trial means the malware activated and altered LiDAR output in the intended direction of the selected effect; for injection trials, we additionally record whether the injected person-like artifacts are taken up as person-class detections by the verification detector of Appendix~\ref{appendix:detector}.

\myparagraph{Setup classes.}
We evaluate three setup classes: indoor static (residential and commercial buildings, including cross-floor geometries), outdoor static, and outdoor mobile drive-by. Fig.~\ref{fig:experiment-setups} summarizes the geometry and attacker-victim arrangement used for each class; Appendix~\ref{appendix:protocol} provides per-class setup composites with photographs of the real test environments together with qualitative galleries. Reported distances are derived from coarse measurements, and reported vehicle speeds are approximate values read from the host vehicle speedometer during drive-by runs.

\myparagraph{Direct and indirect delivery.}
Triggering succeeded with both direct illumination and reflected delivery paths (for example, wall and ceiling reflections). Depending on the scenario, the operator either aimed the emitter approximately toward the victim sensor or deliberately directed it toward reflective surfaces to test indirect delivery. In sampled trials, delivery did not require precise aiming at the sensor aperture.

\input{figures/figure_main_triptych_feasibility}

\myparagraph{Recorded evidence.}
Fig.~\ref{fig:main-feasibility-triptych} shows representative camera and before/after LiDAR views under the schema \emph{Camera | LiDAR Before Attack | LiDAR After Attack}: injection adds a coherent person-like cluster at the targeted range, and suppression removes the targeted returns, leaving apparent free space. Table~\ref{tab:feasibility-configurations} summarizes the sampled configurations and outcomes across indoor static, outdoor static, and mobile settings. These present a subset curated to show the diversity of conditions spanned, not an exhaustive log; they are drawn from a larger set of activation trials, all of which ultimately activated the malware. Appendix~\ref{appendix:protocol} extends the gallery with further static and mobile cases (Figs.~\ref{fig:appendix-static-triptych} and~\ref{fig:appendix-mobile-triptych}), and accompanying videos are provided through the \ArtifactWebsiteLink{artifact website}.

\input{tables/table_feasibility_configurations.tex}

\myparagraph{Static sensor.}
We evaluated activation on a static victim sensor in both indoor and outdoor configurations. Across these conditions, we observed on-demand activation and intended LiDAR-output alteration in same-floor indoor scenes at approximately 150~ft, via wall and ceiling reflection-driven triggering, across stairwell-separated floors, and in outdoor operation at approximately 200~ft and 300~ft standoffs with modest elevation offsets from road slopes. Both suppression and injection were demonstrated in each environment. Figs.~\ref{fig:experiment-setups-indoor} and~\ref{fig:experiment-setups-tripod} illustrate the indoor and outdoor arrangements.

\myparagraph{Mobile drive-by.}
To evaluate robustness to victim motion, we mounted the compromised sensor on the roof rails of a vehicle and executed repeated drive-by passes on urban campus and suburban neighborhood roads. In these tests, the attacker remained at roadside positions while the victim vehicle drove past at varying speeds, and the attacker fired at various angles and positions throughout the experiments. The choice of a static attacker was a simplification for experimental control rather than a requirement of the attack. Across these drive-by trials, we observed successful triggering and intended attack execution while the victim vehicle was in motion, with recorded runs reaching approximately 35~mph. The corresponding mobile geometry is shown in Fig.~\ref{fig:experiment-setups-driveby}.

\myparagraph{Observed failure mode.}
In outdoor mobile testing, we observed occasional unintended activation under ambient optical conditions. This is a direct consequence of the trigger message entropy tradeoff and the randomness of blackbody solar emissions. We did not observe this indoors, and revisit the robustness caveat in Section~\ref{sec:discussion}.

\subsection{Semantic Detectability of Injections}
\label{sec:evaluation-detectability}

We examine whether injected person-like artifacts are taken up at the perception level as person-class detections by a fine-tuned 3D object detector, and whether their point density instead betrays them as synthetic to a low-level check.

\myparagraph{Detector adaptation.}
We selected FCAF3D~\cite{rukhovich2022fcaf3d} for its strong performance on indoor LiDAR benchmarks, initialized it from an open-source ScanNet model~\cite{dai2017scannet}, and adapted it to OS1-128 person detection on a custom person-class dataset of 44 runs and 7{,}438 curated scene samples built with an automated labeling pipeline. Table~\ref{tab:detector-verification-summary} summarizes the dataset and the detector's quality on the benign person-detection task after fine-tuning; Appendix~\ref{appendix:detector} details the dataset-construction and adaptation procedure.

\input{figures/figure_detector_qual_composite}
\input{tables/table_detector_verification_summary.tex}

\myparagraph{Detection under injection.}
With the detector fine-tuned, we then applied it to manipulated point clouds to test attack-time person detection. Figure~\ref{fig:detector-qual-composite} shows a representative replayed inference sequence consisting of a pre-attack baseline frame followed by two attack-time frames, illustrating how the injected person-like artifact emerges and is tracked as a person-class detection longitudinally. This replayed example comes from a significantly different environment than the scenes used in the curated detector-adaptation set. We treat this as bounded semantic verification that the injected artifact survives one downstream perception stage, not a quantitative adversarial benchmark (Appendix~\ref{appendix:detector}).

To assess whether this behavior is specific to a single recording, we replayed a second, independent attack sequence captured in a different environment. In both sequences the injected artifact is taken up as a person-class detection throughout the attack: it is detected in every frame of its active walk phases in the first sequence and in 6{,}148 of 8{,}006 frames (77\%) of the second, longer recording. Its mean detection score is consistent across the two sequences (0.35 and 0.37) and sits above the detector's median score of 0.21 on genuine person detections in the same traces, that is, comparable to a real detection rather than a high-confidence one.

\myparagraph{Point-density consistency.}
Beyond detector uptake, we asked whether the injected returns are separable from genuine returns using point density, the most immediate low-level check available to a downstream pipeline. Within this scene, the injected object carries the point density of a real object at its range: its nearest-neighbor point spacing is 1.3~cm at a range of 2.0~m, scaling with range in the same proportion as the real pedestrians' 2.3~cm spacing at 3.7~m, and its local volumetric density falls within five percent of theirs. Because the malware synthesizes the injected returns directly, such low-level statistics are attacker-controlled: point density, and other per-return properties, can be shaped to match the surrounding scene, so a density-based filter has no reliable separating signal against a deliberately constructed injection.

%% file: figures/figure_victim_platform.tex
\begin{figure}[t]
    \centering
    \begin{subfigure}[b]{0.72\linewidth}
        \centering
        \begin{tikzpicture}
            \node[anchor=south west,inner sep=0] (rig)
                {\includegraphics[width=\linewidth,trim=3.539cm 6.452cm 0.423cm 2.498cm,clip]{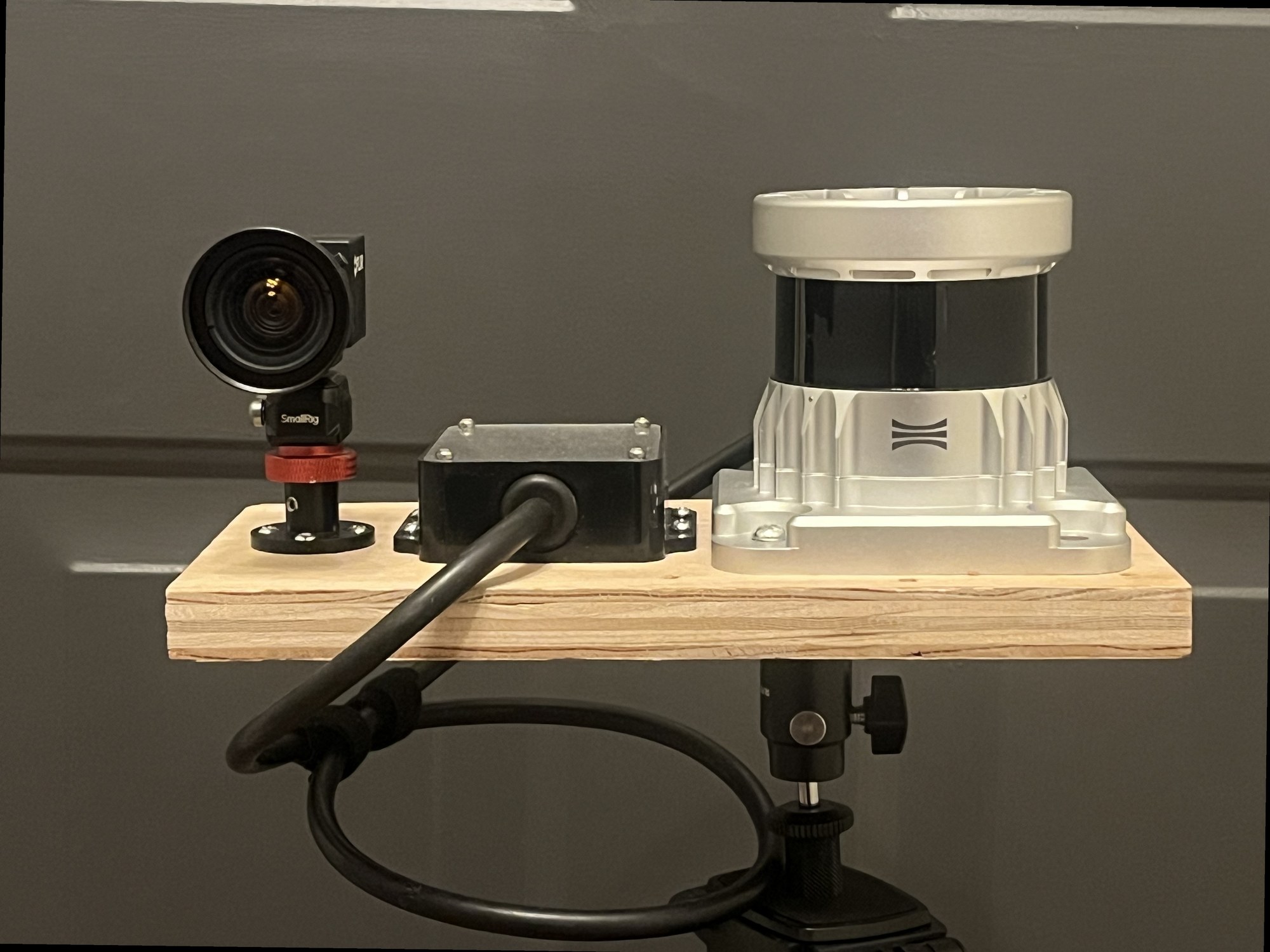}};%
            \begin{scope}[x={(rig.south east)},y={(rig.north west)},
                          tag/.style={anchor=base west, text=ACMDarkBlue, font=\sffamily\scriptsize\bfseries,
                                      fill=white, fill opacity=0.78, text opacity=1,
                                      rounded corners=1pt, inner sep=1.5pt},
                          point/.style={ACMDarkBlue, line width=0.7pt, -{Stealth[length=1.8mm]}}]
                \node[tag] at (0.02,0.92) {camera};%
                \draw[point] (0.080,0.905) -- (0.130,0.805);%
                \node[tag] at (0.31,0.92) {interface box};%
                \draw[point] (0.385,0.905) -- (0.370,0.505);%
                \node[tag] at (0.86,0.60) {LiDAR};%
                \draw[point] (0.855,0.615) -- (0.828,0.657);%
            \end{scope}
        \end{tikzpicture}
        \caption{Sensor platform (tripod, static trials).}
        \label{fig:victim-platform-bench}
    \end{subfigure}

    \vspace{5pt}

    \begin{subfigure}[b]{0.72\linewidth}
        \centering
        \includegraphics[width=\linewidth,trim=0.7cm 1.5cm 0.7cm 0.5cm,clip]{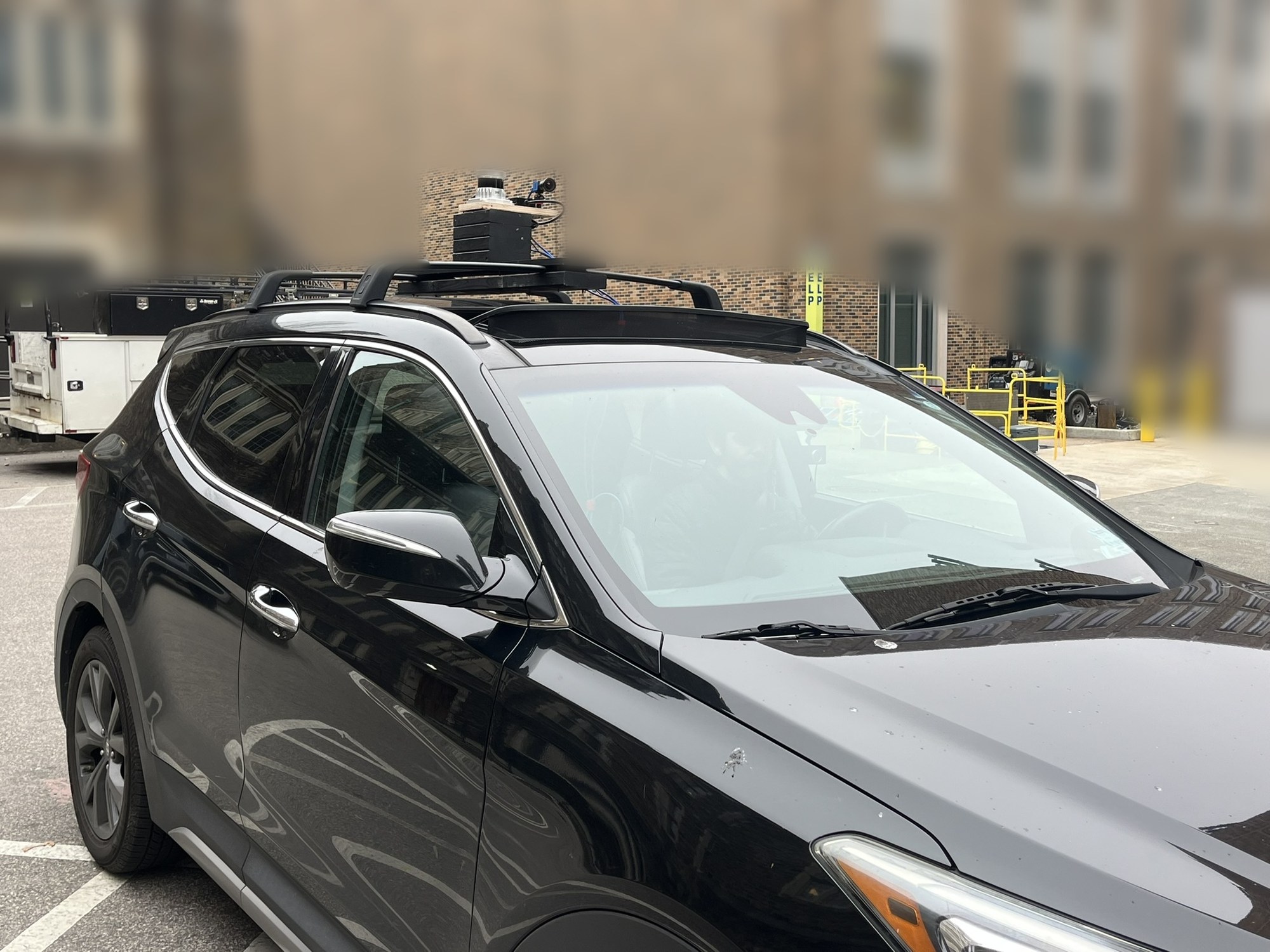}
        \caption{Deployed on the vehicle roof (mobile trials).}
        \label{fig:victim-platform-vehicle}
    \end{subfigure}
    \caption{The victim sensor platform. (a) An Ouster OS1-128 LiDAR (right) is co-mounted with a reference camera (left) and the LiDAR interface box (center) on a shared rigid board, shown on the tripod used for the static trials. (b) The same assembly mounted to the roof of a passenger vehicle for the mobile drive-by trials. The camera records an independent visual reference and is never a target of attack.}
    \label{fig:victim-platform}
\end{figure}

%% file: figures/figure_modulating_laser.tex
\begin{figure}[!t]
    \centering
    \includegraphics[width=0.47\linewidth,height=0.279\linewidth,trim=130bp 0 50bp 0,clip]{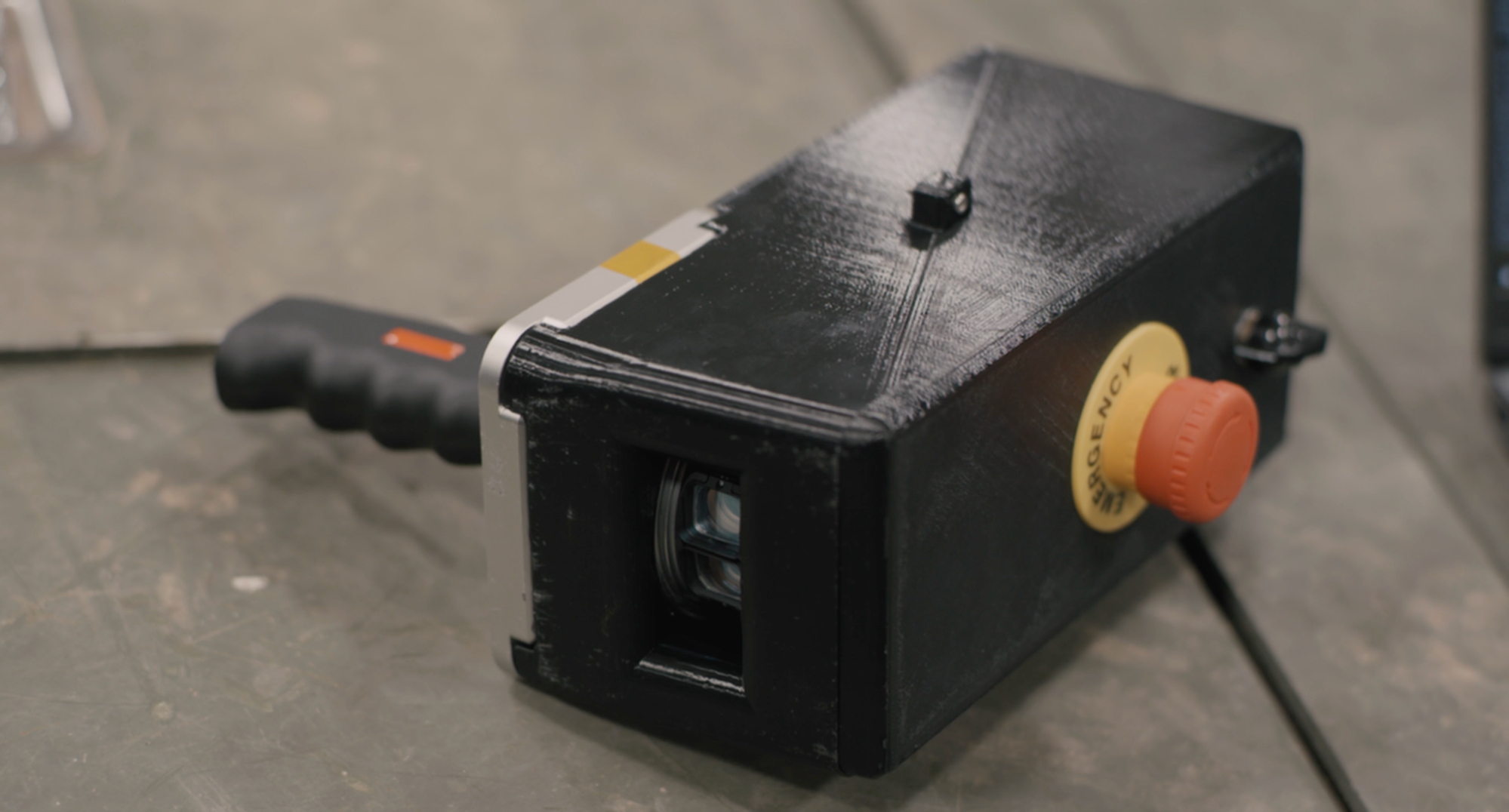}
    \hfill
    \includegraphics[width=0.47\linewidth,height=0.279\linewidth,trim=500bp 112bp 0 112bp,clip]{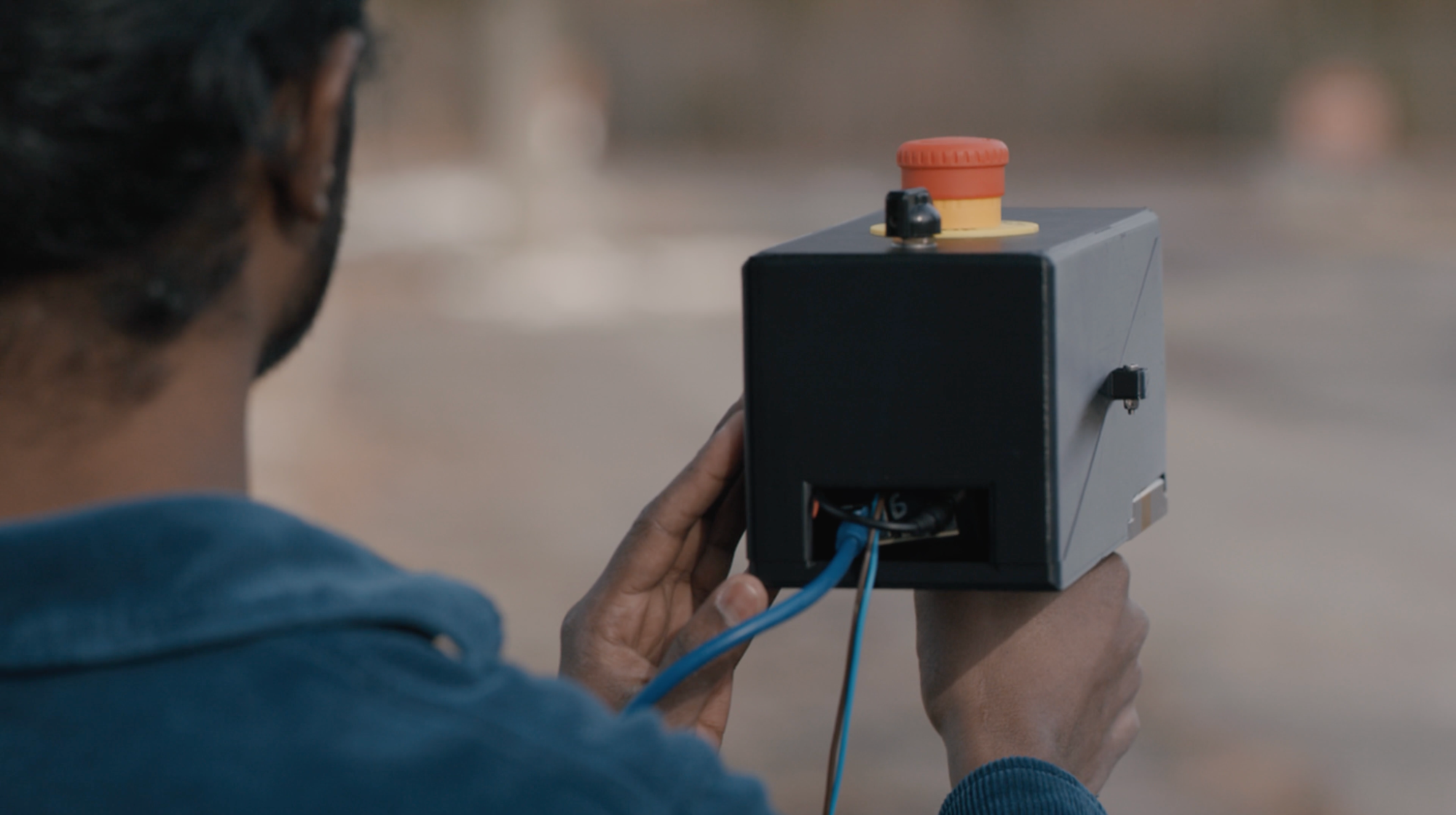}
    \caption{Handheld optical trigger hardware used in the experiments. The emitter is built from a modified commodity optical source configured to transmit along a fixed direction, and a small key interface selects among preconfigured attack modes at runtime. No exotic hardware is required, and the attacker never communicates with the victim over any network.}
    \label{fig:modulating-laser}
\end{figure}

%% file: figures/figure_experiment_setups.tex
\begin{figure*}[t]
    \centering
    \begin{subfigure}[t]{0.32\linewidth}
        \centering
        \includegraphics[width=\linewidth,trim=0 0 0 0,clip]{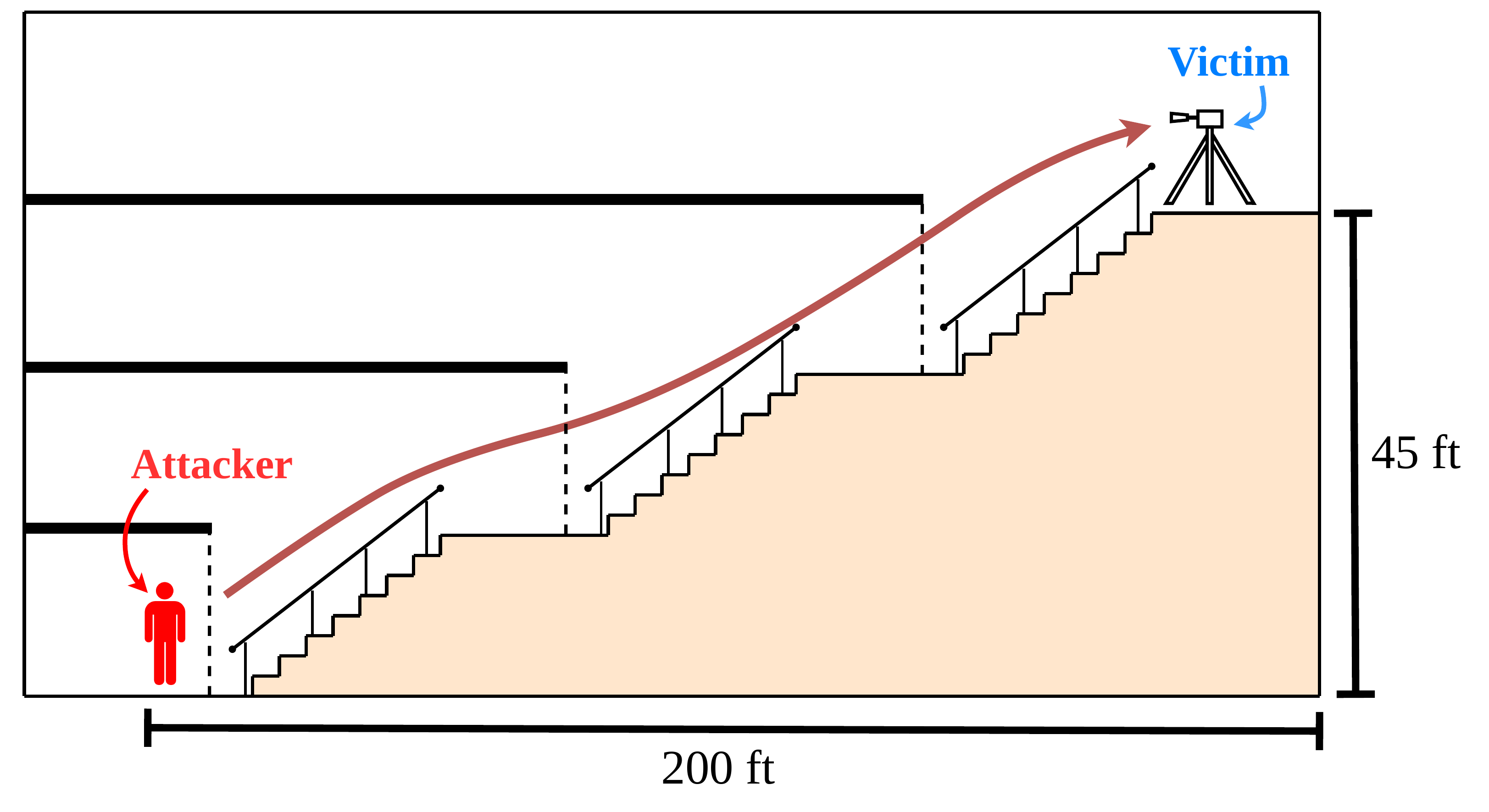}
        \caption{Indoor static setup used for residential and commercial building feasibility trials.}
        \label{fig:experiment-setups-indoor}
    \end{subfigure}
    \hfill
    \begin{subfigure}[t]{0.32\linewidth}
        \centering
        \includegraphics[width=\linewidth,trim=0 0 0 0,clip]{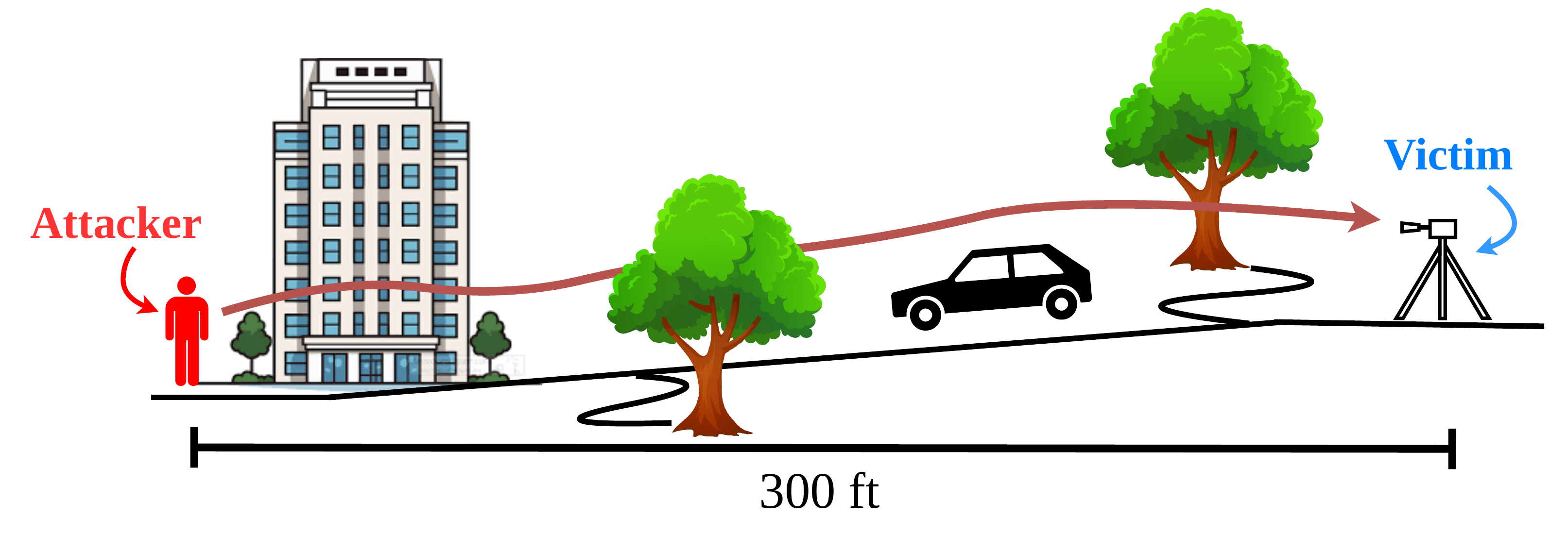}
        \caption{Outdoor static setup used for long-range trigger delivery, including 200 and 300~ft conditions.}
        \label{fig:experiment-setups-tripod}
    \end{subfigure}
    \hfill
    \begin{subfigure}[t]{0.32\linewidth}
        \centering
        \includegraphics[width=\linewidth,trim=0 0 0 0,clip]{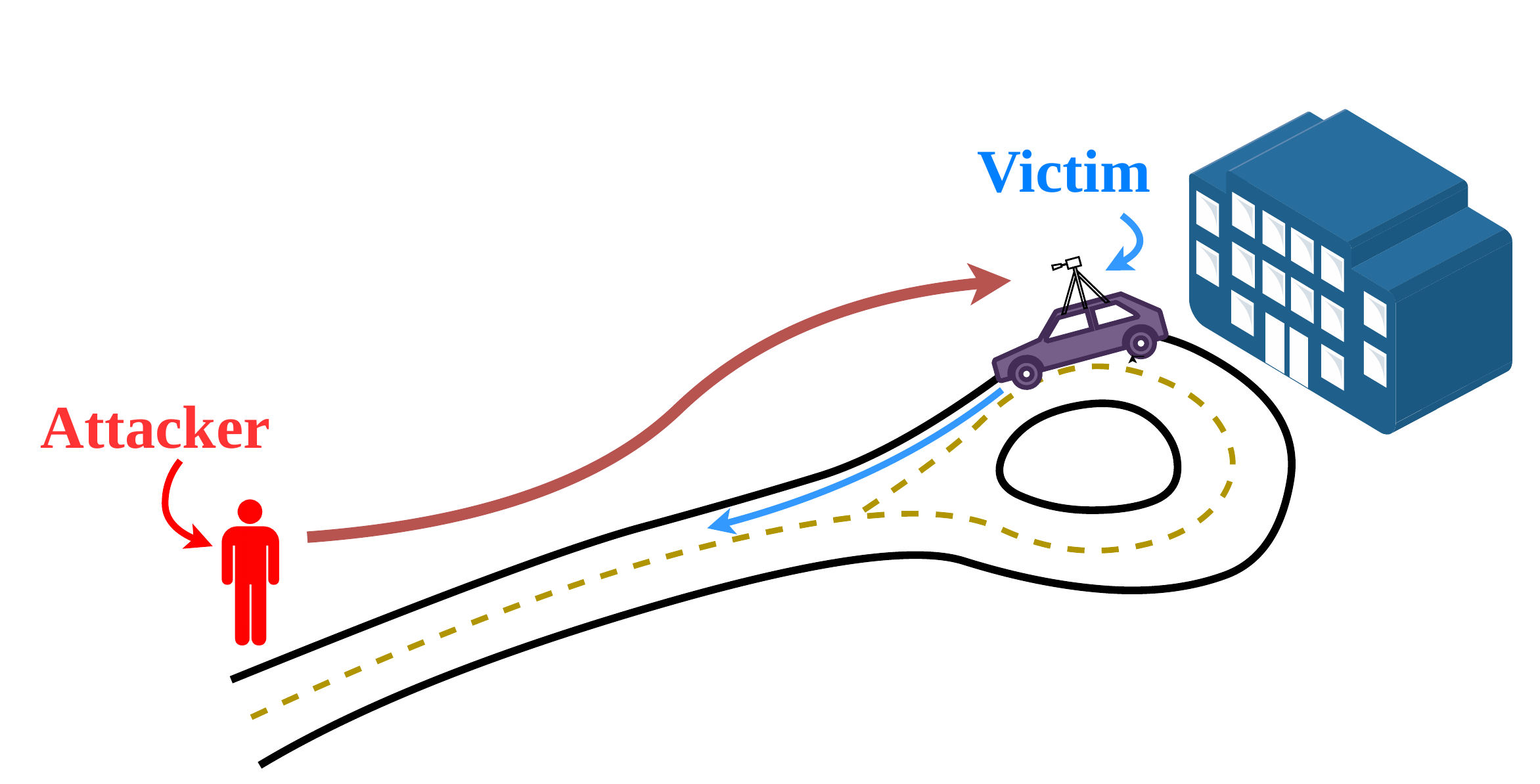}
        \caption{Mobile-sensor drive-by setup used to evaluate triggering under victim motion.}
        \label{fig:experiment-setups-driveby}
    \end{subfigure}
    \caption{Experimental setup diagrams for the three feasibility-envelope classes. Indoor (a) spans same-floor direct and reflected (wall and ceiling) paths and cross-floor delivery in residential and commercial buildings; outdoor (b) is the long-range tripod configuration evaluated at 200 and 300~ft; mobile (c) mounts the victim on a moving vehicle to test triggering under motion. Each diagram fixes the attacker and victim geometry; photographs of the corresponding environments appear in Appendix~\ref{appendix:protocol}.}
    \label{fig:experiment-setups}
\end{figure*}

%% file: figures/figure_main_triptych_feasibility.tex
\begin{figure*}[t]
    \centering
    \setlength{\tabcolsep}{4pt}
    \begin{tabular}{>{\centering\arraybackslash}m{0.14\textwidth} >{\centering\arraybackslash}m{0.27\textwidth} >{\centering\arraybackslash}m{0.27\textwidth} >{\centering\arraybackslash}m{0.27\textwidth}}
        \toprule
        \textbf{Scenario} & \textbf{Camera} & \textbf{LiDAR Before Attack} & \textbf{LiDAR After Attack} \\
        \midrule
        \textbf{\textit{Suppression}} Indoor wall reflection &
        \TriptychCamera{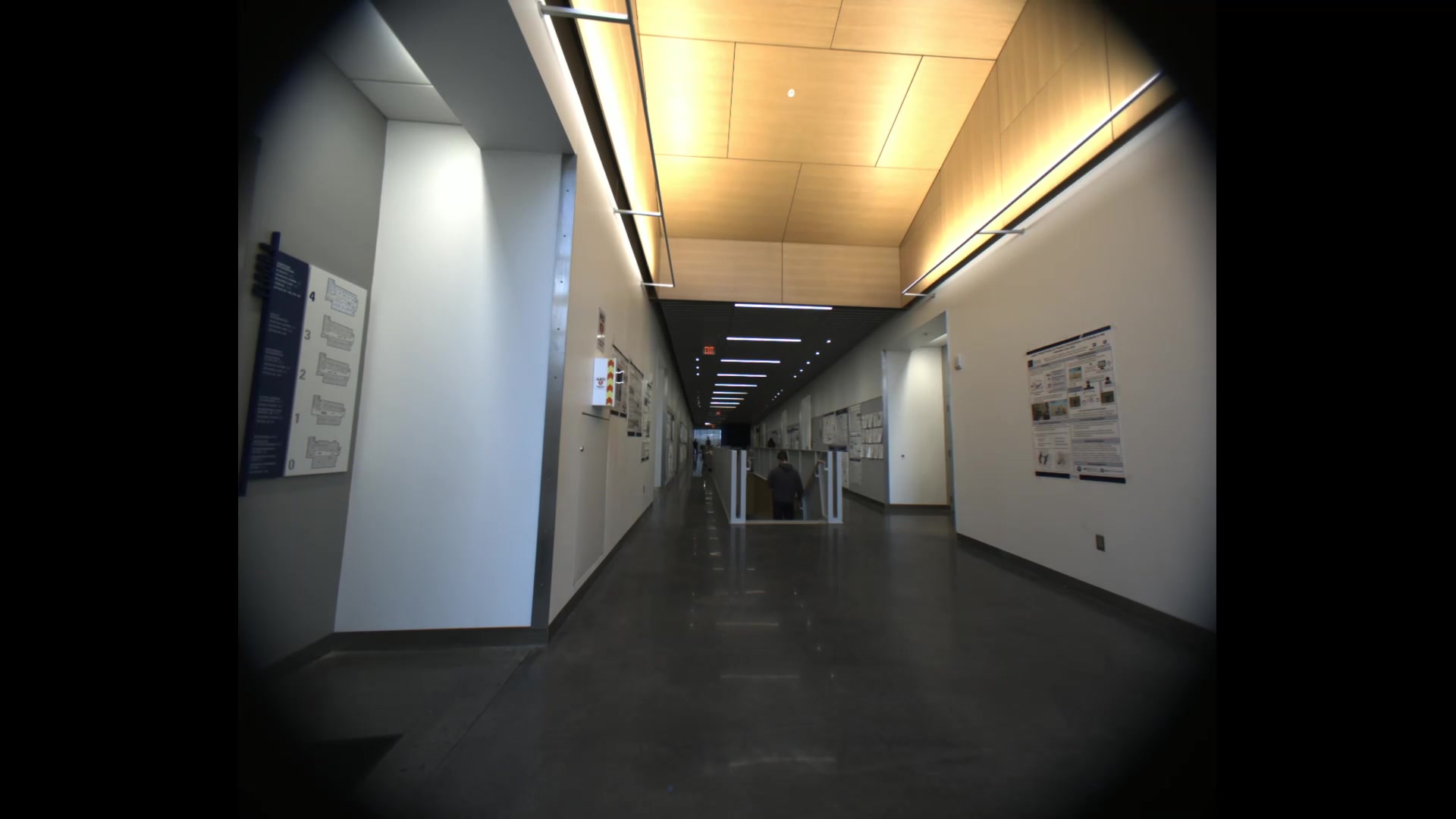} &
        \TriptychLidar{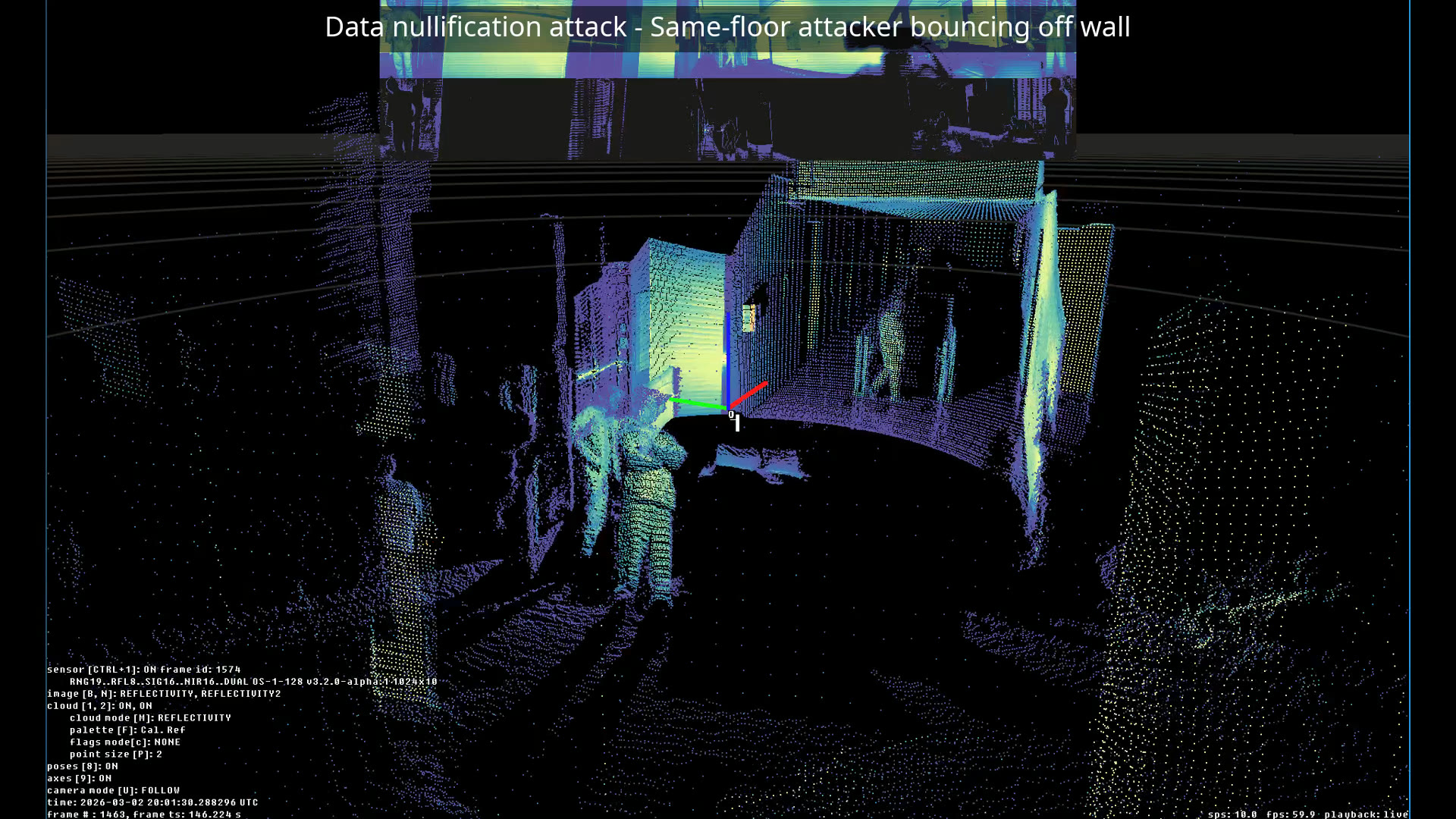} &
        \TriptychLidarSup{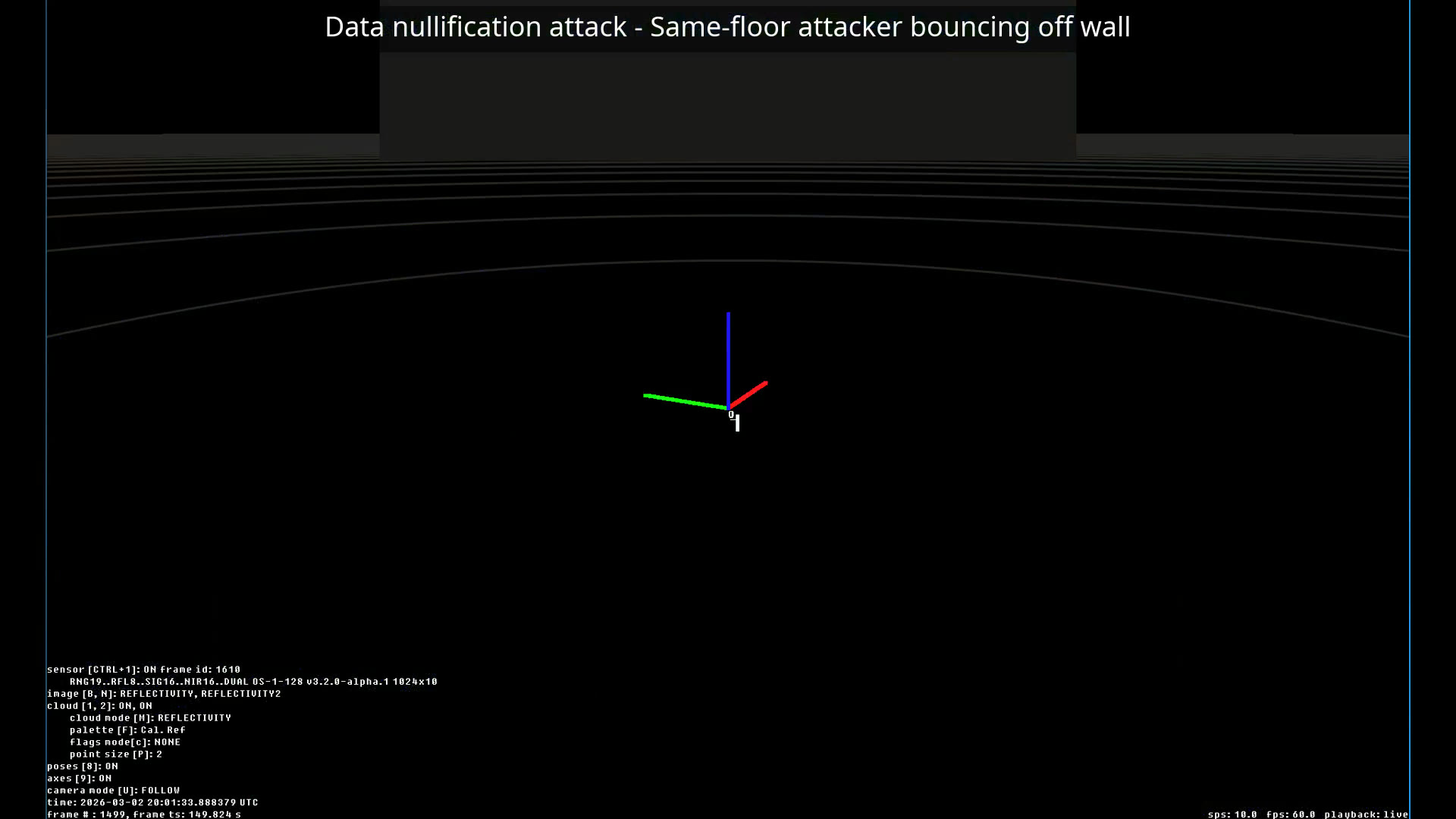} \\

        \textbf{\textit{Injection}} Indoor stairwell (1st floor attacker) &
        \TriptychCamera{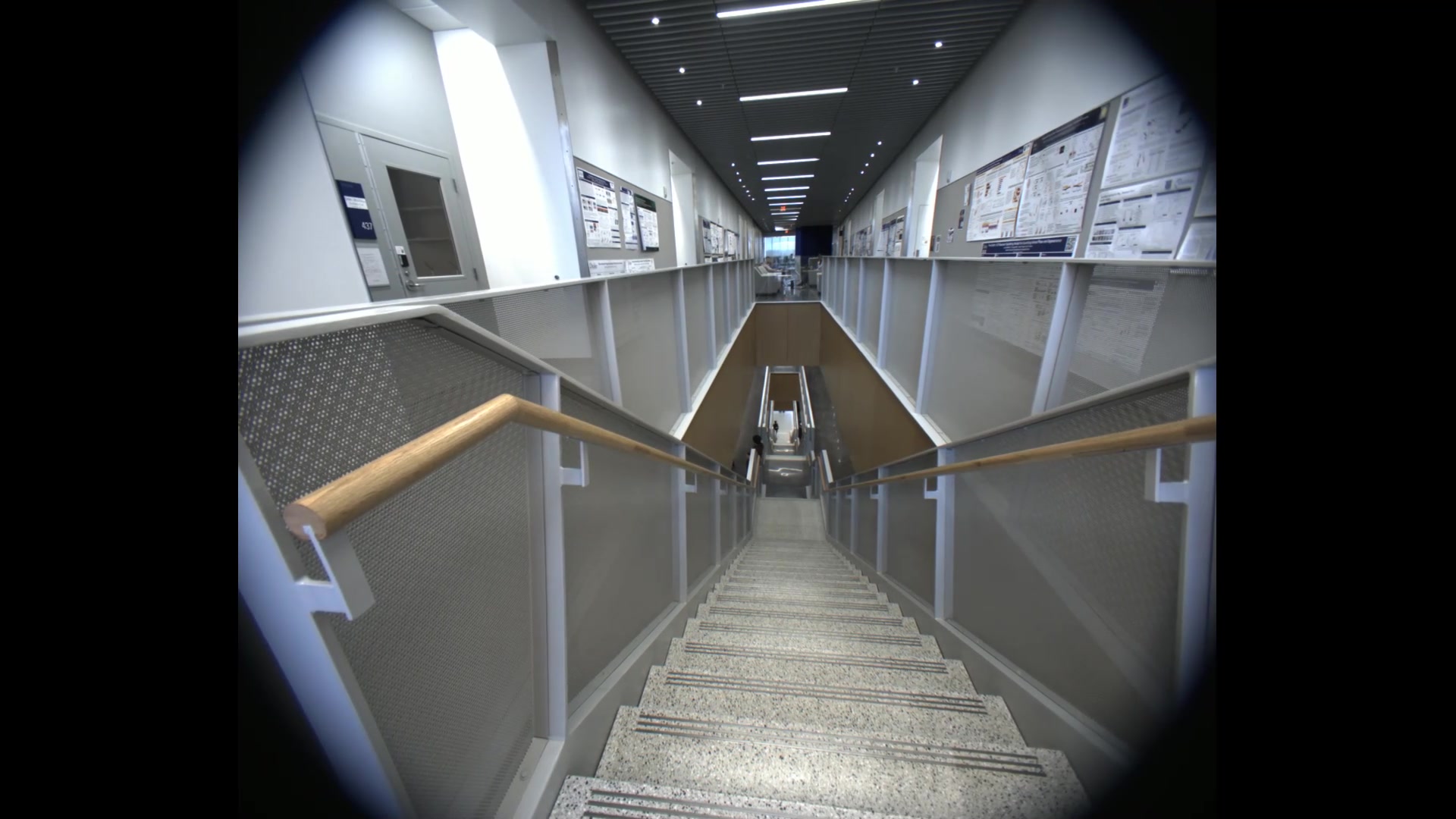} &
        \TriptychLidar{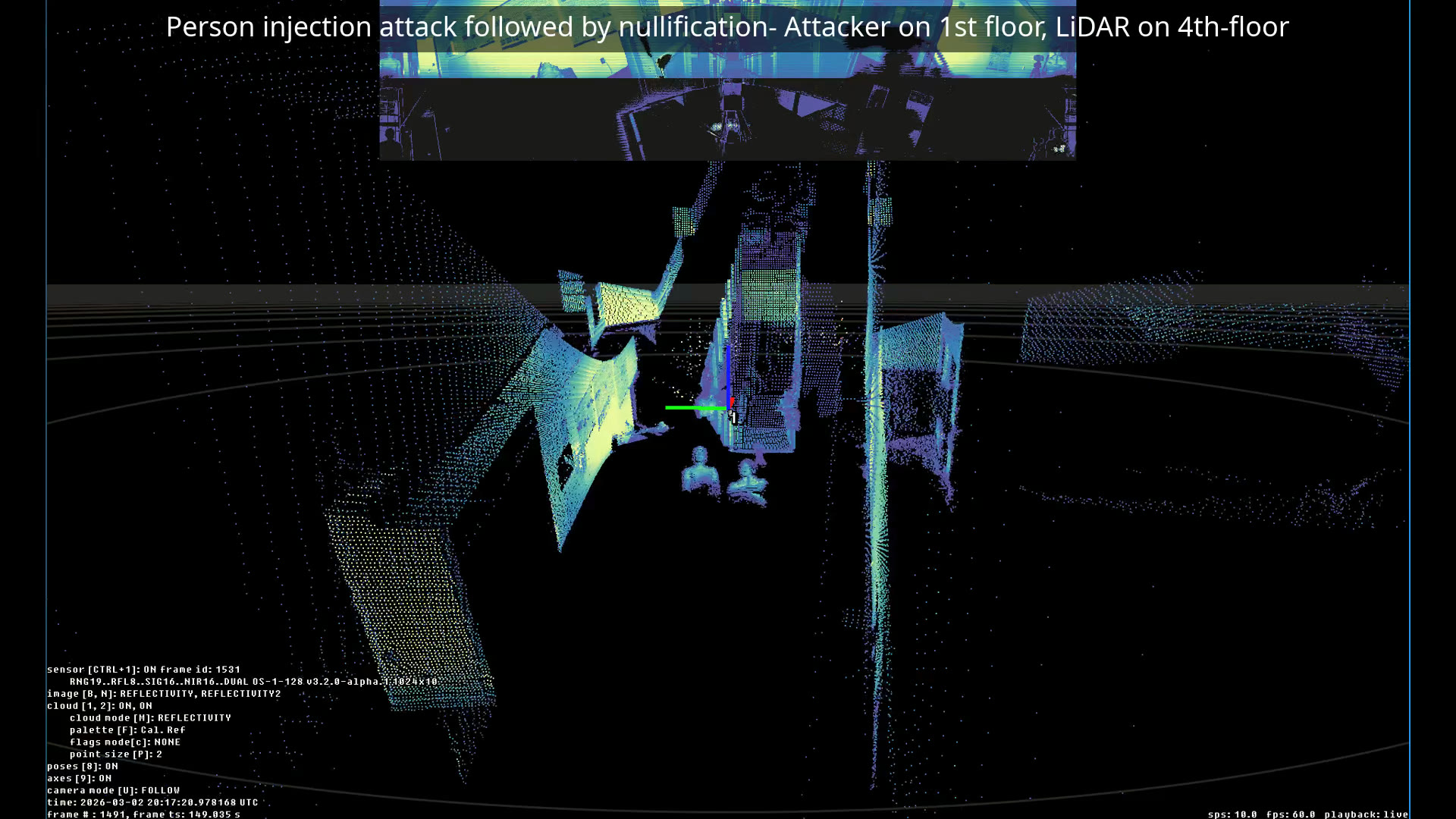} &
        \TriptychLidarInj{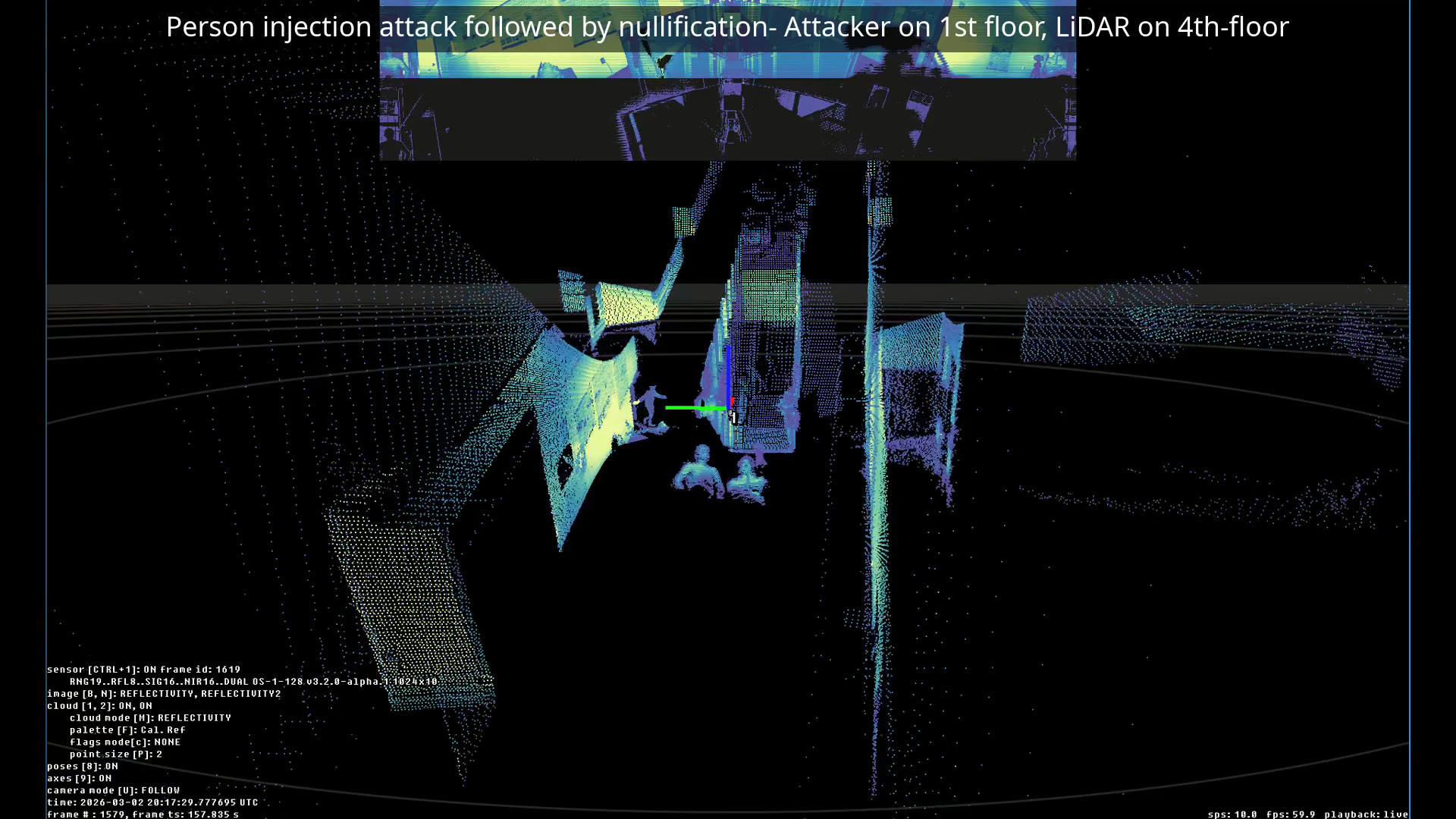}{0.232}{0.375}{0.149}{0.181} \\

        \textbf{\textit{Suppression}} Outdoor static (300~ft) &
        \TriptychCamera{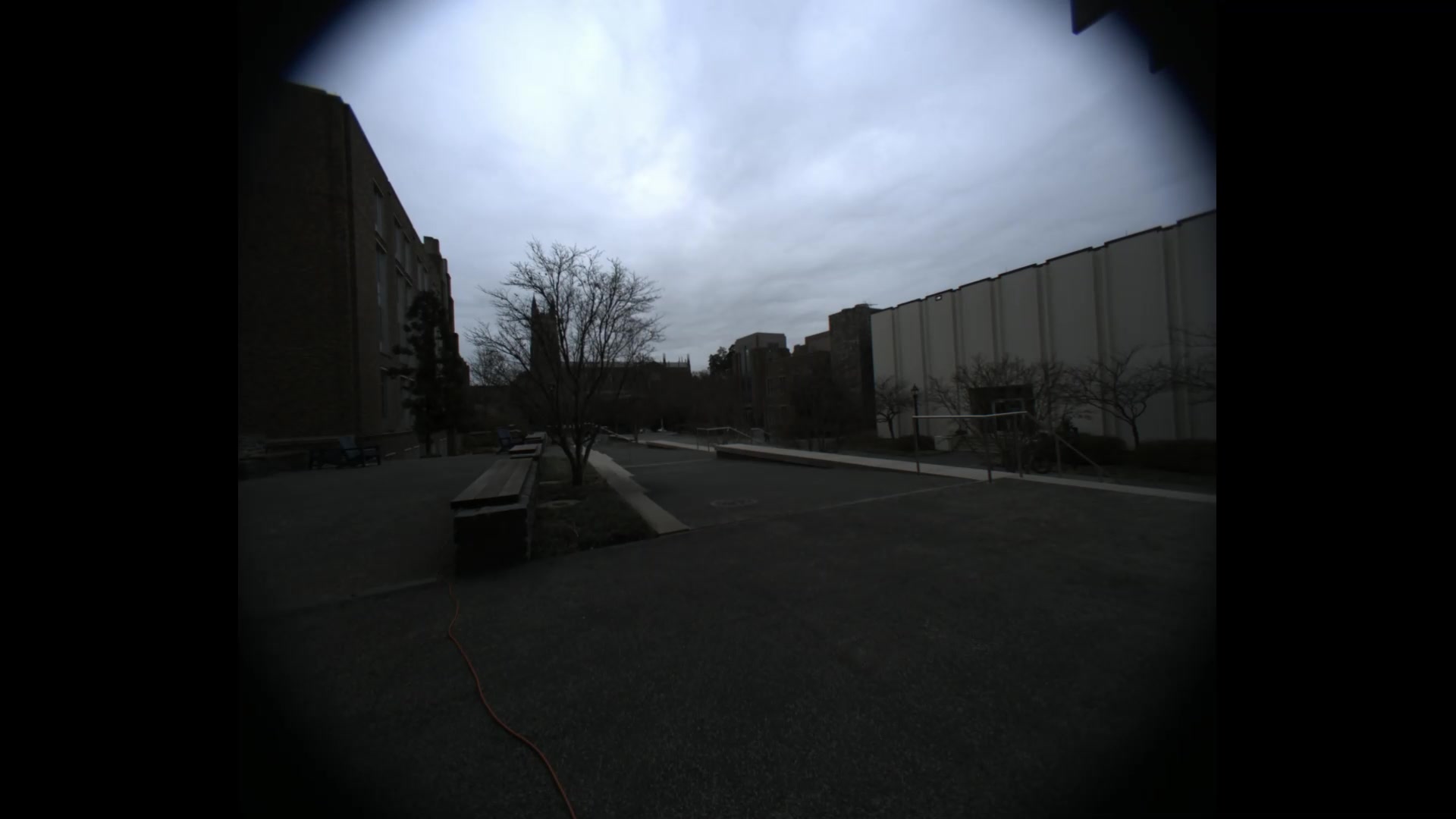} &
        \TriptychLidar{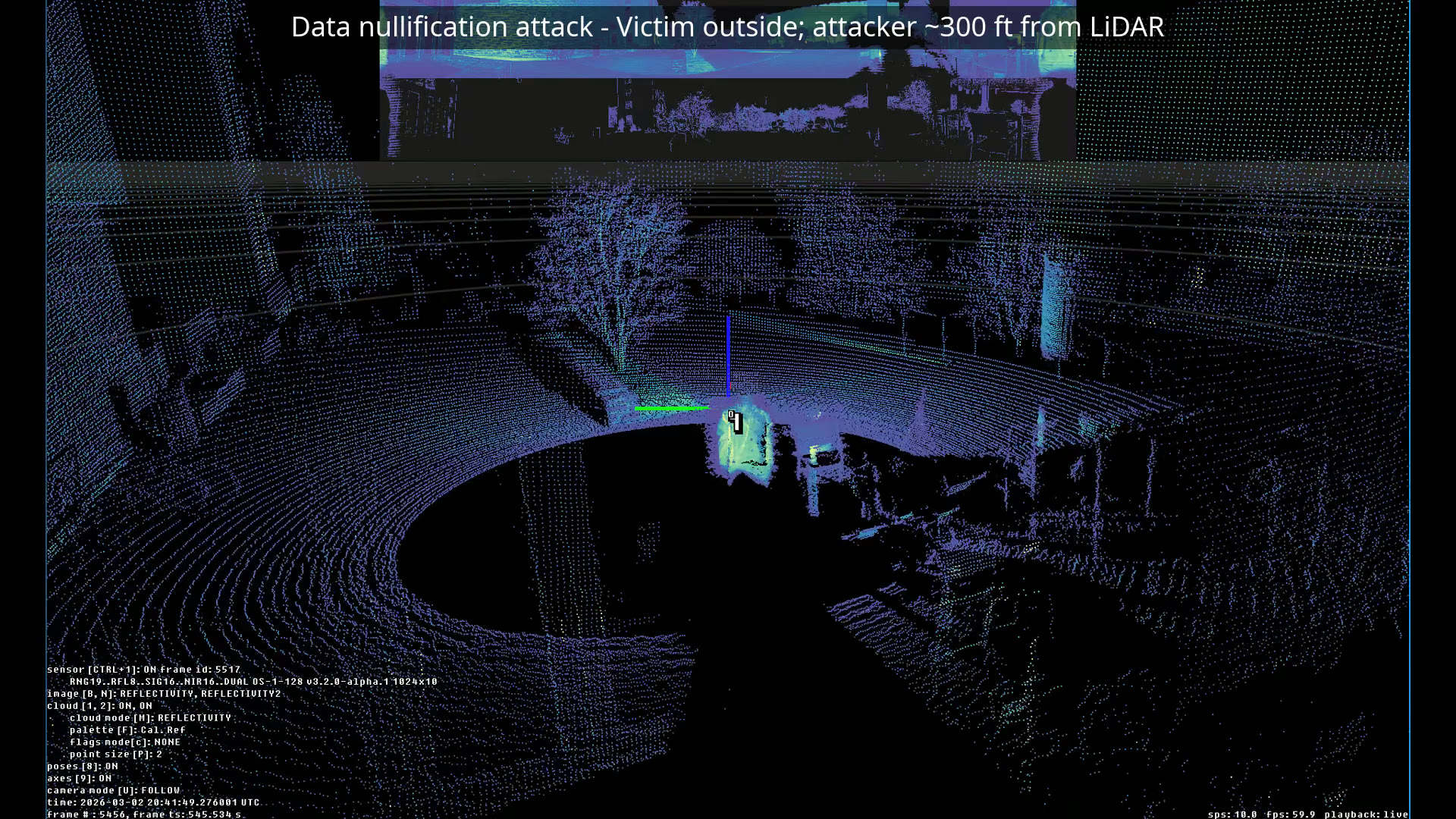} &
        \TriptychLidarSup{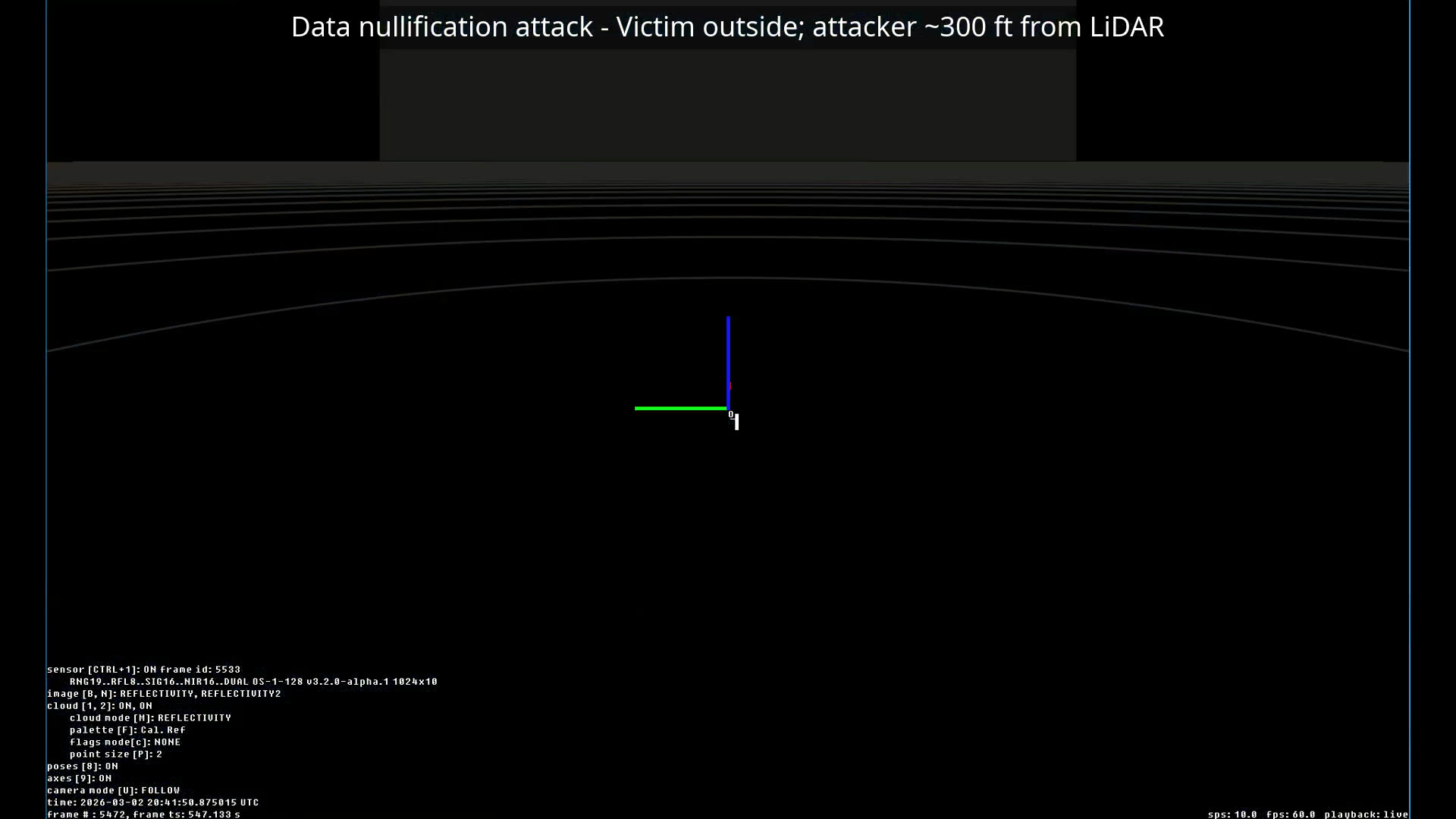} \\

        \textbf{\textit{Injection}} Drive-by &
        \TriptychCamera{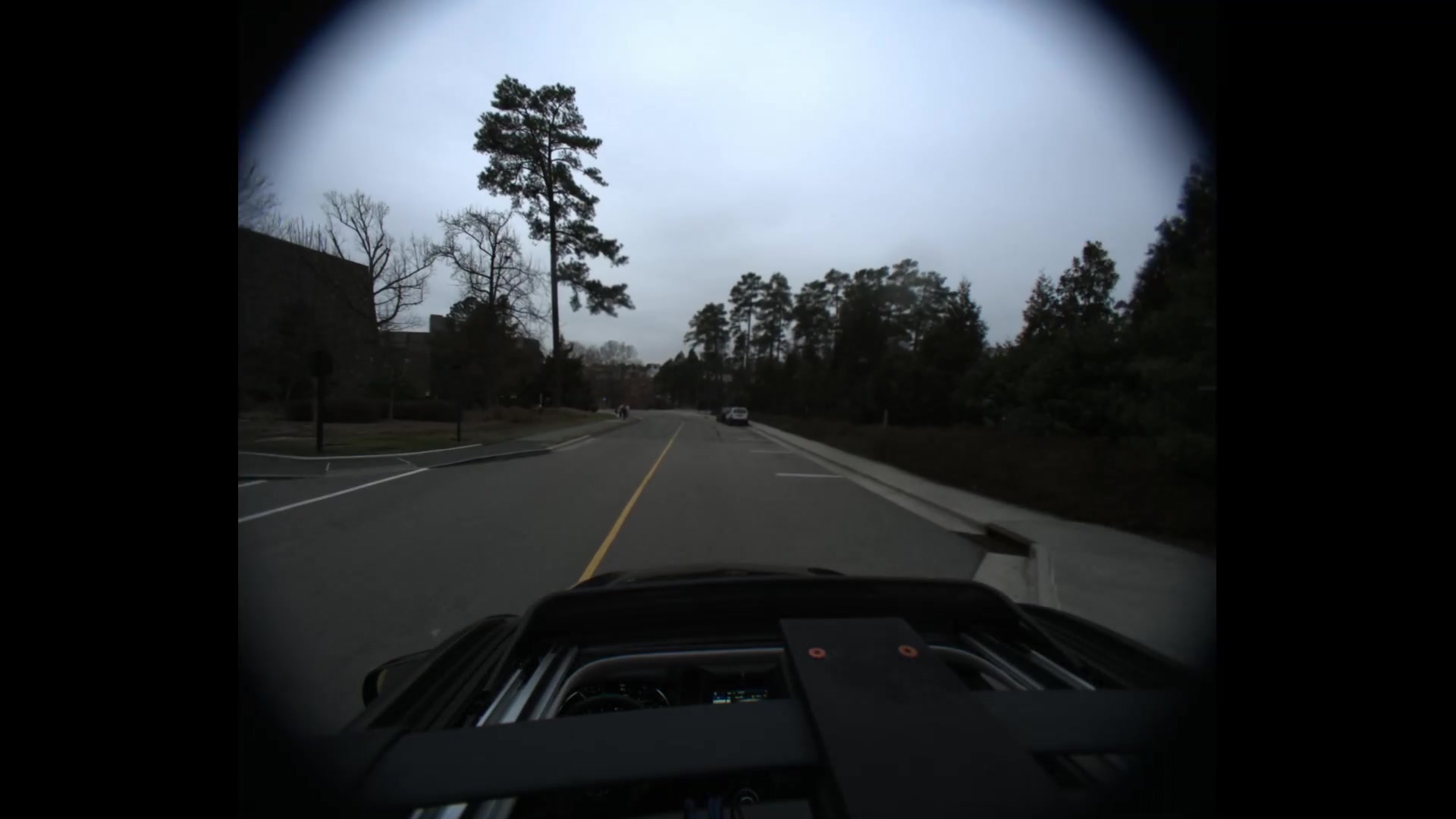} &
        \TriptychLidar{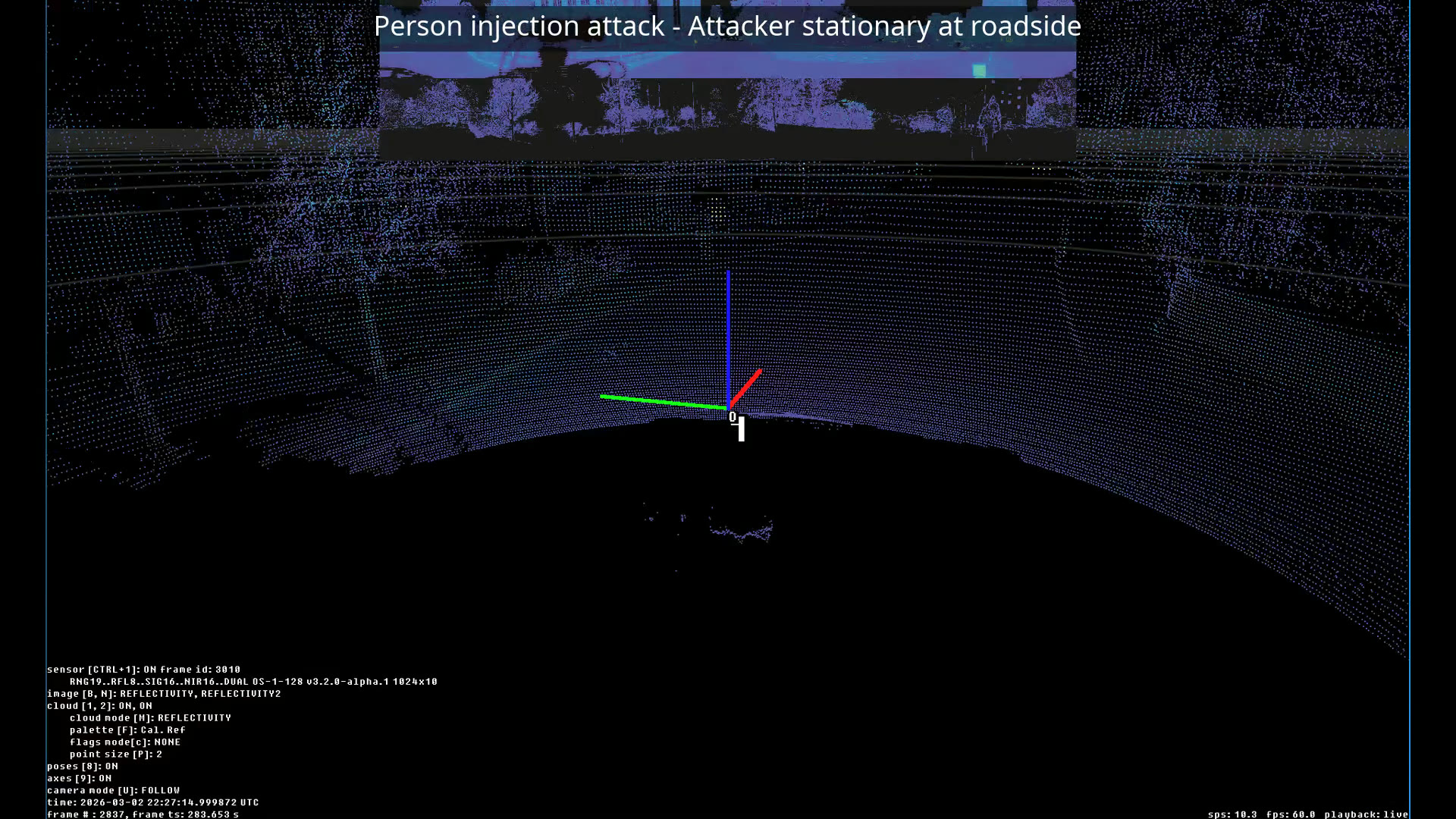} &
        \TriptychLidarInj{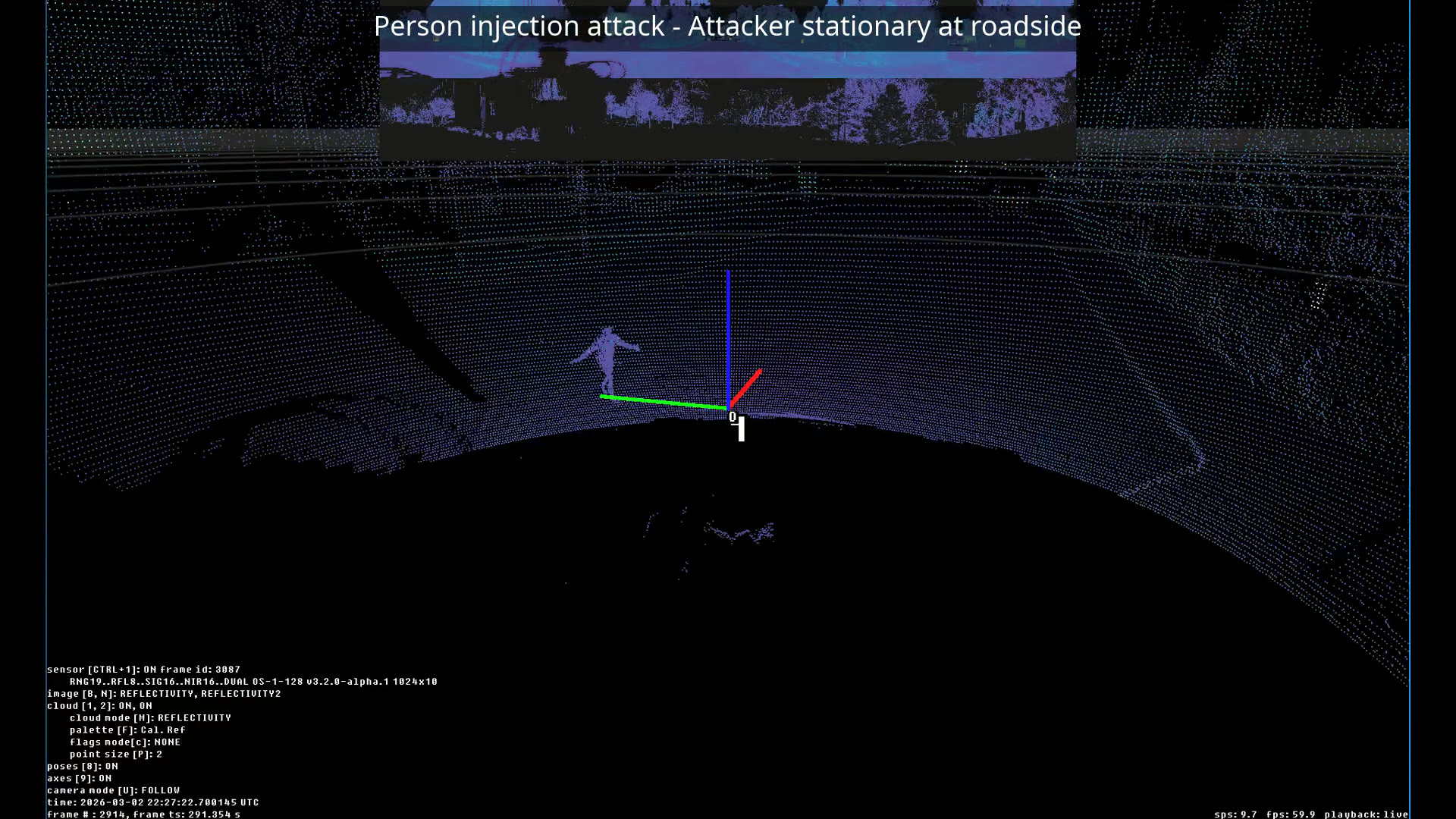}{0.171}{0.464}{0.166}{0.227} \\
        \bottomrule
    \end{tabular}
    \caption{Suppression and injection outcomes across static and mobile tests. Each row shows the camera view, the LiDAR before triggering, and after attack-mode execution. Dashed outlines mark injected returns (author annotations, not detector outputs).}
    \label{fig:main-feasibility-triptych}
\end{figure*}

%% file: tables/table_feasibility_configurations.tex
\begin{table*}[t]
    \caption{Operating envelope of malware activation. Each row is a representative condition, curated from a larger set of activation trials to show the diversity spanned, under which the optical trigger activated the implant and the selected manipulation executed as intended (suppression and injection). Activation ultimately succeeded in every case; the entries report the geometry, standoff, and dynamics spanned rather than a per-trial rate.}
    \label{tab:feasibility-configurations}
    \centering
    \small
    \begin{tabular}{p{0.15\textwidth} p{0.23\textwidth} p{0.30\textwidth} p{0.20\textwidth}}
        \toprule
        Setup class & Condition & Standoff / dynamics & Effects verified \\
        \midrule
        Indoor static & Same-floor, direct & $\approx$150~ft, direct aim & Suppr., Inj. \\
        Indoor static & Same-floor, wall reflection & $\approx$150~ft, no direct aim at sensor & Suppr., Inj. \\
        Indoor static & Same-floor, ceiling reflection & $\approx$150~ft, no direct aim at sensor & Suppr., Inj. \\
        Indoor static & Stairwell, cross-floor & 15--45~ft vertical drop, floors 3/2/1 to 4 & Suppr., Inj. at each floor \\
        Outdoor static & Tripod deployment & $\approx$200~ft and 300~ft standoff, road-slope elevation offset & Suppr., Inj. \\
        Outdoor mobile & Roadside drive-by & victim in motion up to $\approx$35~mph & Suppr., Inj. \\
        \bottomrule
    \end{tabular}
\end{table*}

%% file: figures/figure_detector_qual_composite.tex
\newcommand{\QualPanel}[1]{\includegraphics[width=0.32\textwidth,trim=128 536 449 330,clip]{#1}}
\newcommand{\QualPanelInj}[3]{
  \begin{tikzpicture}
    \node[anchor=south west,inner sep=0] (qp) {\includegraphics[width=0.32\textwidth,trim=128 536 449 330,clip]{#1}};%
    \begin{scope}[x={(qp.south east)},y={(qp.north west)}]
      \node[anchor=south, text=red!85!black, font=\sffamily\scriptsize\bfseries] at (#2,0.9) {injected};%
      \draw[red!85!black, line width=0.7pt, -{Stealth[length=1.6mm]}] (#2,0.88) -- (#3,0.62);%
    \end{scope}
  \end{tikzpicture}}
\begin{figure*}[t]
    \centering
    \QualPanel{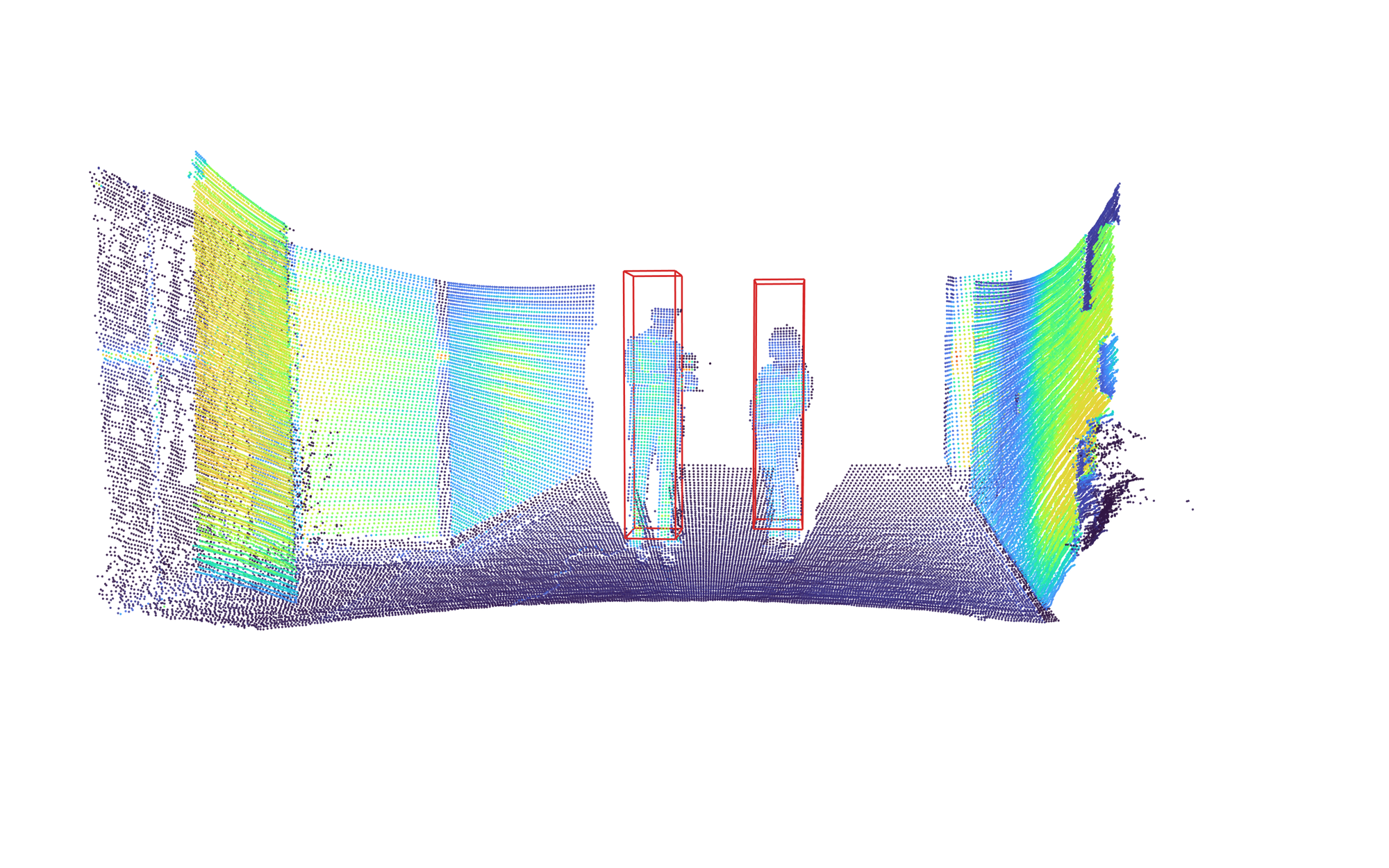}\hfill
    \QualPanelInj{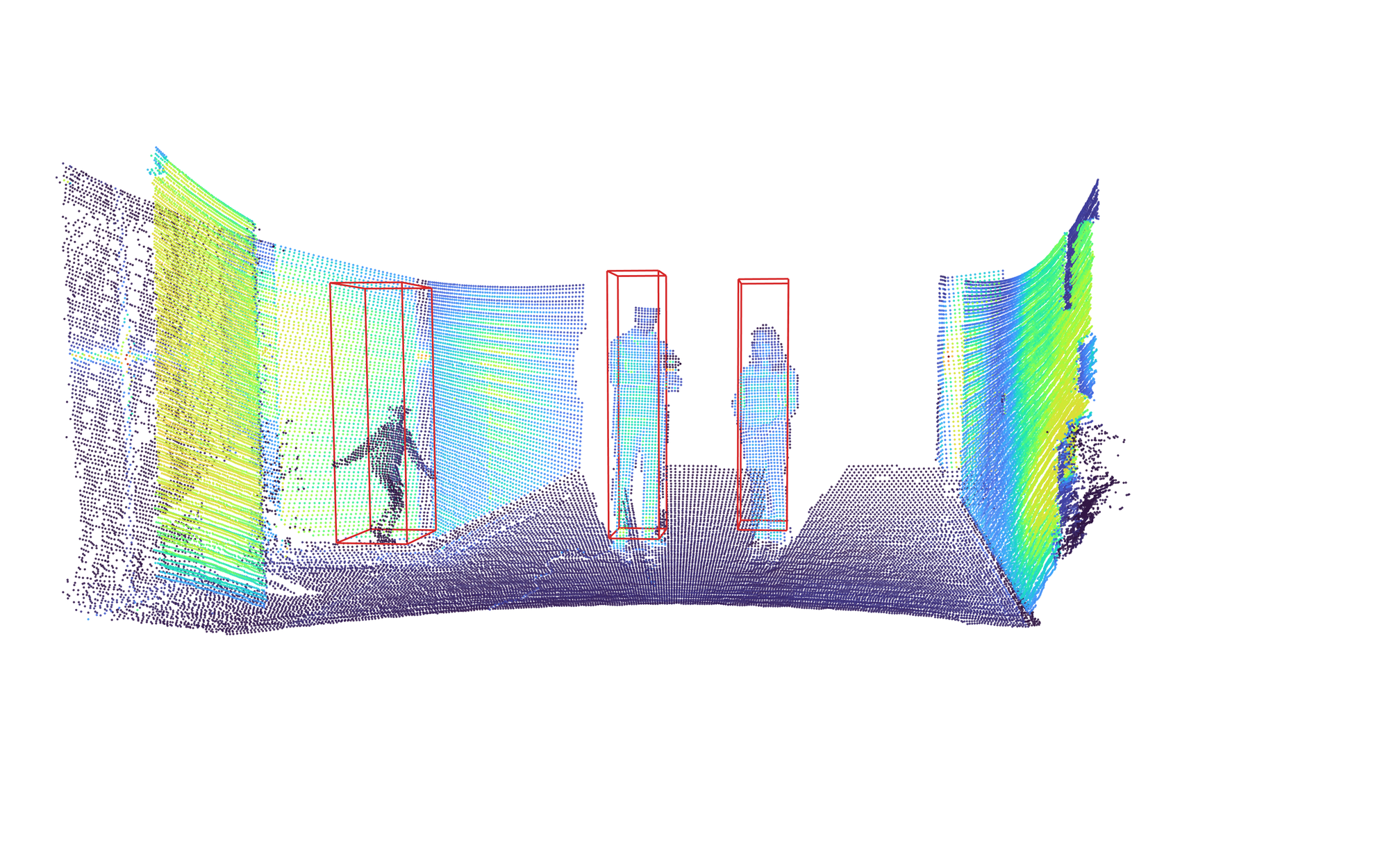}{0.24}{0.27}\hfill
    \QualPanelInj{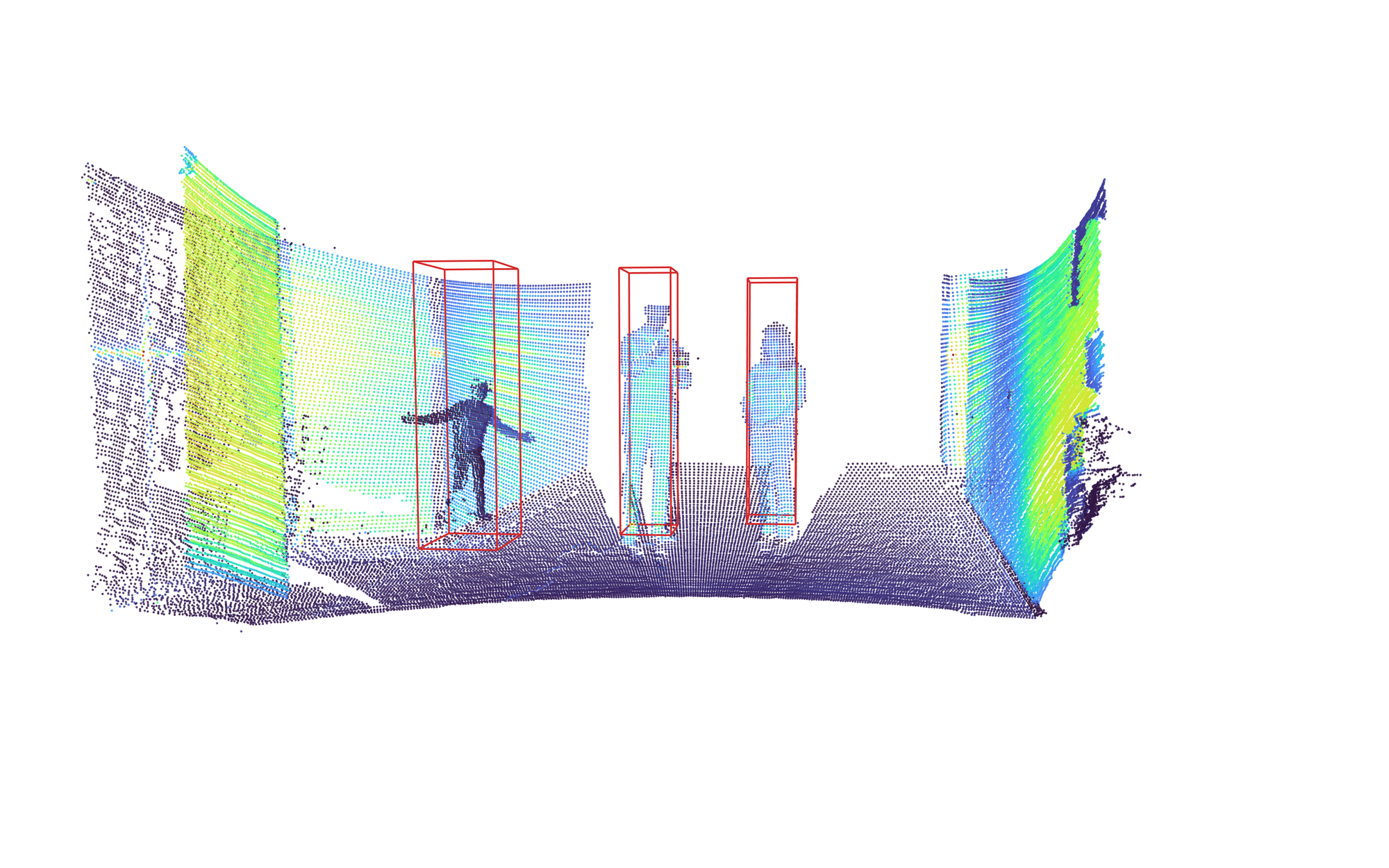}{0.30}{0.33}
    \caption{Inference sequence for person detection under false-data injection, shown left to right in time. Left: the pre-attack baseline, with only the two real people at right detected. Center and right: after activation, the injected artifact at left (\textcolor{red!85!black}{marked}) is detected and tracked with a person-class box identical to those on the real people, comparable to a genuine person detection. A video of the full replayed results is available through the \ArtifactWebsiteLink{artifact submission link}.}
    \label{fig:detector-qual-composite}
\end{figure*}

%% file: tables/table_detector_verification_summary.tex
\begin{table}[t]
    \caption{Summary of benign detector adaptation (dataset size, splits, and benign validation quality). Attack-time behavior on manipulated data is verified qualitatively in Fig.~\ref{fig:detector-qual-composite}.}
    \label{tab:detector-verification-summary}
    \centering
    \small
    \begin{tabular}{p{0.54\linewidth} p{0.32\linewidth}}
        \toprule
        Quantity & Value \\
        \midrule
        Curated benign indoor runs & 44 \\
        Scene samples (pre / post filter) & 8,040 / 7,438 \\
        Train / val / test samples & 5,903 / 753 / 782 \\
        Target class set & person only \\
        Val mAP / mAR @ 0.25 IoU & 0.9897 / 0.9987 \\
        Val mAP / mAR @ 0.50 IoU & 0.8963 / 0.9203 \\
        \bottomrule
    \end{tabular}
\end{table}

%% file: 7-impact.tex
\section{Safety-Critical Effects on a Deployed Vehicle}
\label{sec:impact}
\label{sec:evaluation-tav}

We now bring the attack to a live vehicle: a deployed tactical autonomous vehicle running an unmodified production autonomy stack, with an Ouster OS1-128 LiDAR and cameras for localization, planning, and control. We evaluate whether the triggered manipulations drive it into unsafe behavior.

In normal operation, the TAV autonomously follows a route of GPS waypoints, relying on its LiDAR to detect obstacles ahead and stop or maneuver around them. The three effects below each corrupt that LiDAR to subvert this behavior.

The attacker delivers the trigger from approximately 100~ft, and all tests run under supervised conditions with operator oversight and safety controls. Each effect is a single end-to-end demonstration that the behavior is achievable.

The platform fuses LiDAR and cameras. We do not instrument its internal algorithms; instead, we observe that a manipulated yet plausible point cloud alone drives the behaviors below despite an unmodified camera observing the same scene, consistent with the stack weighting LiDAR heavily for near-field obstacle geometry. Whatever its exact fusion rule, multi-sensor fusion does not save the stack: the compromised LiDAR alone is enough to change the vehicle's behavior.

We group the outcomes by safety consequence. Two are fail-safe, leaving the vehicle intact: error induction triggers an \emph{emergency stop}, and injection forces a \emph{mission abort}. One is fail-dangerous: suppression causes a collision with a real obstacle. We present them in that order of increasing severity.

\subsection{Error Induction}
\label{sec:evaluation-tav-error}

\input{figures/figure_attack_error_mode}

Activation forces the LiDAR into an error state, and point cloud publication ceases. Before activation, the vehicle is proceeding through a benign scene with no nearby obstructions and a healthy LiDAR output (Fig.~\ref{fig:attack-error-mode}a). Shortly after triggering, the platform health report indicates that the LiDAR has stopped publishing data, and the vehicle enters fail-safe emergency-stop behavior even though the companion camera shows no obstacle requiring a stop (Fig.~\ref{fig:attack-error-mode}b). From the vehicle's perspective, the event is indistinguishable from a benign sensor fault without low-level forensic inspection. Malware-triggered sensor failure thus halts a healthy vehicle on demand.

\subsection{Person-like Injection}
\label{sec:evaluation-tav-fdi}

\input{figures/figure_attack_spoof}

The malware injects person-like objects across the vehicle's path while preserving scene structure (Fig.~\ref{fig:attack-spoof}). The injected objects retain scene-consistent geometric cues, including the shadowed regions that would normally appear behind foreground obstacles, which makes them more plausible within the surrounding point cloud. The autonomy stack interprets them as real foreground hazards and responds with evasive maneuvering rather than nominal mission execution. As the malware reinjects them over time, the vehicle cannot clear the perceived blockage and ultimately aborts the mission. This shows that injecting person-like objects redirects autonomy even when the rest of the scene remains nominal.

\subsection{Targeted Suppression}
\label{sec:evaluation-tav-suppression}

\input{figures/figure_attack_suppression}

Suppression removes returns in the forward direction while packet formatting and timing remain nominal. As the vehicle advances along its route toward a traffic barrel in its path, the malware suppresses the returns from that barrel, leaving apparent free space where it stands. The packets are still published normally, only without the obstacle geometry, so the autonomy stack never registers the obstruction and continues forward. Unlike the fail-safe outcomes above, the result is fail-dangerous: the vehicle collides with an otherwise detectable obstacle it would have avoided (Fig.~\ref{fig:attack-suppression}).

\myparagraph{Evaluation summary.}
These results establish both conditions of Section~\ref{sec:vulnerability}. The malware is remotely triggerable across a range of sampled static and mobile conditions via a handheld optical emitter, and its activation produces safety-critical effects at two levels: injected artifacts are taken up as person-class detections by a tuned 3D object detector, and on a deployed TAV the manipulations drive an emergency stop, a mission abort, and a collision. This combined evidence demonstrates that globally sourced sensing components are vulnerable to malware insertion with safety-critical consequences.

%% file: figures/figure_attack_error_mode.tex
\renewcommand{\subfiggraphicswidth}{0.85}
\begin{figure}[t]
    \centering
    \begin{subfigure}[b]{\linewidth}
        \centering
        \includegraphics[width=\subfiggraphicswidth\linewidth,trim=0 0 0 140,clip]{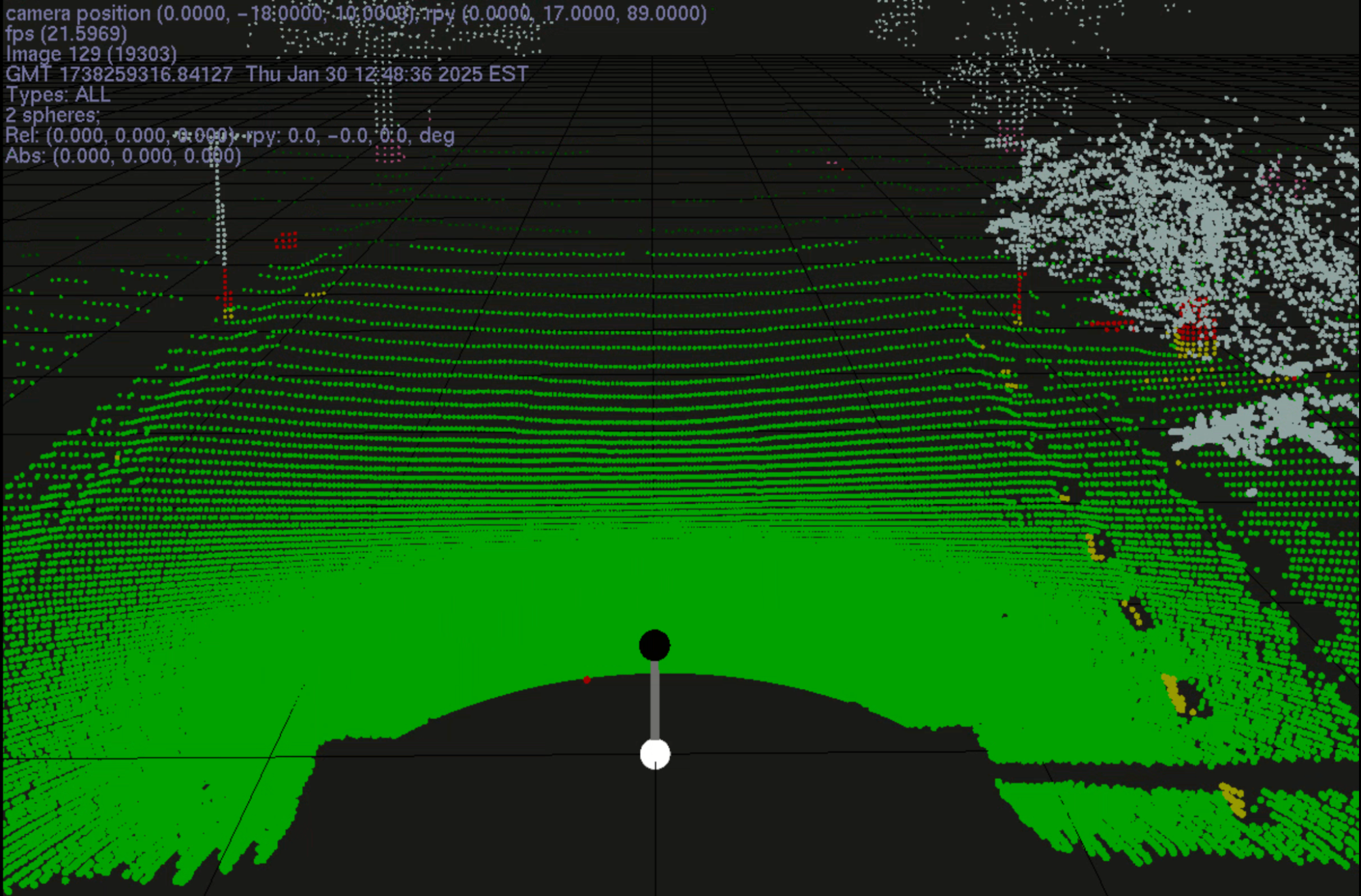}
        \caption{LiDAR point cloud before triggered error-stop mode.}
        \label{fig:attack-error-mode-a}
    \end{subfigure}
    \hfill
    \begin{subfigure}[b]{\linewidth}
        \centering
        \includegraphics[width=\subfiggraphicswidth\linewidth,trim=0 0 0 130,clip]{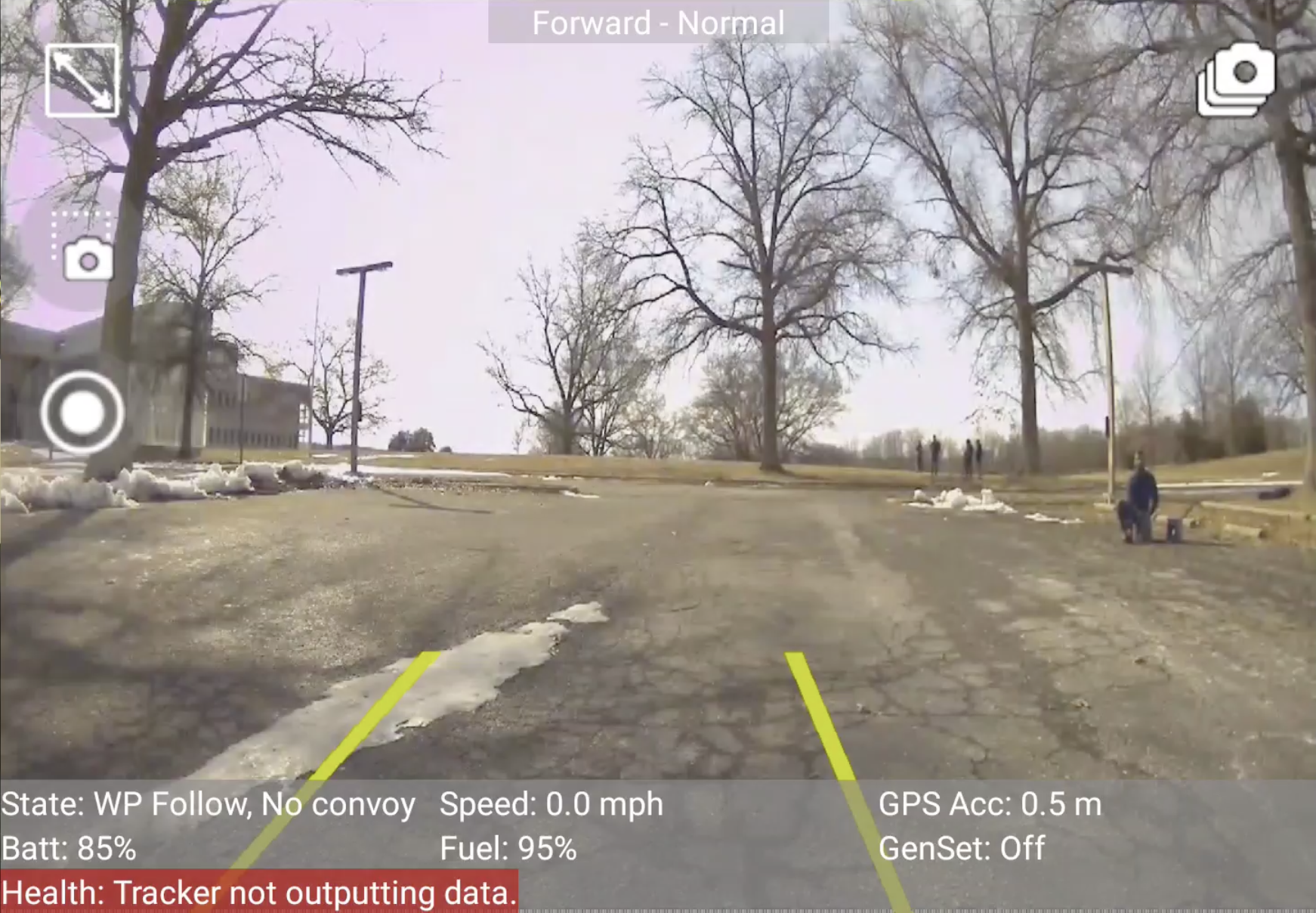}
        \caption{Camera image shows no obstacles, but platform stopped.}
        \label{fig:attack-error-mode-b}
    \end{subfigure}
    \caption{\textbf{Error Induction on the deployed vehicle.} (a) LiDAR yields a healthy point cloud before attack. (b) The camera shows no obstructions, yet the health report indicates the tracking system has stopped reporting. The LiDAR has entered error mode with no indication of an adversary.}
    \label{fig:attack-error-mode}
\end{figure}

%% file: figures/figure_attack_spoof.tex
\renewcommand{\subfiggraphicswidth}{1.0}
\newcommand{\SpoofPC}[3]{%
  \begin{tikzpicture}
    \node[anchor=south west,inner sep=0] (sp) {\includegraphics[width=\subfiggraphicswidth\linewidth,trim=0 0 0 35,clip]{#1}};%
    \begin{scope}[x={(sp.south east)},y={(sp.north west)}]
      \draw[red!85!black,dashed,line width=0.6pt] (#2,#3) ellipse (0.12 and 0.075);%
    \end{scope}
  \end{tikzpicture}}
\begin{figure*}[!t]
    \centering
    \makebox[\linewidth][c]{%
        \foreach \x [count=\i from 1] in {1,3,4}{%
            \begin{subfigure}[b]{0.32\linewidth}
                \centering
                \includegraphics[width=\subfiggraphicswidth\linewidth,trim=0 0 0 0,clip]{images/ouster_cases/spoof_images/t\x-tav__inpaint.png}
            \end{subfigure}%
            \ifnum\i<4\hfill\fi
        }%
    }

    \vspace{4pt}

    \makebox[\linewidth][c]{%
        \foreach \x/\cx/\cy [count=\i from 1] in {1/0.475/0.35, 3/0.485/0.34, 4/0.470/0.345}{%
            \begin{subfigure}[b]{0.32\linewidth}
                \centering
                \SpoofPC{images/ouster_cases/spoof_images/t\x-pc.png}{\cx}{\cy}
            \end{subfigure}%
            \ifnum\i<4\hfill\fi
        }%
    }
    \caption{\textbf{Person-like Injection on the deployed vehicle.} False-data-injection attack on the deployed TAV. Top: external view of the vehicle over time; bottom: the corresponding LiDAR point cloud, with the persistent front-near injected obstacles marked by dashed outlines (author annotations, not detector outputs). The vehicle responds with evasive maneuvering rather than nominal mission execution and, as the false geometry persists in the foreground, ultimately aborts the mission while the scene evolves normally.}
    \label{fig:attack-spoof}
\end{figure*}

%% file: figures/figure_attack_suppression.tex
\begin{figure}[!t]
    \centering
    \begin{subfigure}[b]{0.49\linewidth}
        \centering
        \includegraphics[width=\linewidth,trim=0 0 0 85,clip]{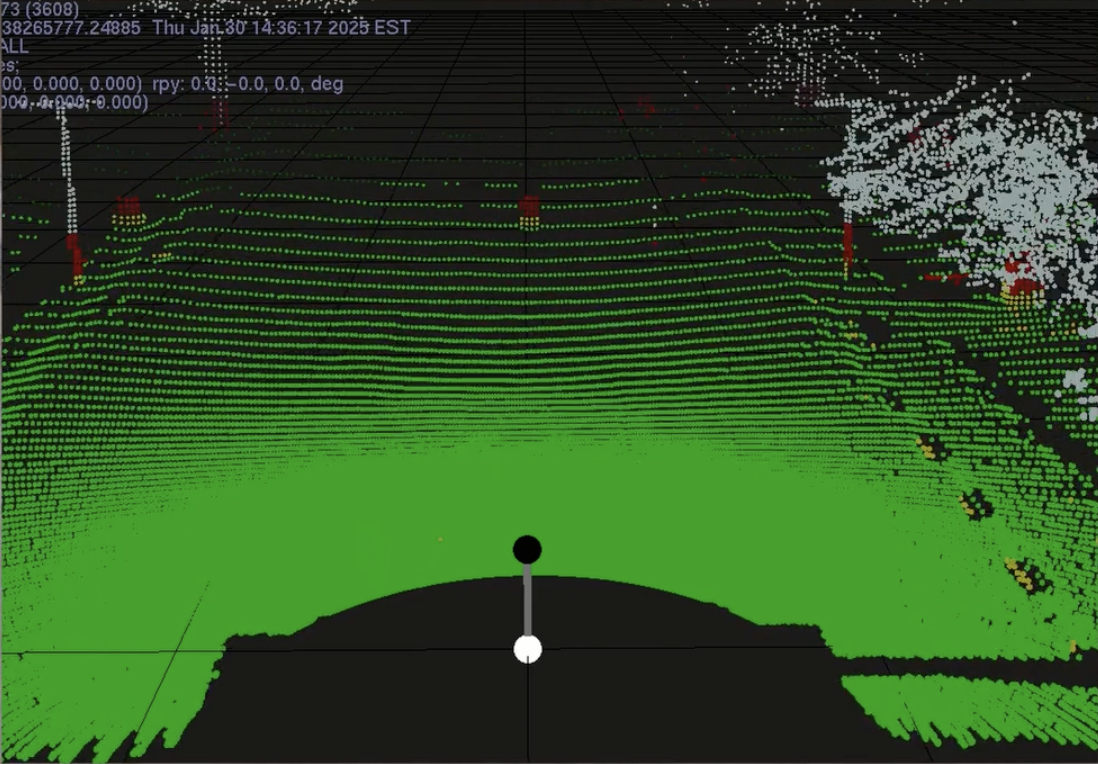}
        \caption{Benign LiDAR: obstacle returns present.}
        \label{fig:attack-suppression-a}
    \end{subfigure}
    \hfill
    \begin{subfigure}[b]{0.49\linewidth}
        \centering
        \includegraphics[width=\linewidth,trim=0 0 0 85,clip]{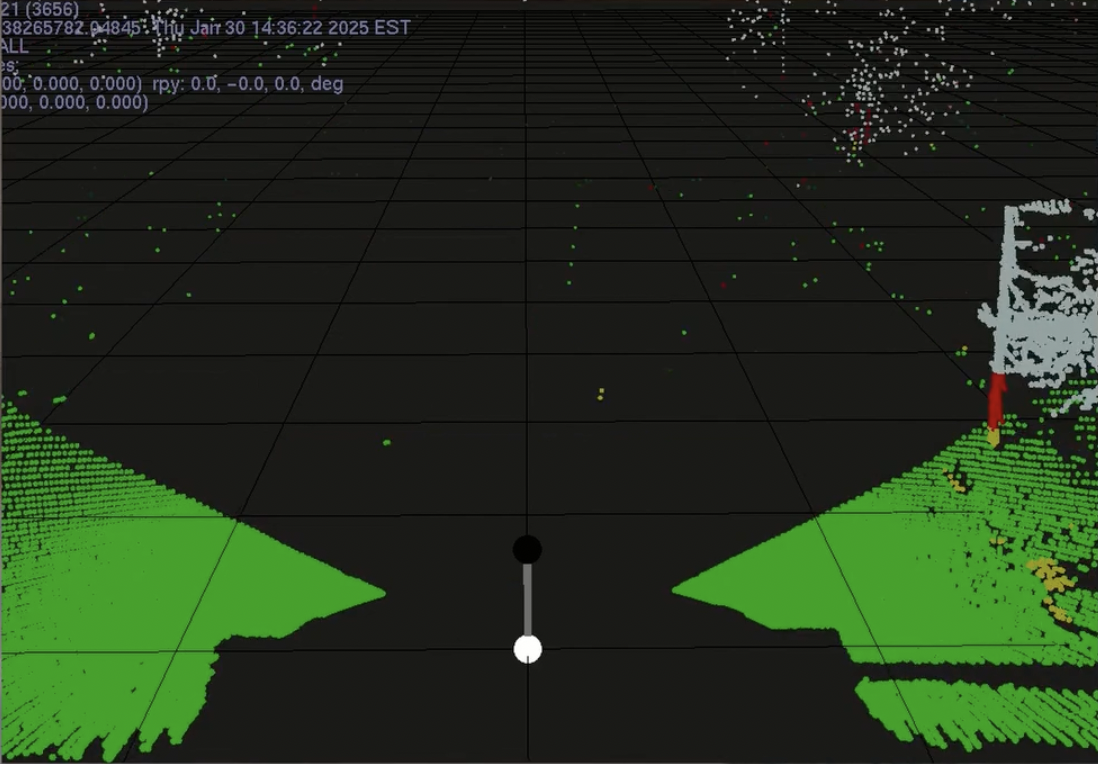}
        \caption{After suppression: obstacle returns removed.}
        \label{fig:attack-suppression-b}
    \end{subfigure}

    \vspace{5pt}

    \begin{subfigure}[b]{0.85\linewidth}
        \centering
        \begin{tikzpicture}
            \node[anchor=south west,inner sep=0] (col) {\includegraphics[width=\linewidth,trim=0 0 0 0,clip]{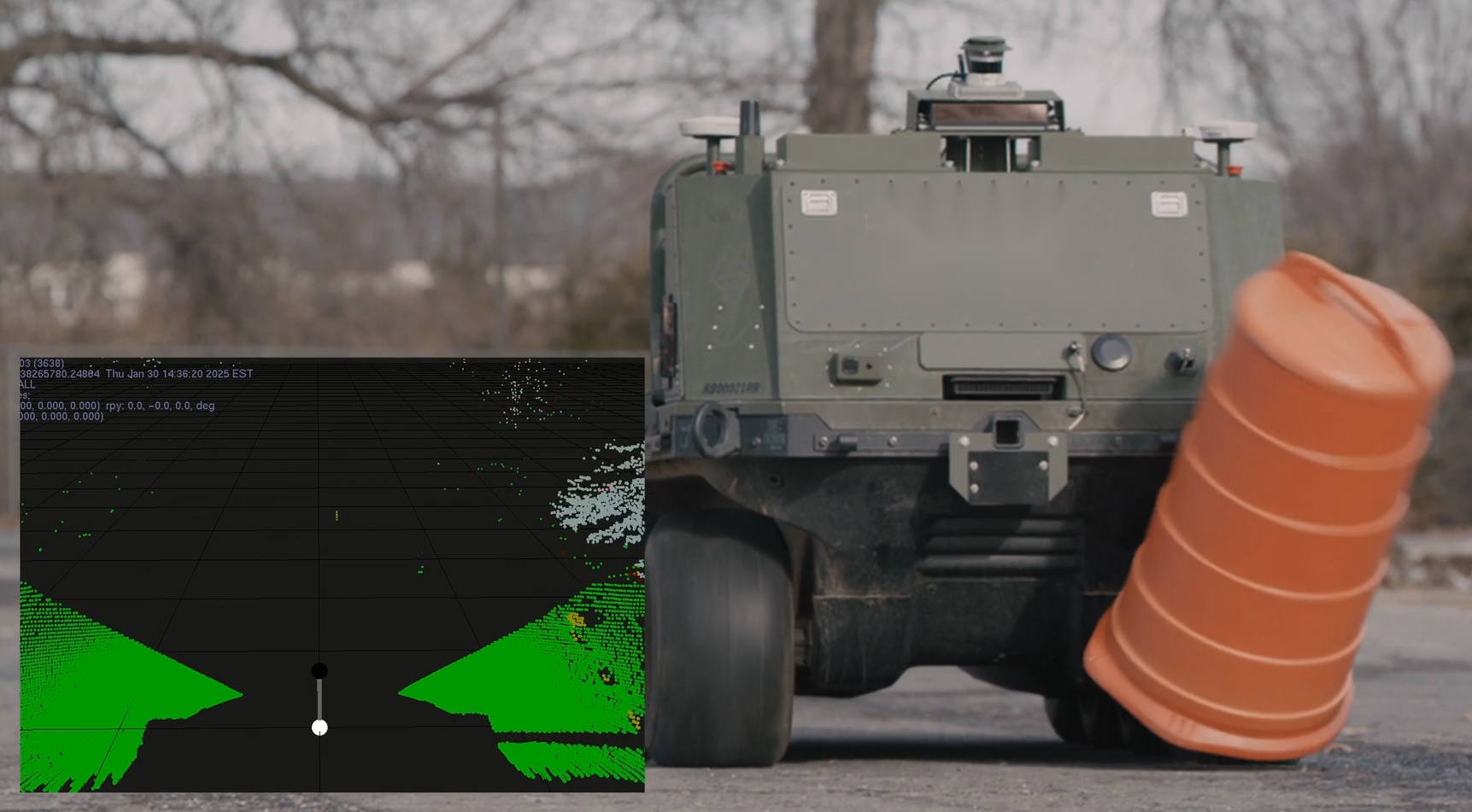}};%
            \begin{scope}[x={(col.south east)},y={(col.north west)}]
              \node[anchor=south,text=red!85!black,font=\sffamily\small\bfseries] at (0.72,0.88) {undetected obstacle};%
              \draw[red!85!black,line width=0.9pt,-{Stealth[length=2.2mm]}] (0.78,0.865) -- (0.852,0.692);%
            \end{scope}
          \end{tikzpicture}
        \caption{TAV reaches the obstacle hidden by suppression.}
        \label{fig:attack-suppression-c}
    \end{subfigure}
    \caption{\textbf{Targeted Suppression on the deployed vehicle.} (a) Benign LiDAR point cloud with obstacle returns. (b) After the suppression attack, returns in the forward region are removed. (c) With the obstacle absent from the manipulated output, the TAV drives into it.}
    \label{fig:attack-suppression}
\end{figure}

%% file: 8-discussion.tex
\section{Discussion and Mitigations}
\label{sec:discussion}

Our results show malware placed in the LiDAR sensing pipeline and triggered post-deployment creates a distinct autonomy risk that existing security analyses miss. This risk spans fail-safe disruption and fail-dangerous behavior: triggered malware can stop vehicles, inject false objects that downstream perception accepts as genuine, and hide real obstacles.

\subsection{Why Existing Defenses Are Insufficient}

Existing defenses are poorly aligned to detect sensor-internal malware that a transient optical signal activates remotely. The threat evades them for two structural reasons: activation comes from a brief external signal that does not itself produce the downstream malicious effect, and once activated, the malware operates inside the sensing pipeline and can induce outcomes that appear either as sensor failures or as structurally valid LiDAR outputs.

\myparagraph{Physical-channel and network defenses.}
Defenses against LiDAR spoofing or jamming look for anomalous optical signals and timing inconsistencies~\cite{petit2015remote,2020sun-spoofing,hallyburton2022security}, and network intrusion detection looks for anomalous communication or process behavior~\cite{cho2016fingerprinting}. Neither assumption holds here. The only external signal is the brief modulated trigger, which does not itself manipulate perception; once activated, all manipulation occurs inside the sensor (Fig.~\ref{fig:lidar-pipeline}), with no network access, external communication, or significant background processes. Both classes therefore see only downstream effects that resemble benign faults, not clear intrusion evidence.

\myparagraph{Perception and fusion-based defenses.}
Multi-sensor fusion and statistical anomaly detection are the most promising lines of defense against manipulated point clouds, since injected or suppressed returns would be expected to leave irregular densities, implausible reflectance, or temporal discontinuities~\cite{hallyburton2022security}. Two properties of this threat weaken that expectation. First, because manipulation occurs inside the pipeline after native signal processing, the outputs remain syntactically valid and temporally plausible, and injected objects are constructed to be scene-consistent (Section~\ref{sec:evaluation-tav}). Second, without a sufficiently large sensor set or an independent trust anchor, cross-modal disagreement may be insufficient to identify which sensor is compromised~\cite{hallyburton2025security}. Our deployed-platform results bear this out: the vehicle fused camera and LiDAR yet still suffered the suppression and injection attacks, consistent with the stack weighting LiDAR heavily for obstacle geometry (Section~\ref{sec:evaluation-tav}).

A further difficulty is that any residual low-level signature is attacker-controlled: because the malware synthesizes returns directly, per-point properties such as density, reflectance, and inter-frame continuity can be shaped to match the scene, and the injected object already reproduces the point density of a genuine return at its range (Section~\ref{sec:evaluation-detectability}). This is why durable protection rests on establishing trust in the sensor rather than on post-hoc filtering of its output. These defenses are misaligned with a sensor-internal adversary rather than defeated by it; a quantitative evaluation of anomaly-based and cross-modal detectors against this attack is an important direction for future work.

\subsection{Implications and Mitigations}

LiDAR sensors should be treated as active computing elements, not trusted measurement devices: once attacker-controlled logic runs inside the sensing pipeline, the assumption that sensors are trustworthy and threats arrive from outside breaks down. Trusting a sensor means trusting the manufacturers, update paths, and maintenance channels behind it, not only downstream perception software. The deployed TAV results sharpen the consequence: even systems that handle explicit sensor loss conservatively, halting when the point cloud drops, still forward manipulated geometry that appears operationally valid, leaving the fail-dangerous case unguarded.

The mitigations that follow establish trust in the sensor itself rather than only checking its outputs downstream. We prioritize them by leverage: first protect sensor integrity, then verify sensor state, and finally treat downstream consistency checks as a secondary layer.

\myparagraph{Sensor integrity.}
The most direct defense is stronger integrity protection inside the sensor itself: secure boot, firmware signing, and hardware roots of trust address the point at which the attack originates. These technical mechanisms should be paired with lifecycle safeguards such as supplier vetting, diversified sourcing, and independent firmware auditing, since trustworthy integrity guarantees still depend on manufacturing, update, and key-management processes.

\myparagraph{Sensor state verification.}
Runtime attestation and verification can help downstream systems determine whether the LiDAR is running approved code and remains in an expected state~\cite{nunes2019vrased, surminski2021realswatt}. This is especially relevant for a threat model in which the sensor continues publishing plausible-looking outputs after compromise. The challenge is to provide this assurance without introducing prohibitive latency, brittle dependencies, or additional points of failure.

\myparagraph{Downstream consistency checks.}
Explicit cross-sensor consistency checks may still catch some manipulations~\cite{hallyburton2025security,hallyburton2025trust,quinonez2020savior}, but they should be treated as secondary mitigations rather than primary ones. The suppression and injection examples in this paper illustrate why downstream algorithmic checks alone are not a reliable trust anchor when corrupted outputs remain structurally plausible.

\subsection{Limitations and Future Work}
We do not exhaustively enumerate attack strategies or empirically evaluate specific defenses. Future work could extend the analysis to additional deployed platforms, formalize accidental activation, and investigate hardware-assisted mechanisms for enforcing trust in sensing pipelines.

\myparagraph{Ambient triggering.}
In outdoor trials, we observed occasional unintended trigger events under ambient optical conditions, indicating that the trigger encoding and matching rules require careful engineering for field robustness. Higher-entropy attack messages would mitigate this, which we leave to future work. More generally, encoding strategies introduced to improve robustness against ambient interference could expand the trigger-design space available to an attacker. We also observed that attempting to trigger the malware through a commercial building window failed, indicating that some materials attenuate the relevant optical band.

\myparagraph{Coordinated and long-range triggering.}
The same threat model motivates future work on coordinated and ultra-long-range triggering. Because the trigger is a broadcast optical key rather than a targeted spoof, any compromised sensor that observes the signal and holds the key decodes it, so activation is one-to-many by construction: a single, sufficiently powerful emitter could in principle activate malware across many victim sensors at once, or from much longer standoff distances than we tested, including airborne delivery platforms.

%% file: 9-conclusion.tex
\section{Conclusion}
\label{sec:conclusion}

This work shows that sensor-internal compromise of LiDAR is a distinct and underexplored autonomy risk. We demonstrate that dormant malware embedded in the LiDAR sensing pipeline can be remotely activated after deployment using a brief modulated optical signal, without network access or sustained interaction with the victim. Empirical evidence spanned three parts of the attack chain: sampled feasibility across static and mobile conditions, injected person-like artifacts taken up as person-class detections, and deployed TAV end-to-end impact.

Sensor-internal compromise challenges the trust boundary that autonomous sensing pipelines assume. Once malicious logic is inside the sensor, downstream autonomy consumes outputs that remain operationally plausible while encoding fail-safe disruption or fail-dangerous manipulation. Securing autonomy therefore requires stronger guarantees about the integrity of the sensing infrastructure on which perception algorithms depend, not robust perception algorithms alone.

%% file: 9-acks.tex
\section*{Acknowledgments}

The authors thank David Hunt for assistance with experimental data collection and Robert Hallyburton for engineering the vehicle mounting platform. The authors thank Ouster, Inc. for providing the LiDAR sensors used in this study for academic research purposes to illuminate potential risks in the sensing supply chain through a controlled proof-of-concept. To facilitate this security analysis, Ouster engineers provided technical assistance in a specialized, closed test environment to intentionally modify sensor firmware with the dormant malware routines described in this paper. The vulnerabilities demonstrated were manually introduced into a specific test-bench version of the firmware to support defensive engineering analysis; they do not indicate inherent vulnerabilities in Ouster’s commercially available products or standard supply chain.

This work is sponsored in part by the ONR under agreement N00014-23-1-2206, AFOSR under the award number FA9550-19-1-0169, and by the NSF under NAIAD Award 2332744 as well as the National AI Institute for Edge Computing Leveraging Next Generation Wireless Networks, Grant CNS-2112562.

%% file: 99-appendix-open-science.tex
\section*{Open Science}

We provide an artifact website with the detector and dataset \ArtifactWebsiteLink{source code} and a \ArtifactWebsiteLink{video collection} of the feasibility experiments (Section~\ref{sec:evaluation-feasibility}), covering the materials needed to evaluate the paper's claims.

\myparagraph{Artifacts provided.}
The release provides the source code, trained model, and datasets behind the detector-training and detector-verification results. The dataset is the curated benign OS1-128 indoor person corpus of 7,438 labeled samples across 44 scene runs spanning multiple indoor environments (Appendix~\ref{appendix:detector}), together with the automated background-subtraction pipeline that generates its bounding-box annotations, the run-level train/validation/test splits, and the export tooling that produces detector-ready inputs. Both the dataset and the labeling procedure can therefore be regenerated and audited independently.

\myparagraph{Artifacts not provided.}
We do not release the firmware or modified firmware images, whose disclosure would increase misuse risk and expose proprietary intellectual property, or the deployed TAV experiment video, which is subject to partner confidentiality obligations.

%% file: 99-appendix-A-experimental-protocol.tex
\section{Feasibility Supporting Materials}
\label{appendix:protocol}

This appendix collects the setup-variant figures and extended qualitative galleries that support the feasibility results in Section~\ref{sec:evaluation}. The trial-annotation and evidence-grounding conventions behind those results are stated in Section~\ref{sec:evaluation-feasibility}.

Figures~\ref{fig:appendix-setup-indoor} and~\ref{fig:appendix-setup-outdoor-mobile} give setup-specific diagram and photo composites that expand the geometry overview of Fig.~\ref{fig:experiment-setups}: Fig.~\ref{fig:appendix-setup-indoor} covers the indoor same-floor, reflected-path, and cross-floor cases, while Fig.~\ref{fig:appendix-setup-outdoor-mobile} covers the outdoor tripod and mobile drive-by arrangements. Figures~\ref{fig:appendix-static-triptych} and~\ref{fig:appendix-mobile-triptych} extend the qualitative gallery of Fig.~\ref{fig:main-feasibility-triptych} under the same \emph{Camera | LiDAR Before Attack | LiDAR After Attack} schema, adding further static and tripod cases and further mobile cases, respectively.

%% file: 99-appendix-B-detector-verification.tex
\section{Detector Verification}
\label{appendix:detector}

This section describes the workflow used to evaluate whether injected person-like LiDAR artifacts are taken up as person-class detections by a 3D object detector. The detector-verification workflow adapts a general-purpose indoor LiDAR detector to a person detection task on Ouster OS1-128 data, validates that adaptation on a curated benign dataset, and then applies the adapted detector to replayed manipulated point cloud data. The implementation builds on perception-model scaffolding from AVstack~\cite{hallyburton2023avstack}.

\subsection{Dataset Creation}
\label{appendix:detector-dataset}

Adapting a general-purpose indoor LiDAR object detector to a person detection task requires a specialized fine-tuning dataset. To support this task, we collected the benign OS1-128 indoor data used for detector adaptation across 44 scene runs together with separate background-reference captures, then organized the resulting data run-by-run for later labeling, curation, and export.

Benign data collection was organized scene-by-scene. For each collection branch, the LiDAR geometry was held fixed while we first recorded a background-reference scene and then multiple scene runs containing people under varied positions, poses, and visibility conditions. We also recorded run-level metadata such as person count and short scenario descriptions to support later curation. The raw pcap captures were then standardized into binary sweep samples.

\subsection{Automated Labeling}
\label{appendix:detector-labeling}

The fine-tuning task requires labeled positive detection examples alongside the raw data. For the person detection task, we designed an automated workflow that yields bounding boxes from standardized LiDAR data rather than labeling the scenes by hand. The workflow includes background estimation and removal, candidate box identification, temporal filtering, object pruning, and dataset collation.

From the background reference capture, we first build a static voxel-based occupancy model. In subsequent scene runs, points close to background structure are removed and novel points persist as ``foreground'' points. Remaining foreground structures are clustered into candidate objects and filtered using temporal and spatial person-like survival criteria. Applied across the detector-adaptation dataset, this procedure produced the labels used for detector training.

To improve the quality of the resulting fine-tuning dataset, we curated the collated dataset using metadata recorded during collection. Each run carries an expected person count, and only samples whose generated box count matches that expected count are retained in the default clean training set. This filtering step removes transitional or weakly labeled frames while preserving the more reliable scenes needed for detector fine-tuning. The final detector dataset therefore contains 44 curated runs, yielding 8,040 scene samples before filtering and 7,438 retained scene samples after count-based curation, with run-level train/validation/test splits of 5,903 / 753 / 782 samples.

\subsection{Detector Training and Adaptation}
\label{appendix:detector-training}

We fine-tune an FCAF3D-based 3D object detector on the curated benign dataset described above. The base detector is initialized from a pretrained indoor model on ScanNet~\cite{dai2017scannet} and then adapted to the single-class person-detection task on OS1-128 data. Detector-ready artifacts are exported from the custom dataset pipeline into the training framework, and setup metadata is used to normalize LiDAR height consistently across scenes. We use run-level train/validation/test partitioning so that near-duplicate frames from the same capture run do not straddle data splits.

We trained the model for 12 epochs with a cyclic learning rate optimization strategy commonly used for point-cloud detectors~\cite{hallyburton2023avstack}. We combined multiple loss functions, using cross entropy loss for the center point detection, axis-aligned IoU loss for the bounding box determination, and focal loss for the classifier. Fig.~\ref{fig:appendix-detector-training-loss} illustrates each loss function's contribution over training while Table~\ref{tab:detector-verification-summary} summarizes the validation-side AP/AR. The resulting detector adapts well to the benign indoor person-detection task, which is the prerequisite for the attack-time semantic verification used in the main paper.

\input{figures/figure_appendix_detector_training_loss}
\input{tables/table_appendix_detector_dataset_training.tex}

\subsection{Verification Protocol}
\label{appendix:detector-protocol}

For person-like injection trials, we run the adapted detector on manipulated point cloud outputs and record whether person detections appear in the injected region. Our replay tooling decodes the recorded LiDAR capture into sweep files, applies a run-level canonicalization transform, executes detector inference frame-by-frame, logs predictions to JSONL and CSV, and exports selected keyframes for manuscript figures.

We use this workflow as targeted semantic verification, not as a full adversarial benchmark. Person-class detections persisted in the injected region across the replayed attack progression, supporting the claim that the injected artifact can survive one downstream perception stage as a person-like hypothesis. Because we did not construct a labeled adversarial corpus, we report no attack-time AP/AR, precision, or recall: the benign metrics in Table~\ref{tab:detector-verification-summary} establish that the detector adaptation succeeded, and the replay result stands as bounded verification of artifact uptake.

%% file: figures/figure_appendix_detector_training_loss.tex
\begin{figure}[t]
    \centering
    \includegraphics[width=\linewidth]{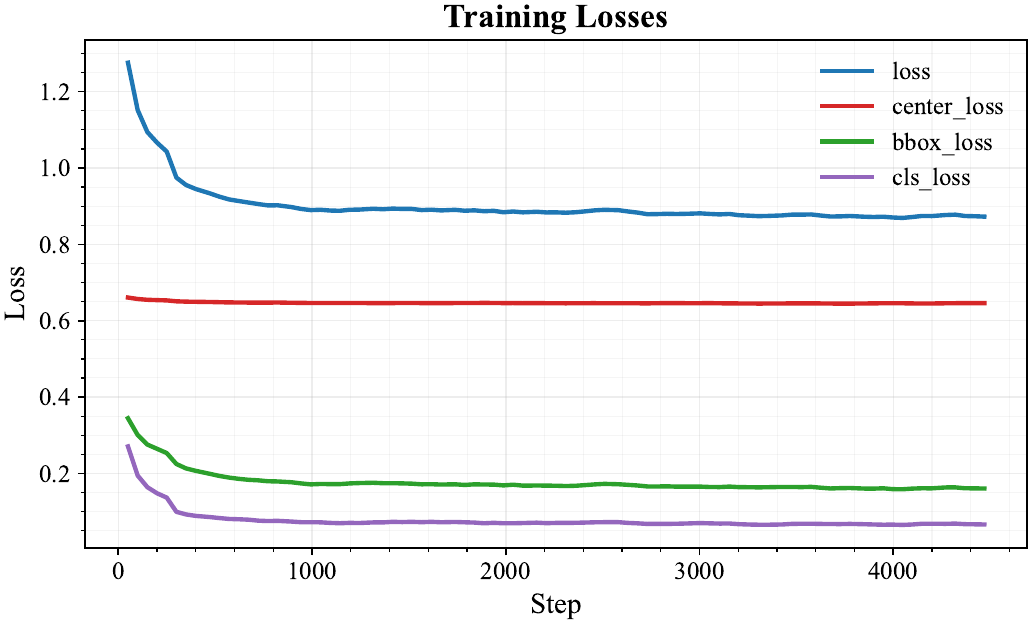}
    \caption{Detector fine-tuning loss curves for the FCAF3D-based person detector adaptation run on 12 epochs. The training losses plateau within the 12-epoch schedule; the corresponding validation AP/AR are reported in Tab.~\ref{tab:detector-verification-summary}.}
    \label{fig:appendix-detector-training-loss}
\end{figure}

%% file: tables/table_appendix_detector_dataset_training.tex
\begin{table}[t]
    \caption{Detector-adaptation configuration for the FCAF3D-based indoor person detector used in Section~\ref{sec:evaluation-detectability}. Dataset sizes and validation metrics are summarized in Tab.~\ref{tab:detector-verification-summary}.}
    \label{tab:appendix-detector-dataset-training}
    \centering
    \small
    \begin{tabular}{@{}p{0.34\columnwidth} p{0.58\columnwidth}@{}}
        \toprule
        Quantity & Value \\
        \midrule
        Coordinate frame & depth-coordinates (x right, y forward, z up, yaw zero along x) \\
        Count-filter policy & strict equality between expected person count and generated boxes \\
        Split seed / partition rule & 42 / run-level ratios plus background reference \\
        Detector architecture & FCAF3D-based indoor LiDAR 3D detector~\cite{rukhovich2022fcaf3d} \\
        Initialization checkpoint & pretrained indoor checkpoint on ScanNet~\cite{dai2017scannet} \\
        Training schedule & cyclic learning rate, 12 epochs \\
        Learning rates & $\left[0.0018, 0.018, 0.00018\right]$ \\
        \bottomrule
    \end{tabular}
\end{table}

%% file: figures/figure_appendix_setup_indoor.tex
\newcommand{\BeamPanel}[5]{%
  \begin{tikzpicture}
    \node[anchor=south west,inner sep=0] (bp) {\includegraphics[width=\linewidth,trim=3cm 2cm 3cm 0.8cm,clip]{#1}};%
    \begin{scope}[x={(bp.south east)},y={(bp.north west)}]
      \draw[red!85!black,dashed,line width=1pt,-{Stealth[length=2.2mm]}] (#2,#3) -- (#4,#5);%
    \end{scope}
  \end{tikzpicture}}
\begin{figure*}[p]
    \centering
    \begin{subfigure}[t]{0.46\linewidth}
        \centering
        \vspace{0pt}
        \includegraphics[width=\linewidth,trim=0 0 0 0,clip]{diagrams/experiment-diagram-indoor.drawio.pdf}
        \caption{Indoor static setup diagram.}
        \label{fig:appendix-setup-indoor-diagram}
    \end{subfigure}
    \hfill
    \begin{subfigure}[t]{0.26\linewidth}
        \centering
        \vspace{0pt}
        \includegraphics[width=\linewidth,trim=3cm 2.82cm 8cm 4cm,clip]{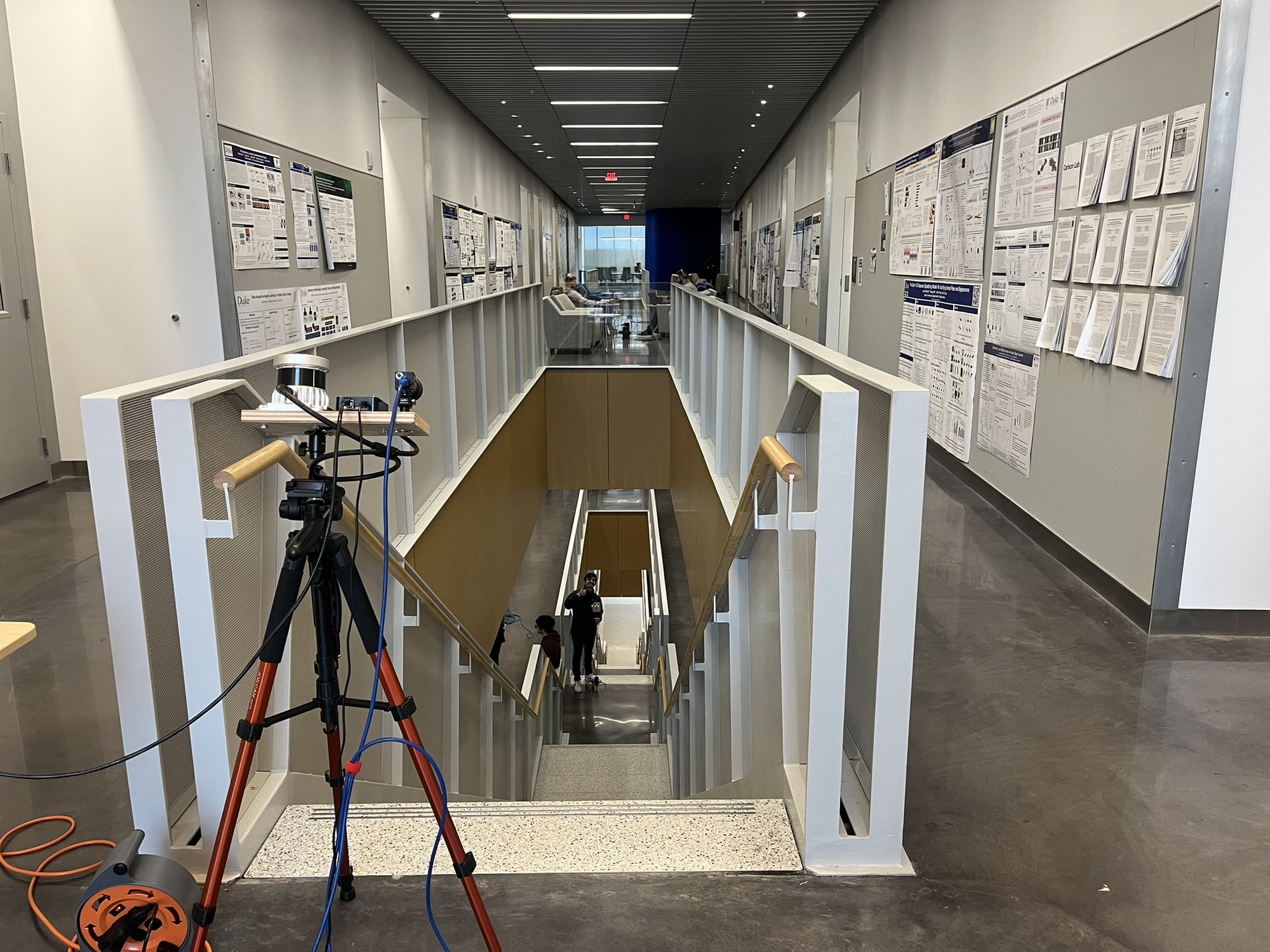}
        \caption{Cross-floor stairwell victim.}
    \end{subfigure}
    \hfill
    \begin{subfigure}[t]{0.26\linewidth}
        \centering
        \vspace{0pt}
        \includegraphics[angle=90,width=\linewidth,trim=9cm 3cm 6cm 3cm,clip]{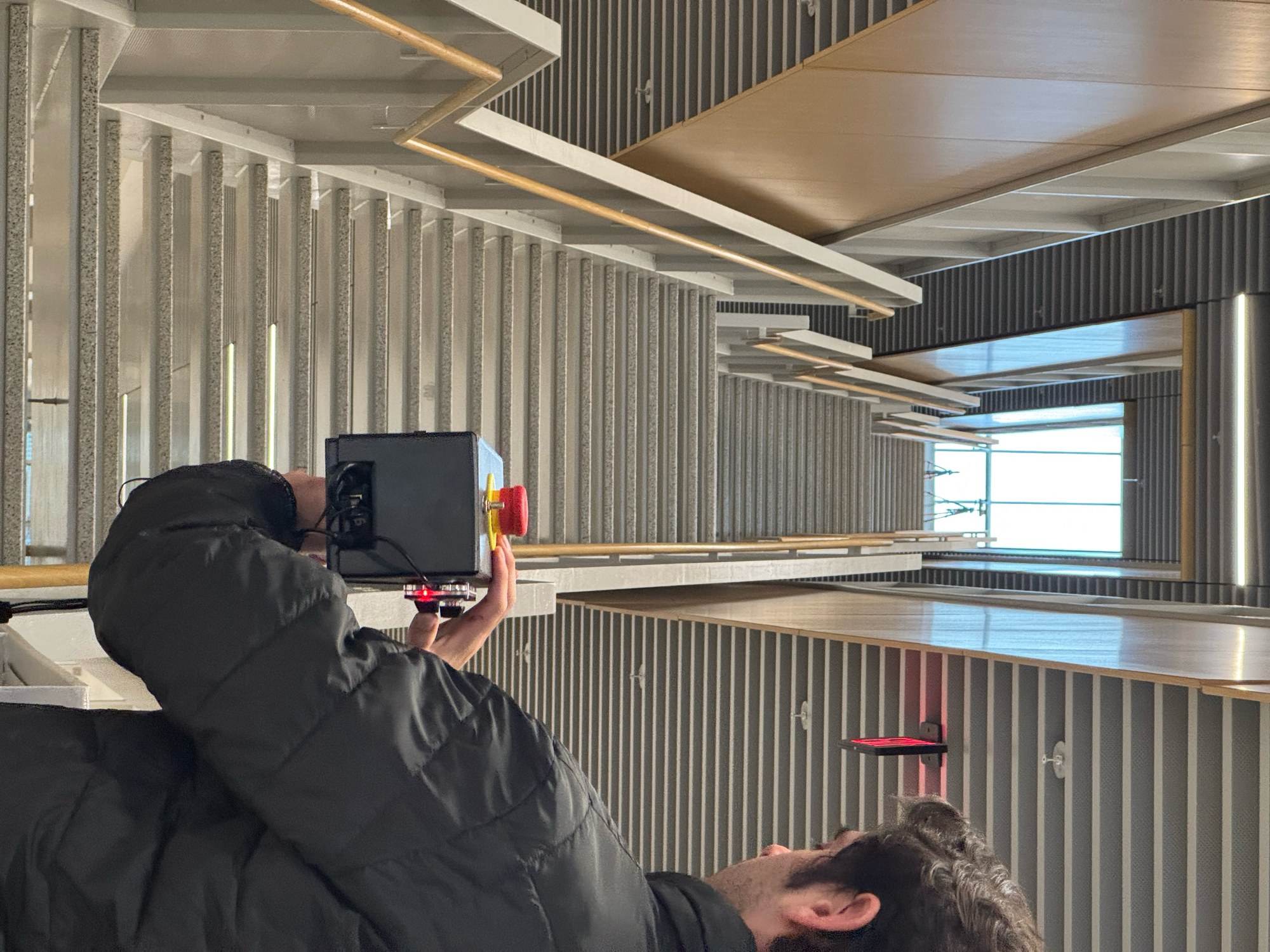}
        \caption{Cross-floor stairwell attacker.}
    \end{subfigure}

    \vspace{4pt}

    \begin{subfigure}[t]{0.32\linewidth}
        \centering
        \vspace{0pt}
        \BeamPanel{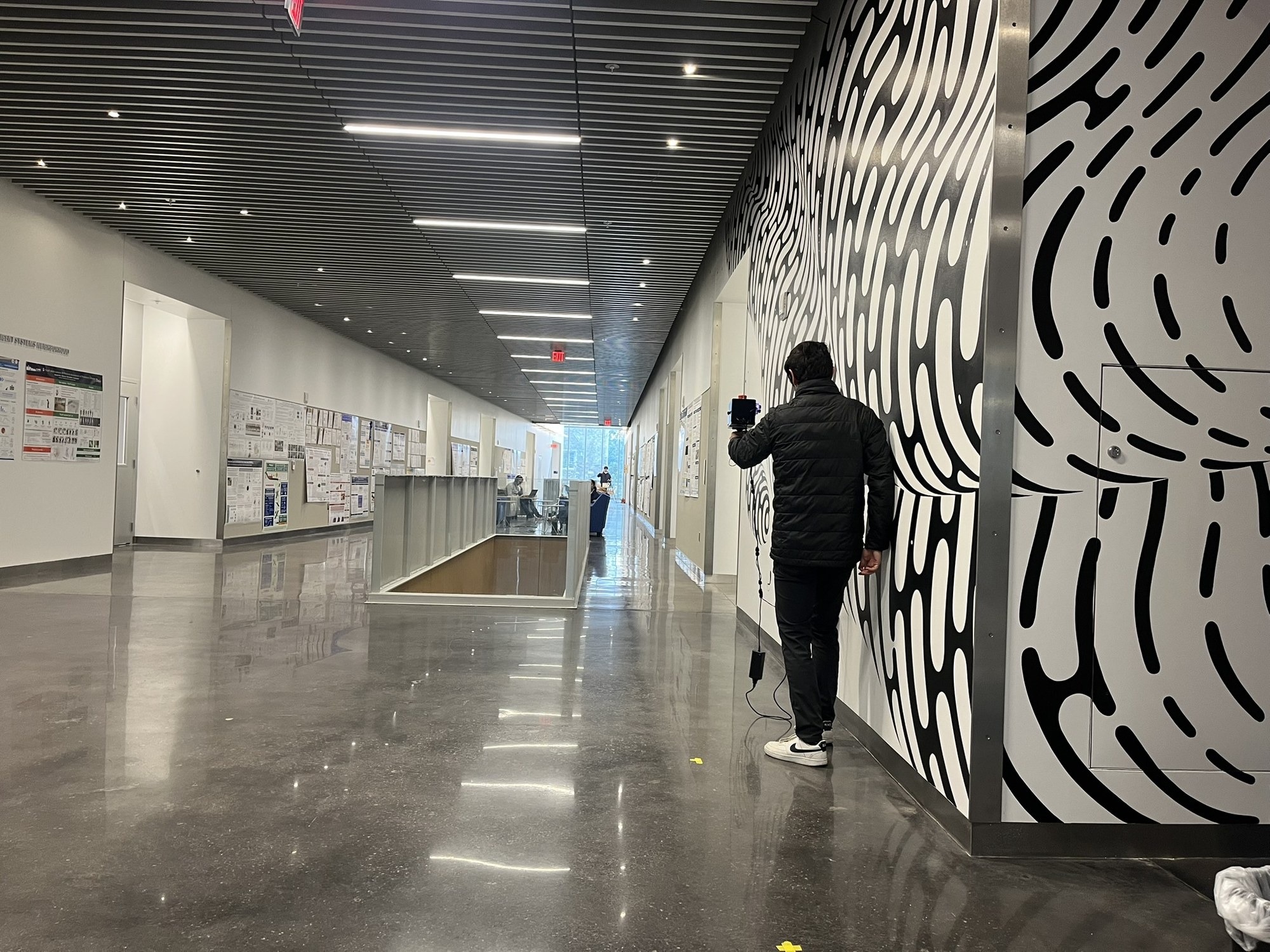}{0.600}{0.522}{0.475}{0.450}
        \caption{Direct same-floor delivery.}
    \end{subfigure}
    \hfill
    \begin{subfigure}[t]{0.32\linewidth}
        \centering
        \vspace{0pt}
        \BeamPanel{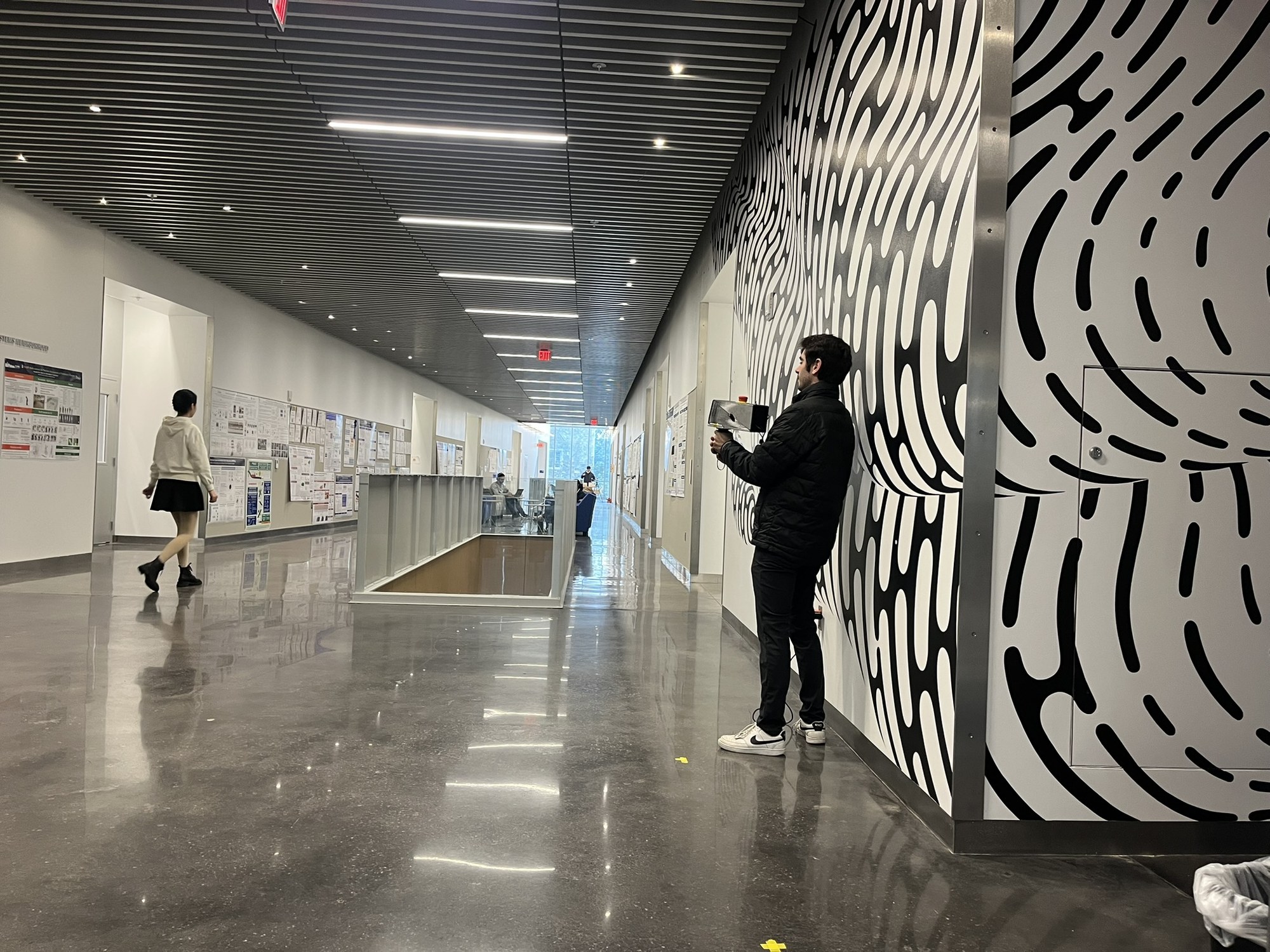}{0.583}{0.527}{0.280}{0.470}
        \caption{Wall-reflection delivery.}
    \end{subfigure}
    \hfill
    \begin{subfigure}[t]{0.32\linewidth}
        \centering
        \vspace{0pt}
        \BeamPanel{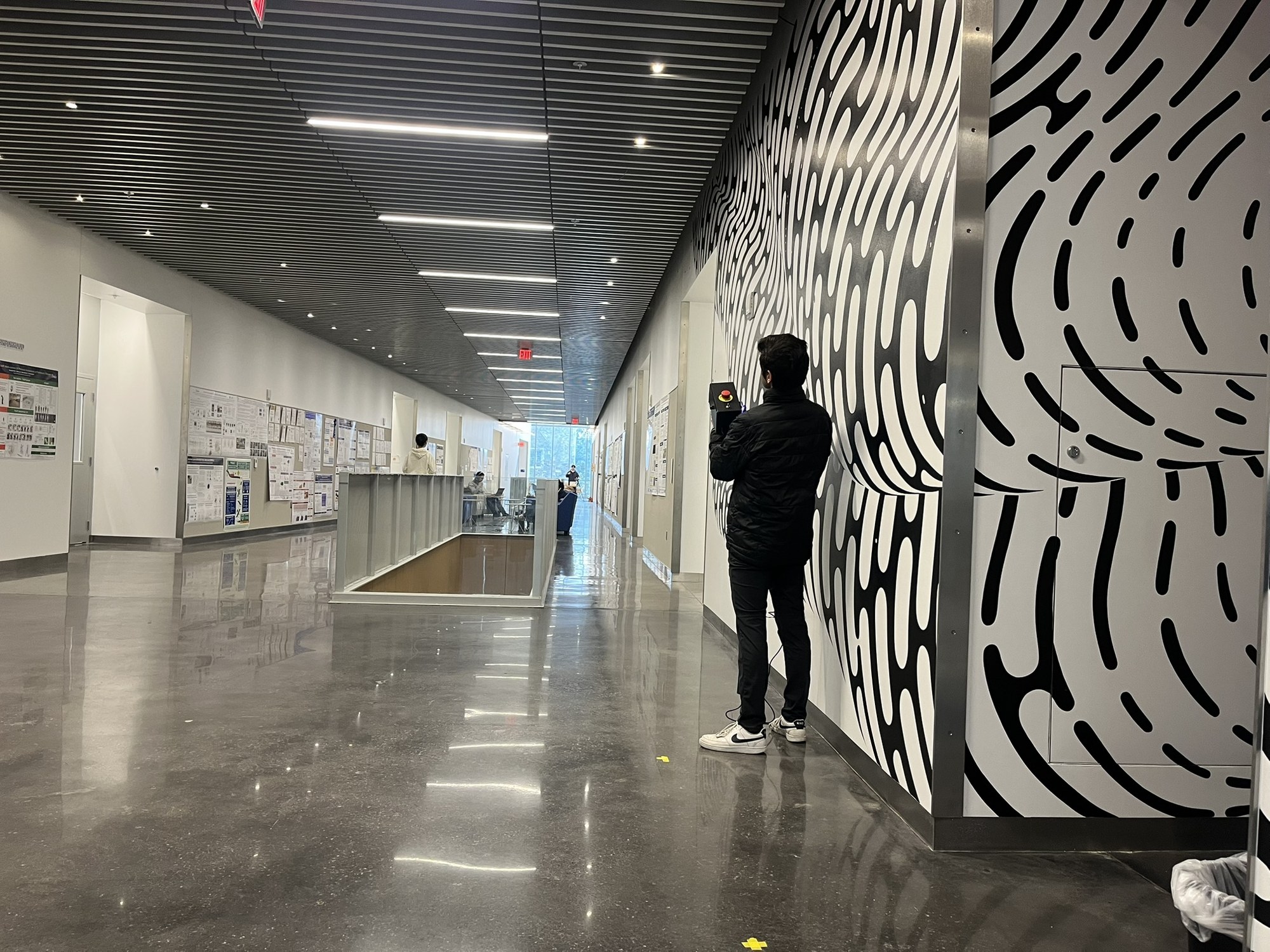}{0.600}{0.585}{0.500}{0.895}
        \caption{Ceiling-reflection delivery.}
    \end{subfigure}
    \caption{Indoor setup variants used in the feasibility-envelope experiments. The diagram summarizes the shared geometry, while the photographs illustrate cross-floor and same-floor delivery conditions, including direct and reflected paths. Red dashed arrows indicate the emitter's aim direction.}
    \label{fig:appendix-setup-indoor}
\end{figure*}

%% file: figures/figure_appendix_setup_outdoor_mobile.tex
\begin{figure*}[p]
    \centering
    \begin{subfigure}[t]{0.46\linewidth}
        \centering
        \vspace{0pt}
        \includegraphics[width=\linewidth,trim=0 0 0 0,clip]{diagrams/experiment-diagram-outdoor-tripod.drawio.pdf}
        \caption{Outdoor tripod setup diagram.}
        \label{fig:appendix-setup-outdoor-diagram}
    \end{subfigure}
    \hfill
    \begin{subfigure}[t]{0.24\linewidth}
        \centering
        \vspace{0pt}
        \includegraphics[angle=-90,width=\linewidth,trim=11.5cm 3cm 6cm 0,clip]{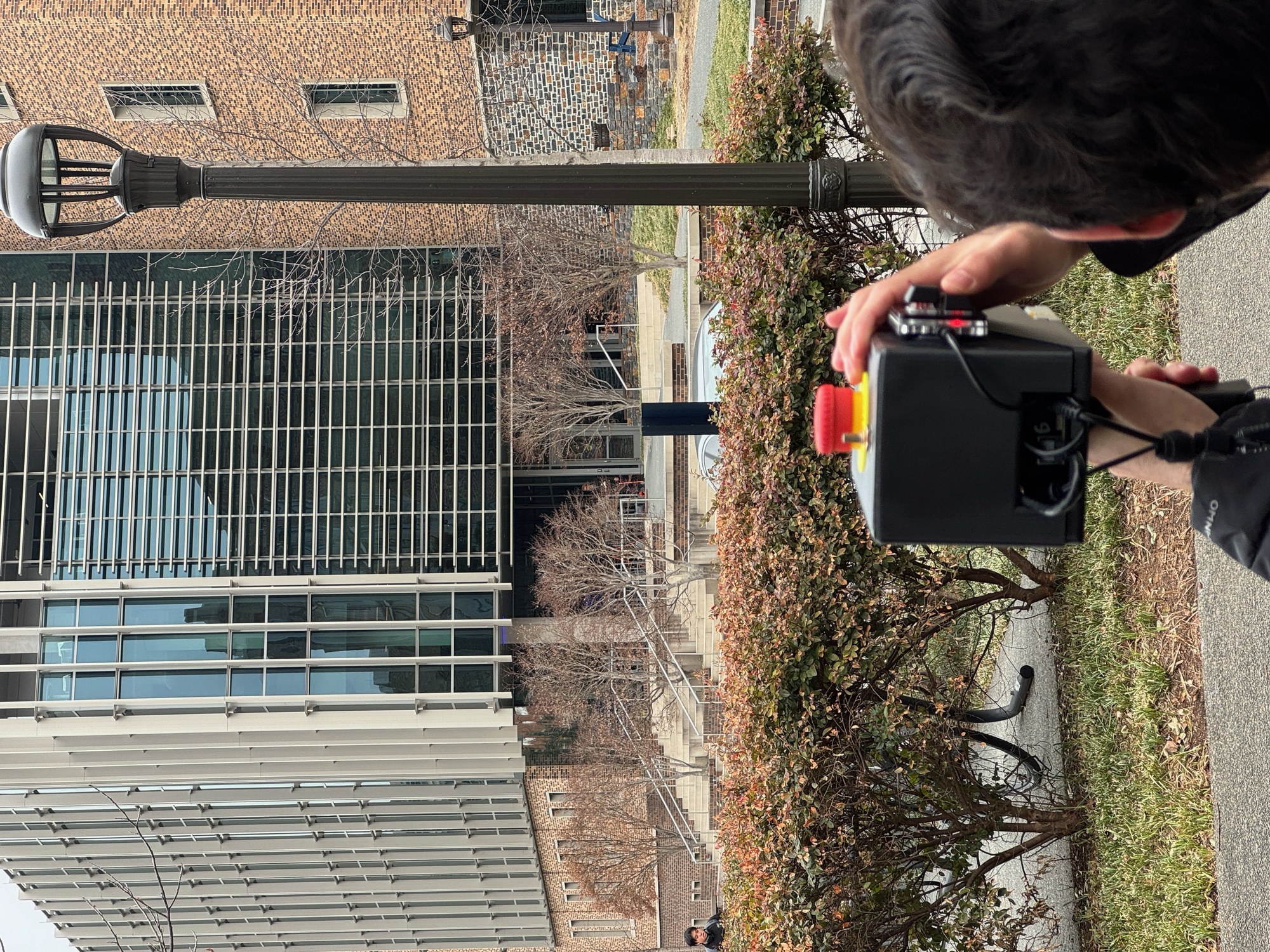}
        \caption{Attacker at 300~ft.}
    \end{subfigure}
    \hfill
    \begin{subfigure}[t]{0.24\linewidth}
        \centering
        \vspace{0pt}
        \includegraphics[width=\linewidth,trim=4cm 1cm 1cm 3cm,clip]{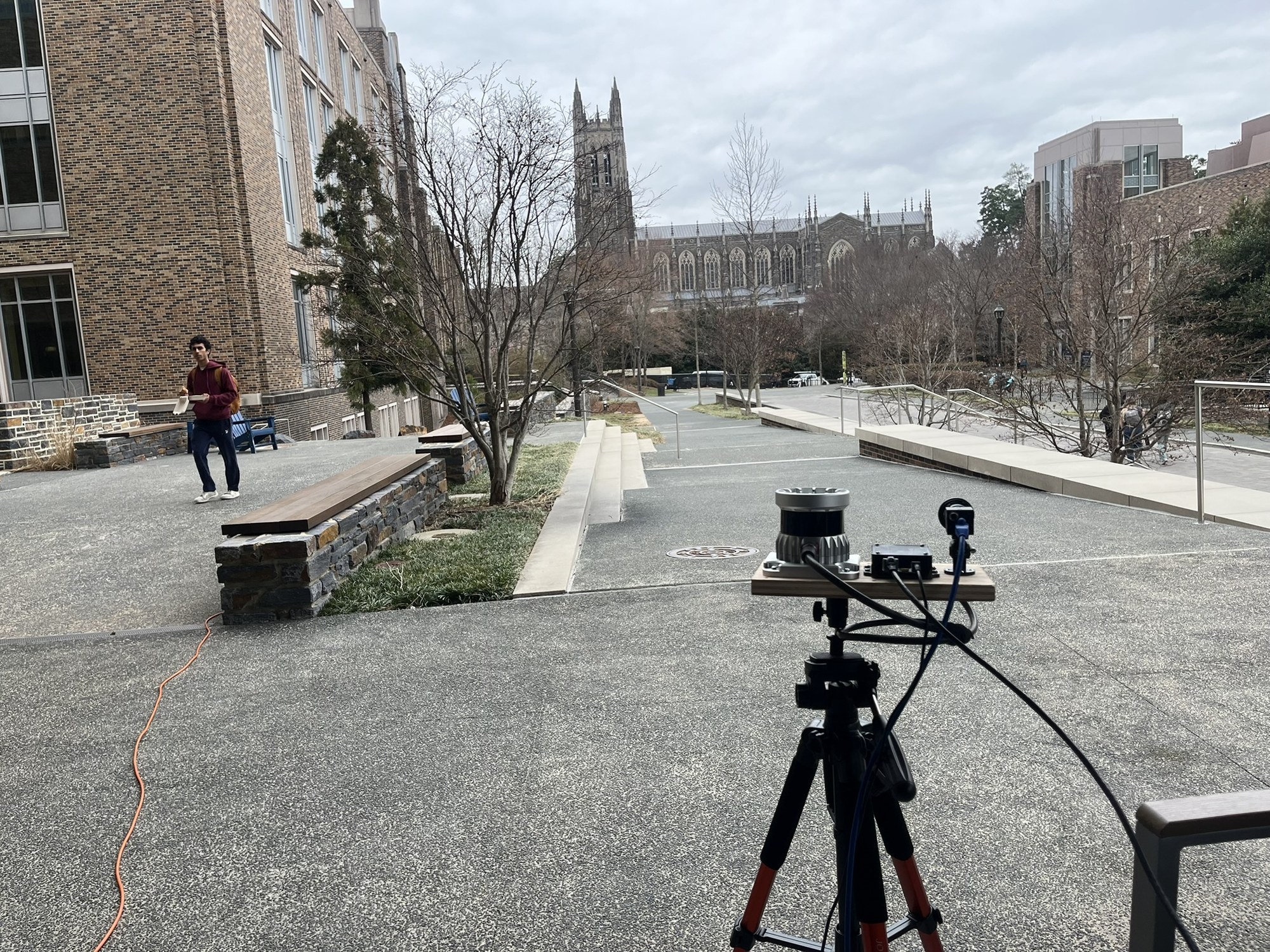}
        \caption{Victim tripod at 300~ft.}
    \end{subfigure}

    \vspace{4pt}

    \begin{subfigure}[t]{0.46\linewidth}
        \centering
        \vspace{0pt}
        \includegraphics[width=\linewidth,trim=0 3cm 0 2.5cm,clip]{diagrams/experiment-diagram-outdoor-car.drawio.pdf}
        \caption{Mobile drive-by setup diagram.}
        \label{fig:appendix-setup-mobile-diagram}
    \end{subfigure}
    \hfill
    \begin{subfigure}[t]{0.24\linewidth}
        \centering
        \vspace{0pt}
        \includegraphics[width=\linewidth,trim=4cm 1.5cm 0 2cm,clip]{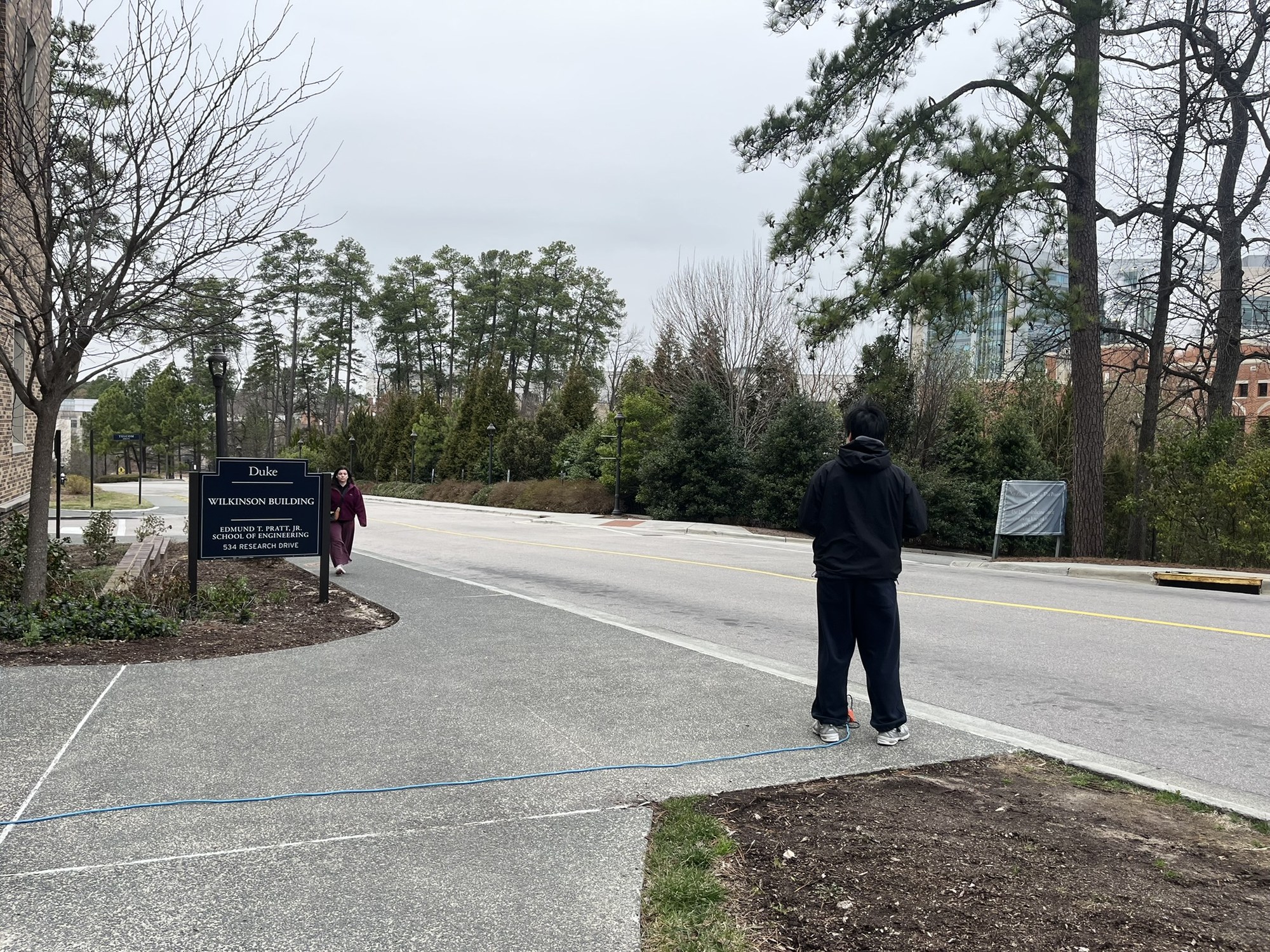}
        \caption{Roadside attacker position.}
    \end{subfigure}
    \hfill
    \begin{subfigure}[t]{0.24\linewidth}
        \centering
        \vspace{0pt}
        \includegraphics[width=\linewidth,trim=1cm 1.9cm 1cm 0,clip]{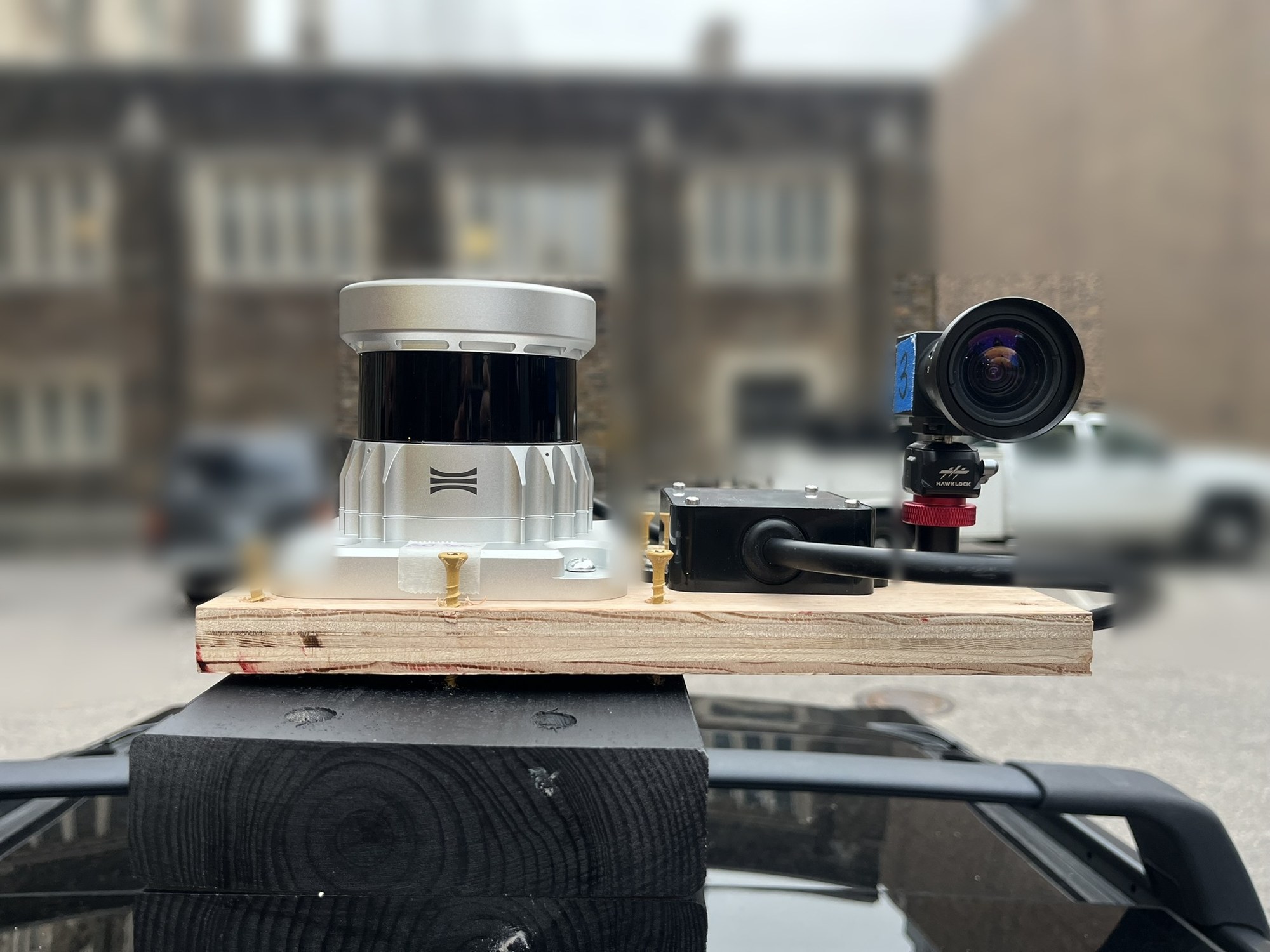}
        \caption{Vehicle-mounted victim sensor.}
    \end{subfigure}
    \caption{Outdoor static and mobile setup variants used in the feasibility-envelope experiments. The top row illustrates the 300~ft tripod configuration, and the bottom row illustrates the mobile drive-by configuration.}
    \label{fig:appendix-setup-outdoor-mobile}
\end{figure*}

%% file: figures/figure_appendix_triptych_static.tex
\begin{figure*}[t]
    \centering
    \setlength{\tabcolsep}{4pt}
    \begin{tabular}{>{\centering\arraybackslash}m{0.14\textwidth} >{\centering\arraybackslash}m{0.27\textwidth} >{\centering\arraybackslash}m{0.27\textwidth} >{\centering\arraybackslash}m{0.27\textwidth}}
        \toprule
        \textbf{Scenario} & \textbf{Camera} & \textbf{LiDAR Before Attack} & \textbf{LiDAR After Attack} \\
        \midrule
        \textbf{\textit{Injection}} Indoor direct &
        \TriptychCamera{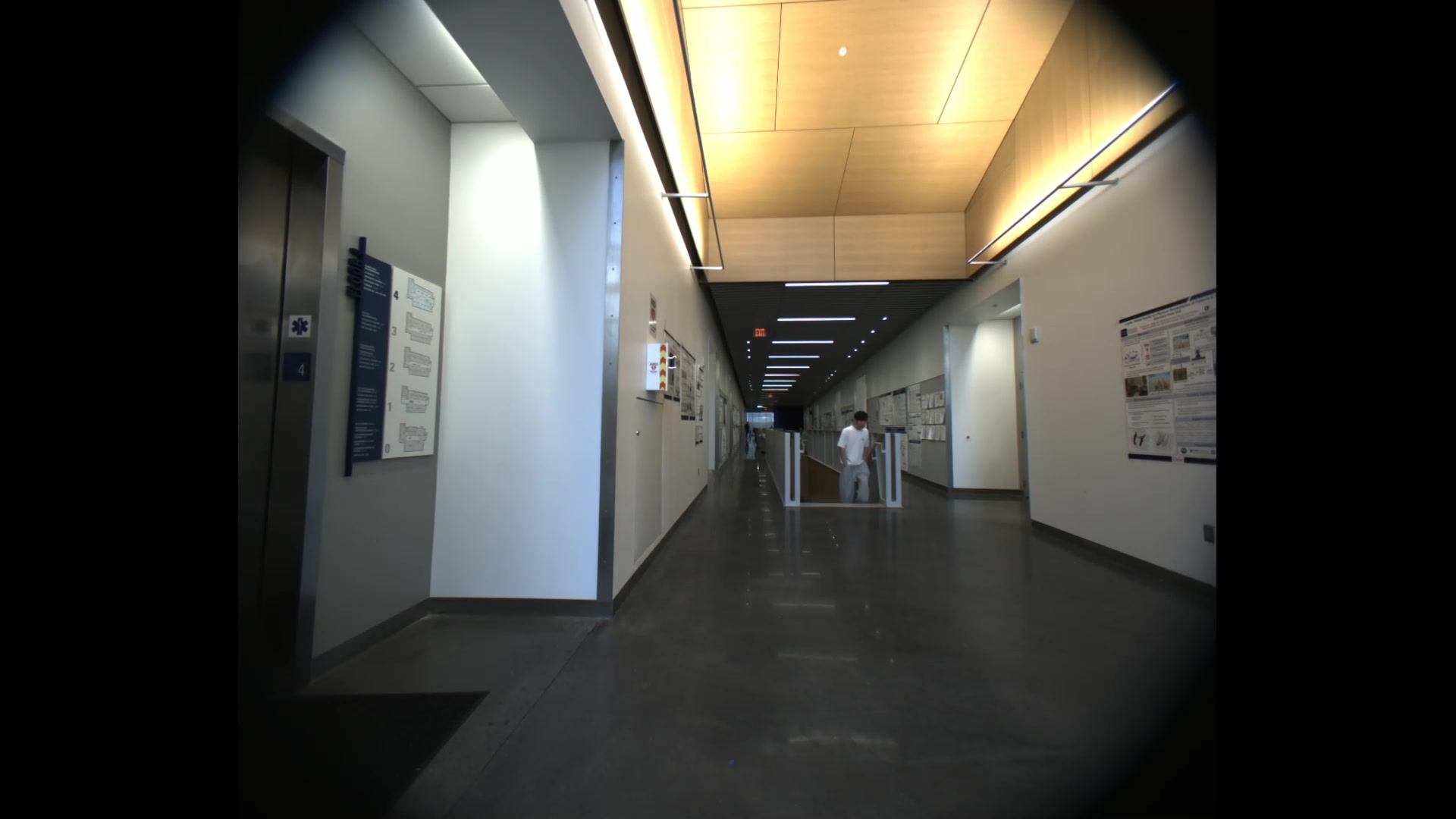} &
        \TriptychLidar{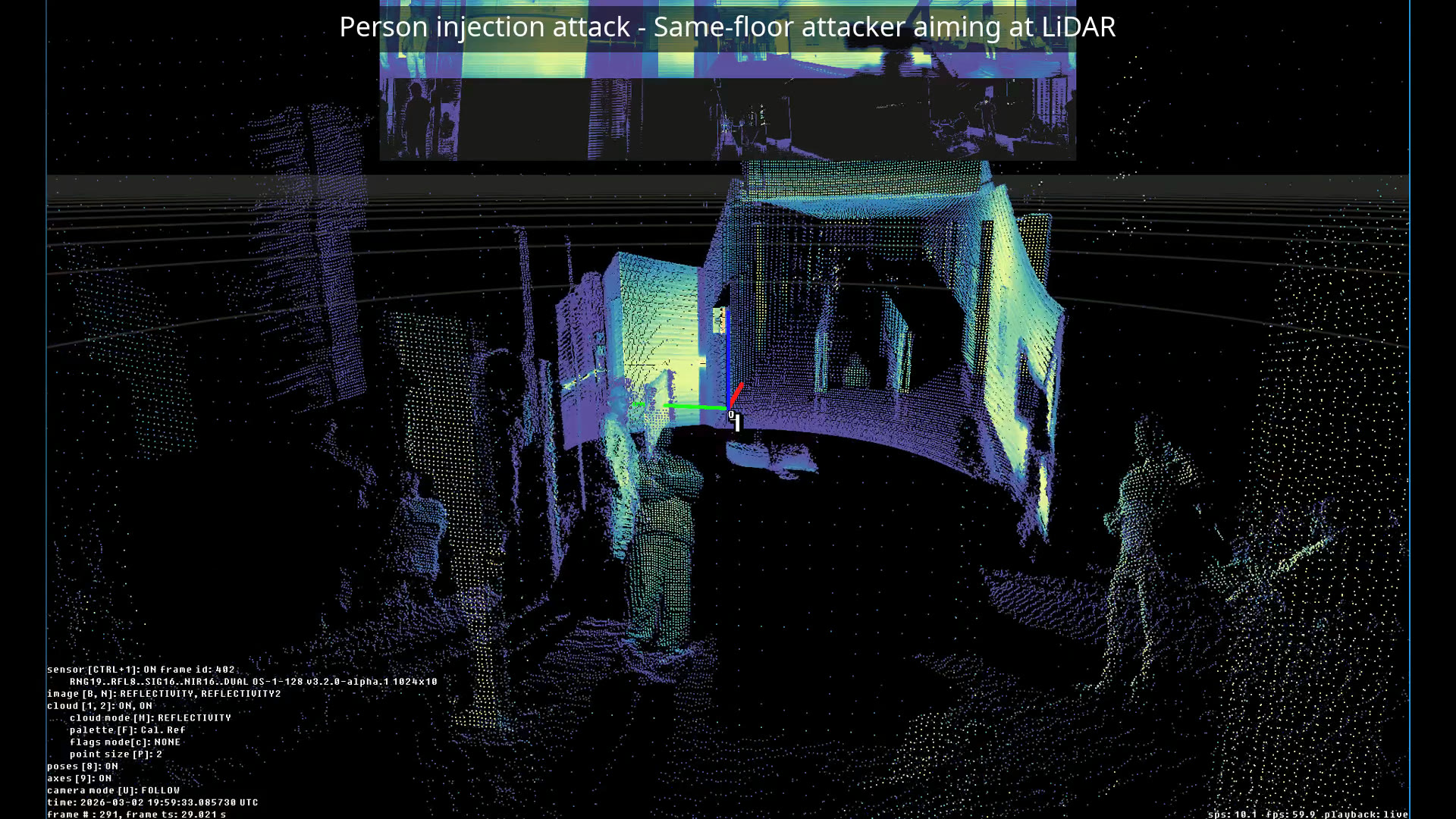} &
        \TriptychLidarInj{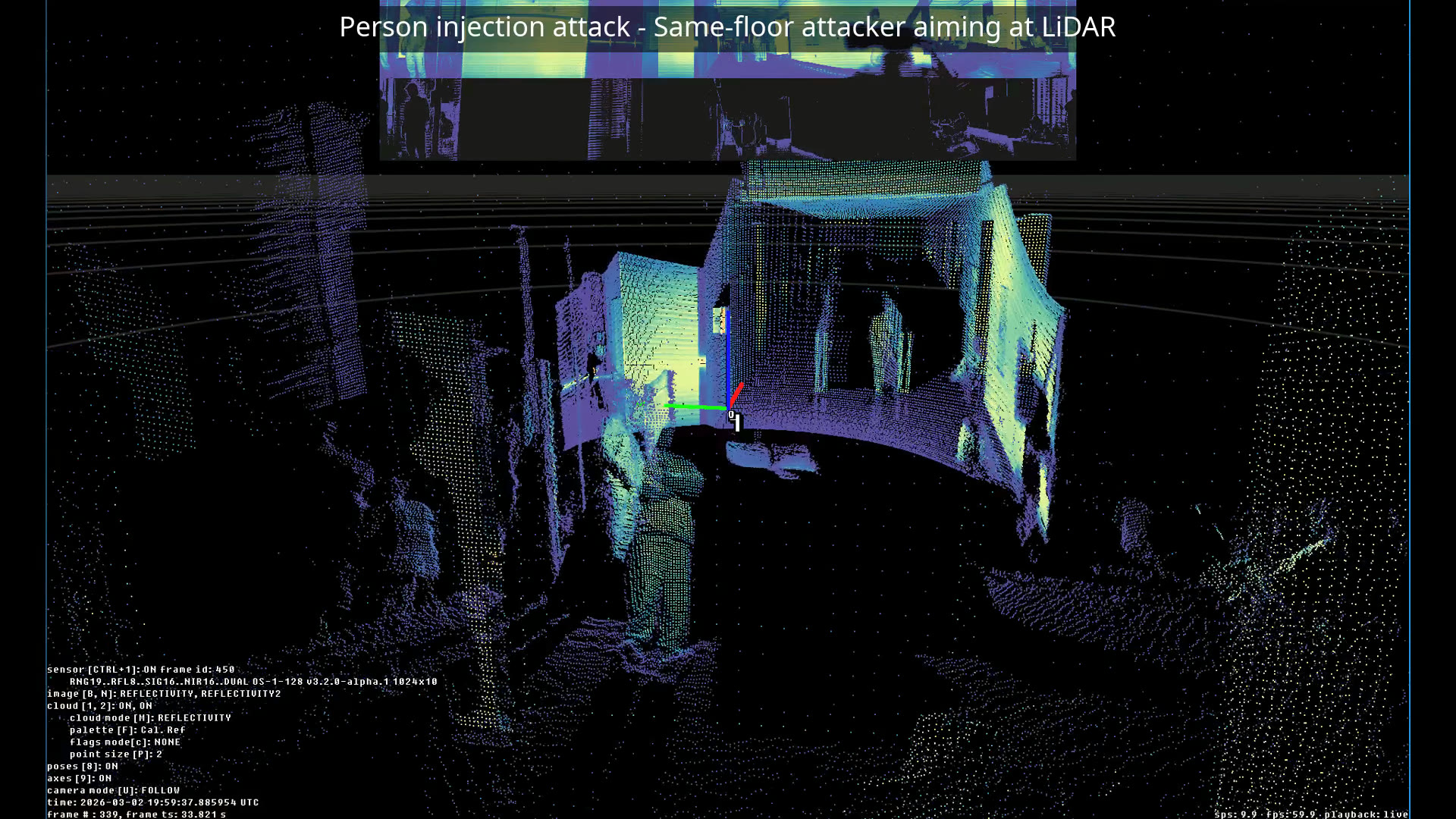}{0.171}{0.397}{0.199}{0.272} \\

        \textbf{\textit{Suppression}} Indoor wall reflection &
        \TriptychCamera{images/triptych_curated/run-indoor-lab-2/camera.jpg} &
        \TriptychLidar{images/triptych_curated/run-indoor-lab-2/lidar_before_unboxed.jpg} &
        \TriptychLidarSup{images/triptych_curated/run-indoor-lab-2/lidar_after_unboxed.jpg} \\

        \textbf{\textit{Suppression}} Indoor ceiling reflection &
        \TriptychCamera{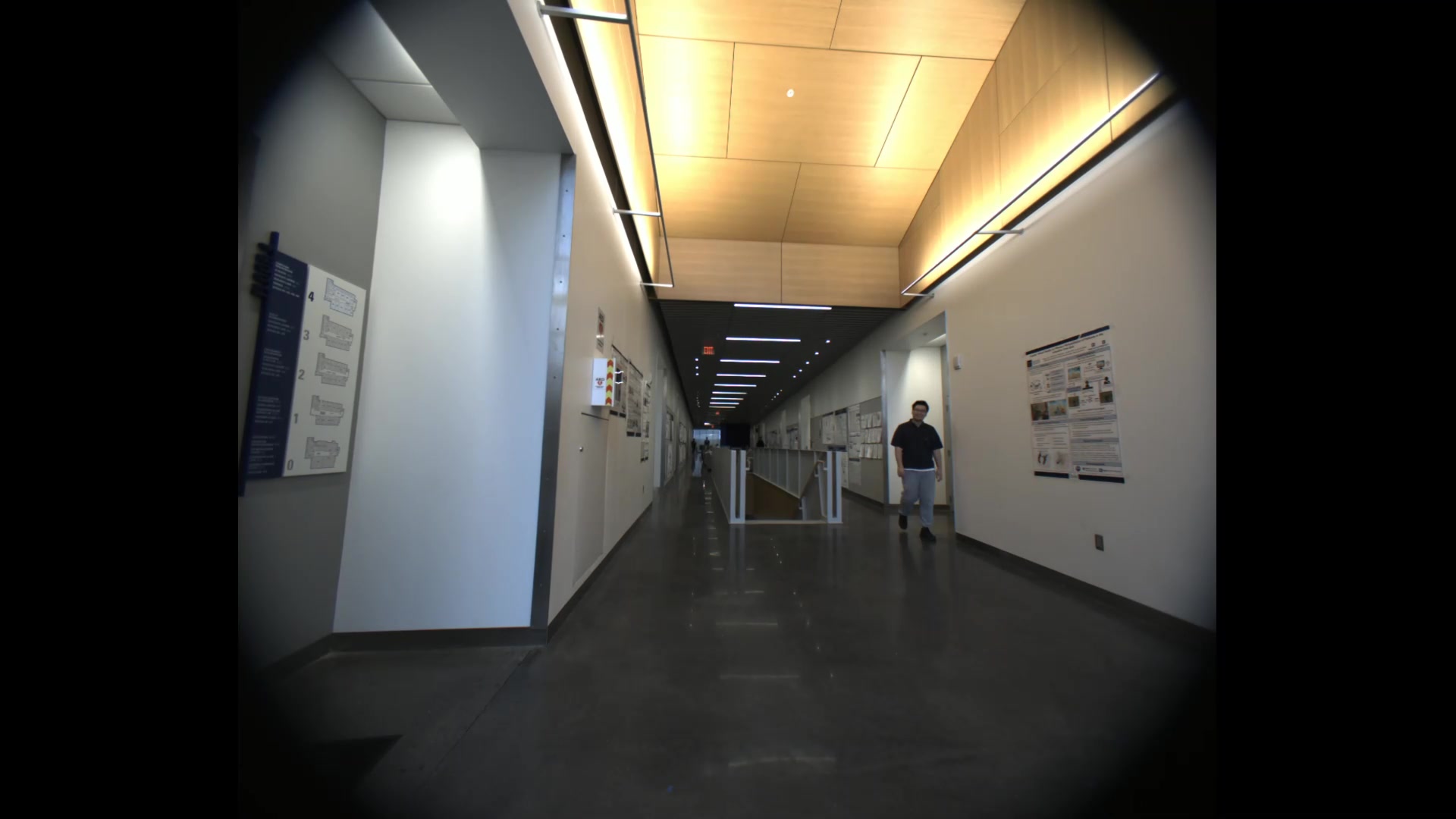} &
        \TriptychLidar{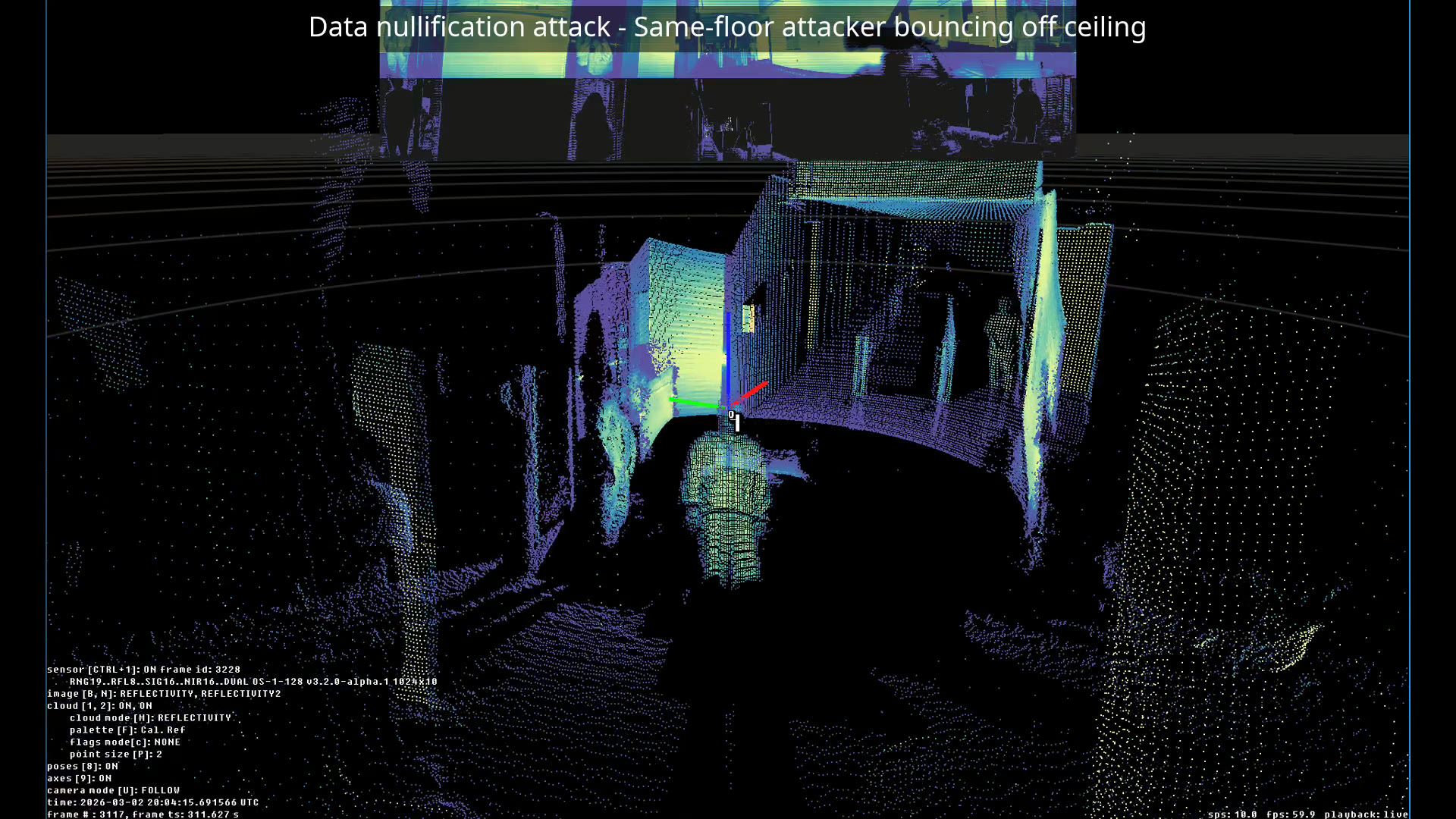} &
        \TriptychLidarSup{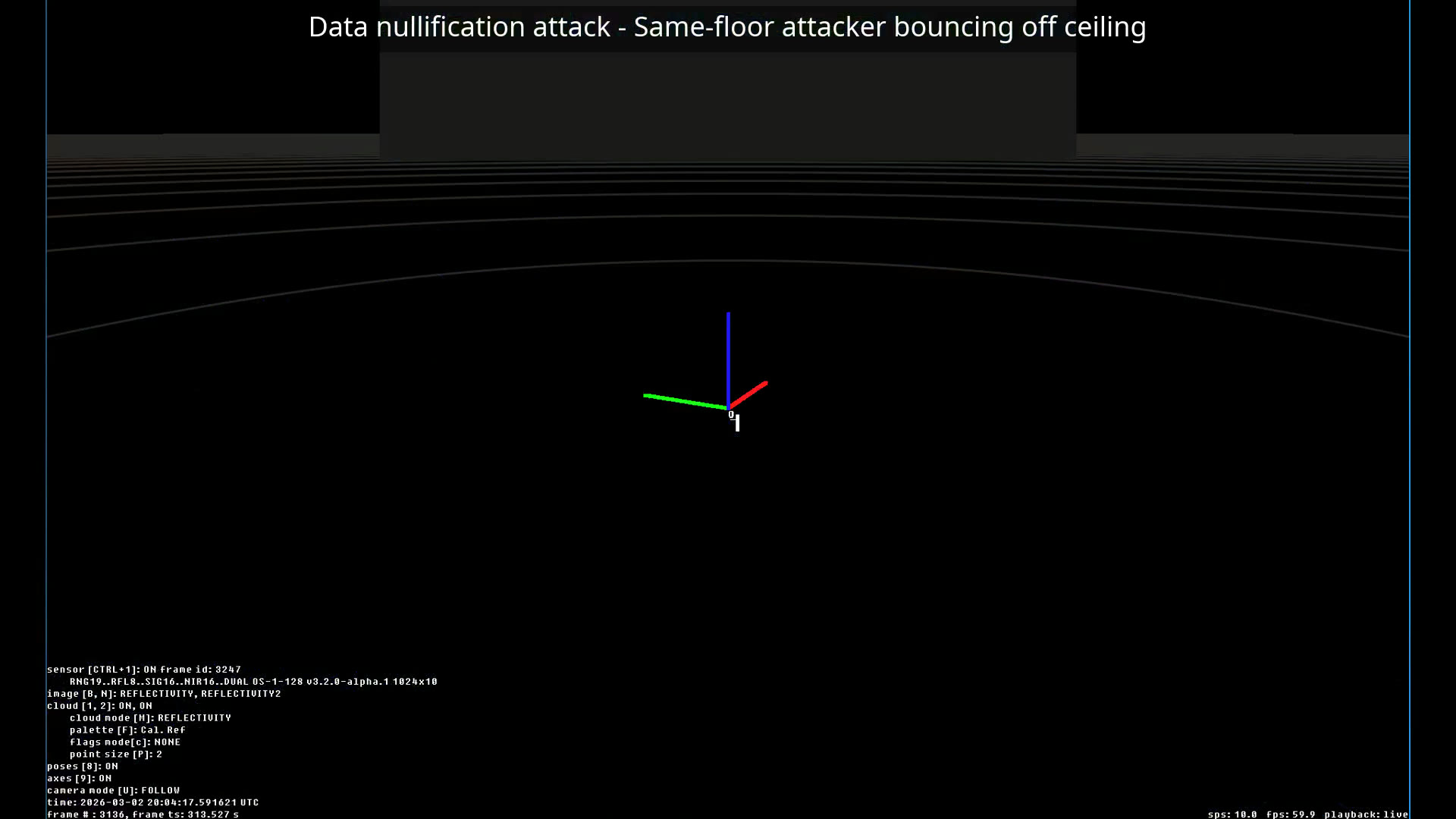} \\

        \textbf{\textit{Injection}} Indoor stairwell (2nd floor attacker) &
        \TriptychCamera{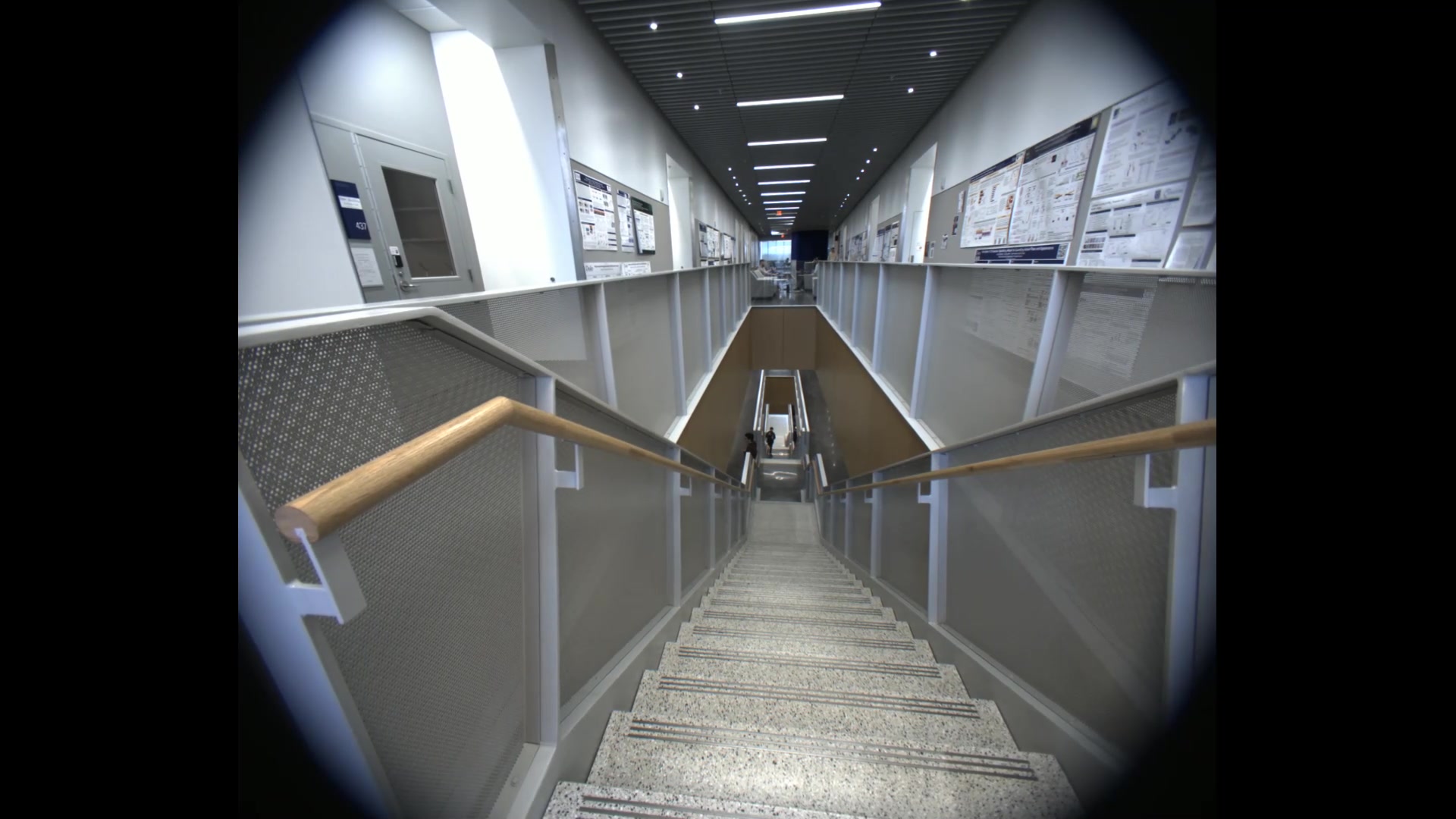} &
        \TriptychLidar{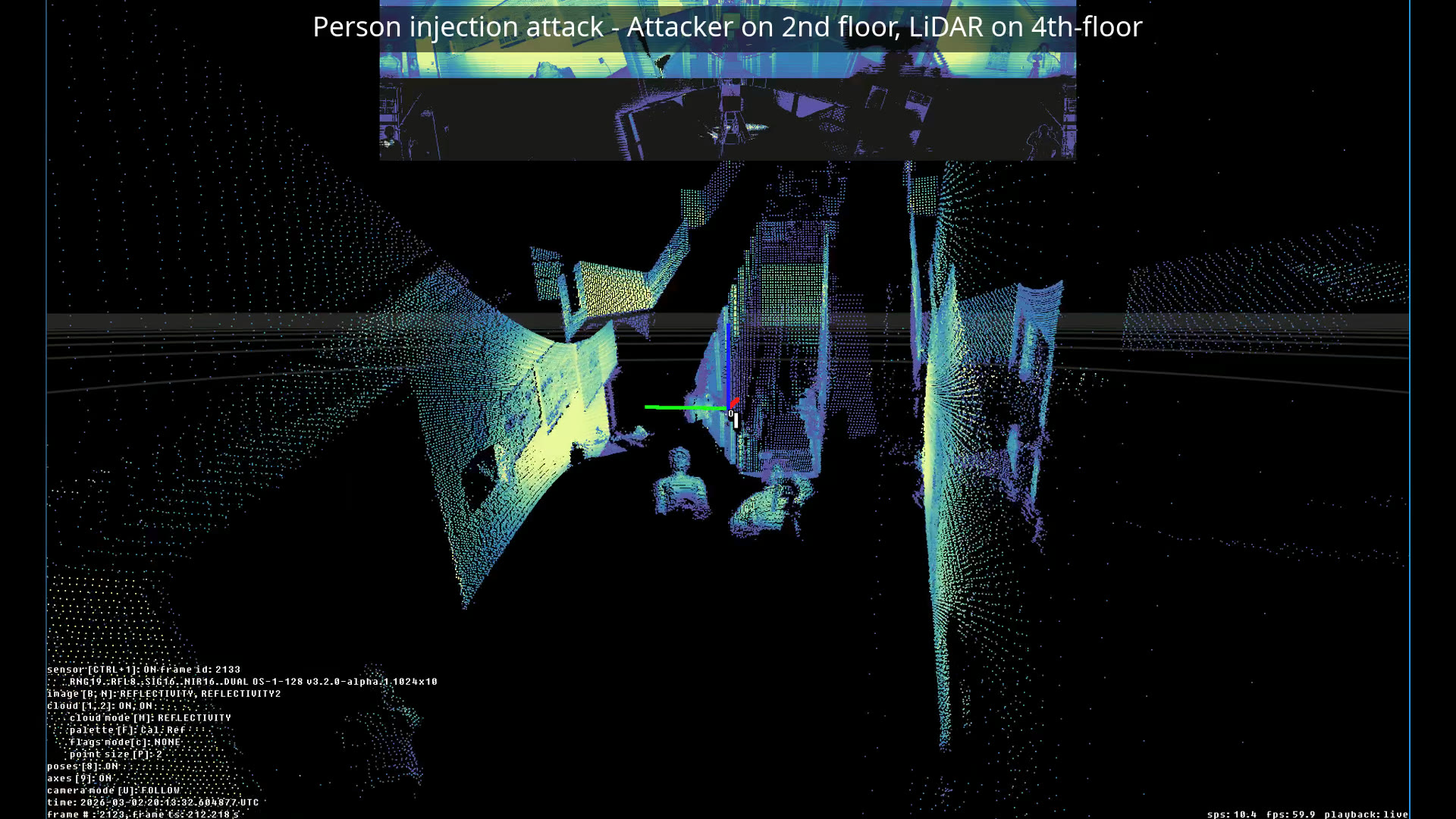} &
        \TriptychLidarInj{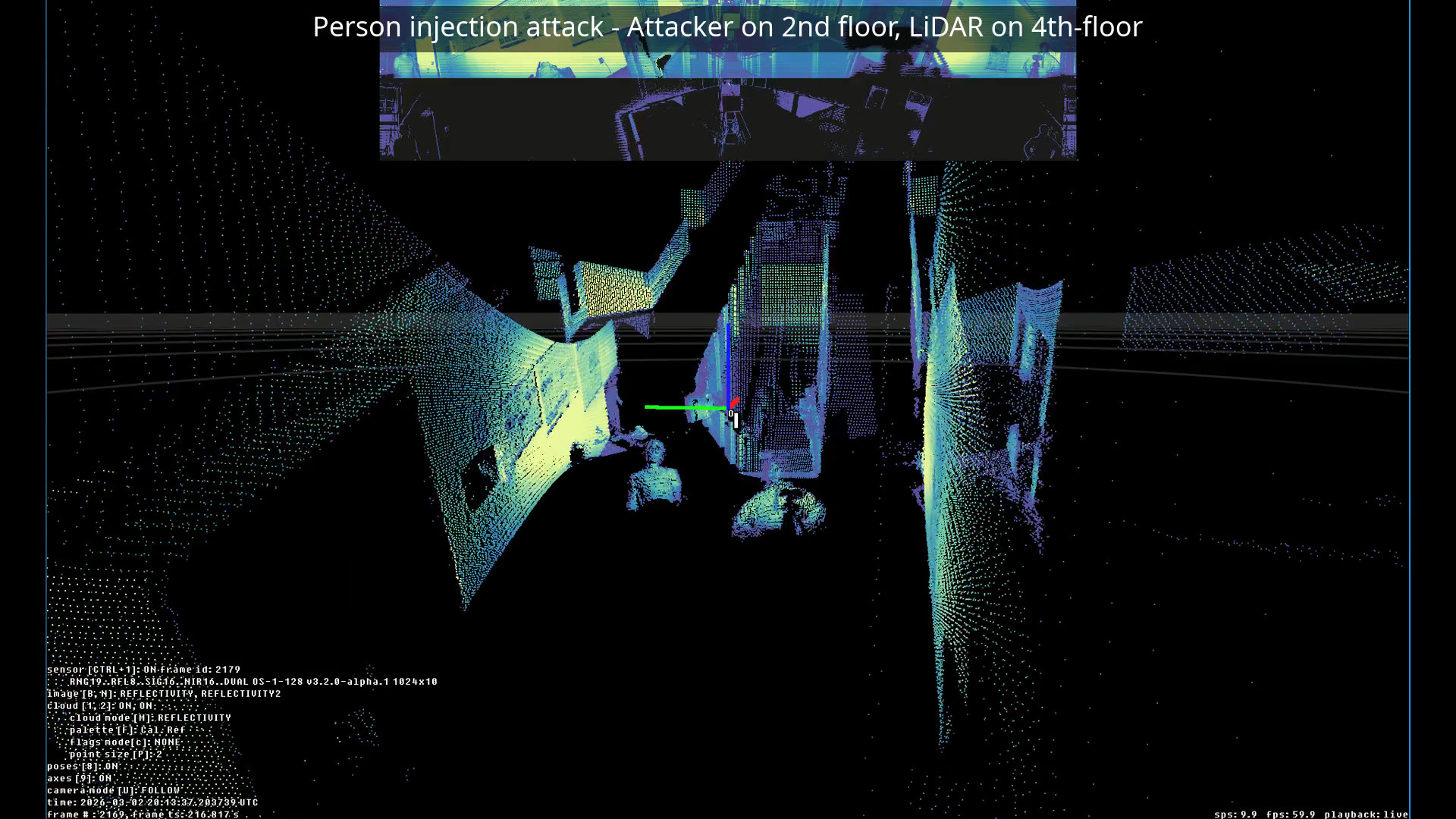}{0.232}{0.375}{0.149}{0.181} \\

        \textbf{\textit{Suppression}} Indoor stairwell (2nd floor attacker) &
        \TriptychCamera{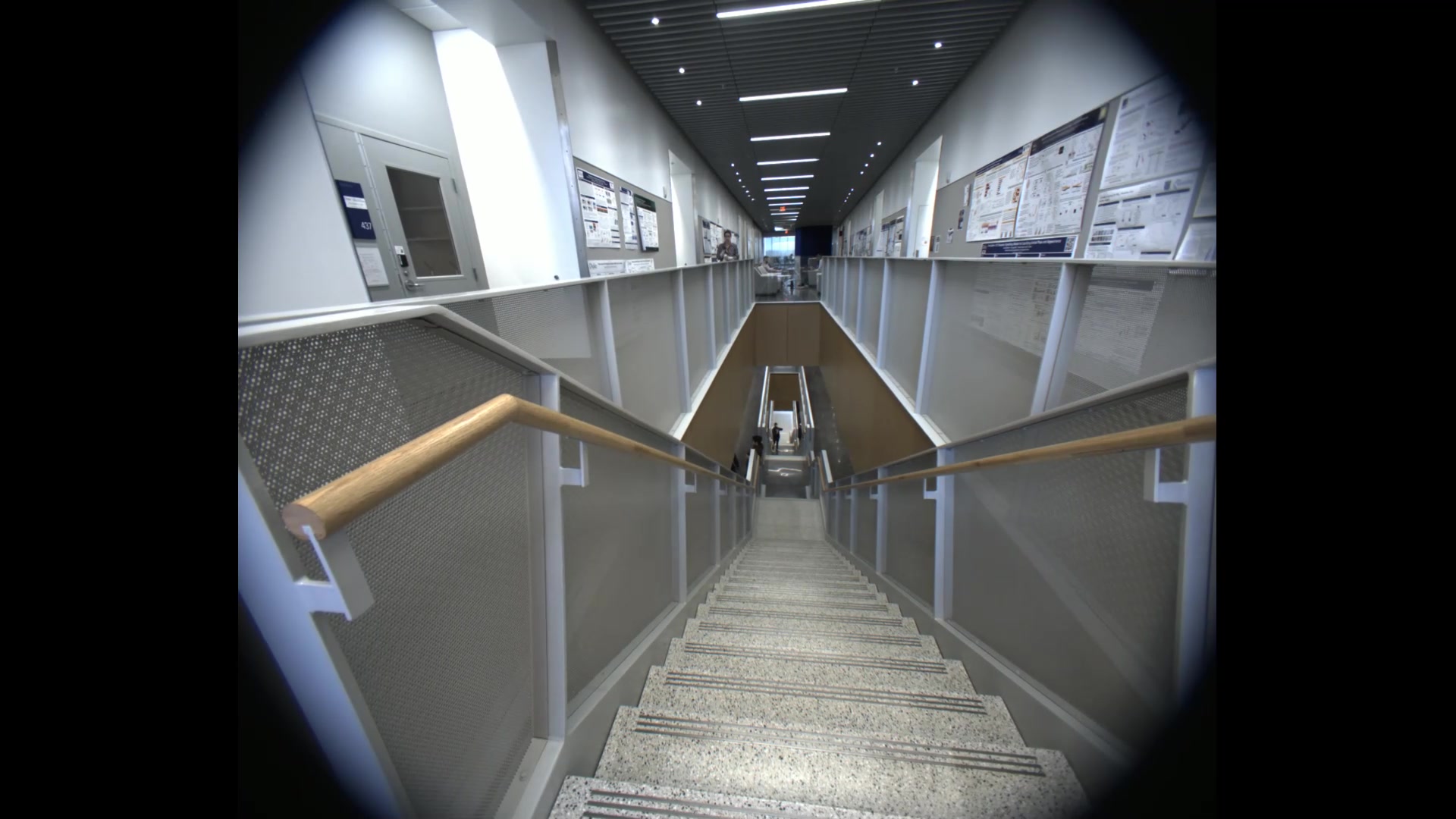} &
        \TriptychLidar{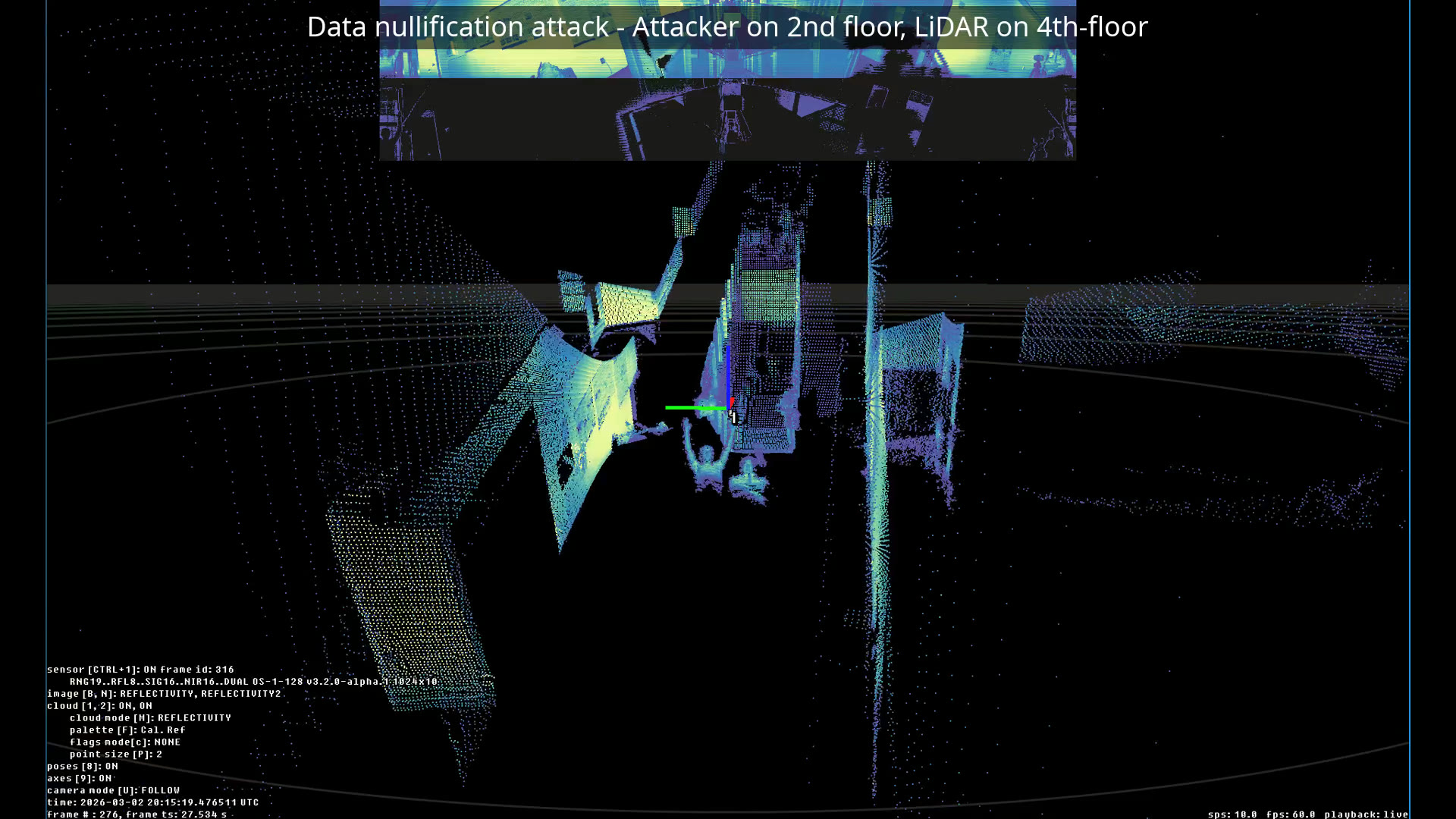} &
        \TriptychLidarSup{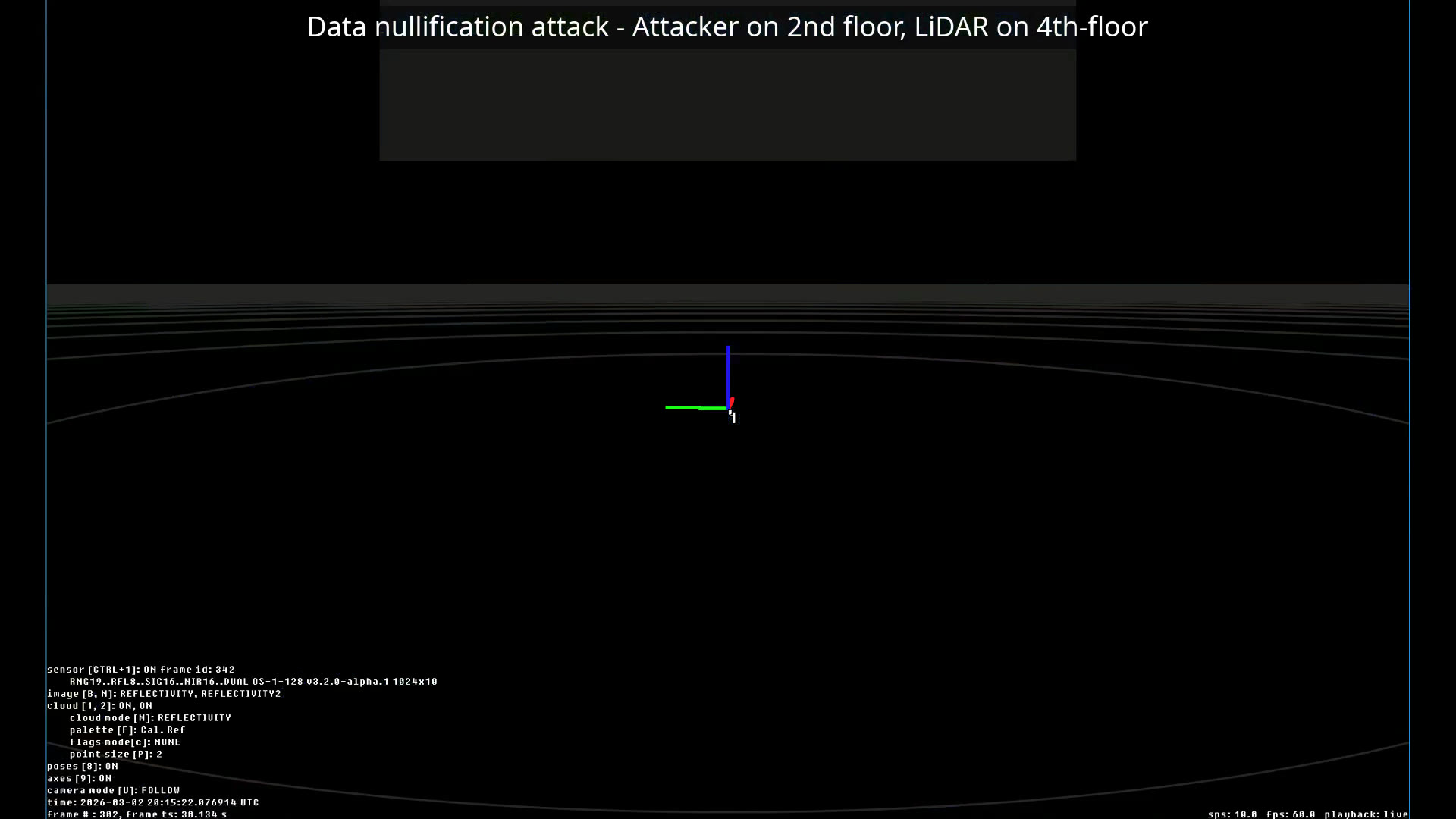} \\

        \textbf{\textit{Suppression}} Outdoor static (300~ft) &
        \TriptychCamera{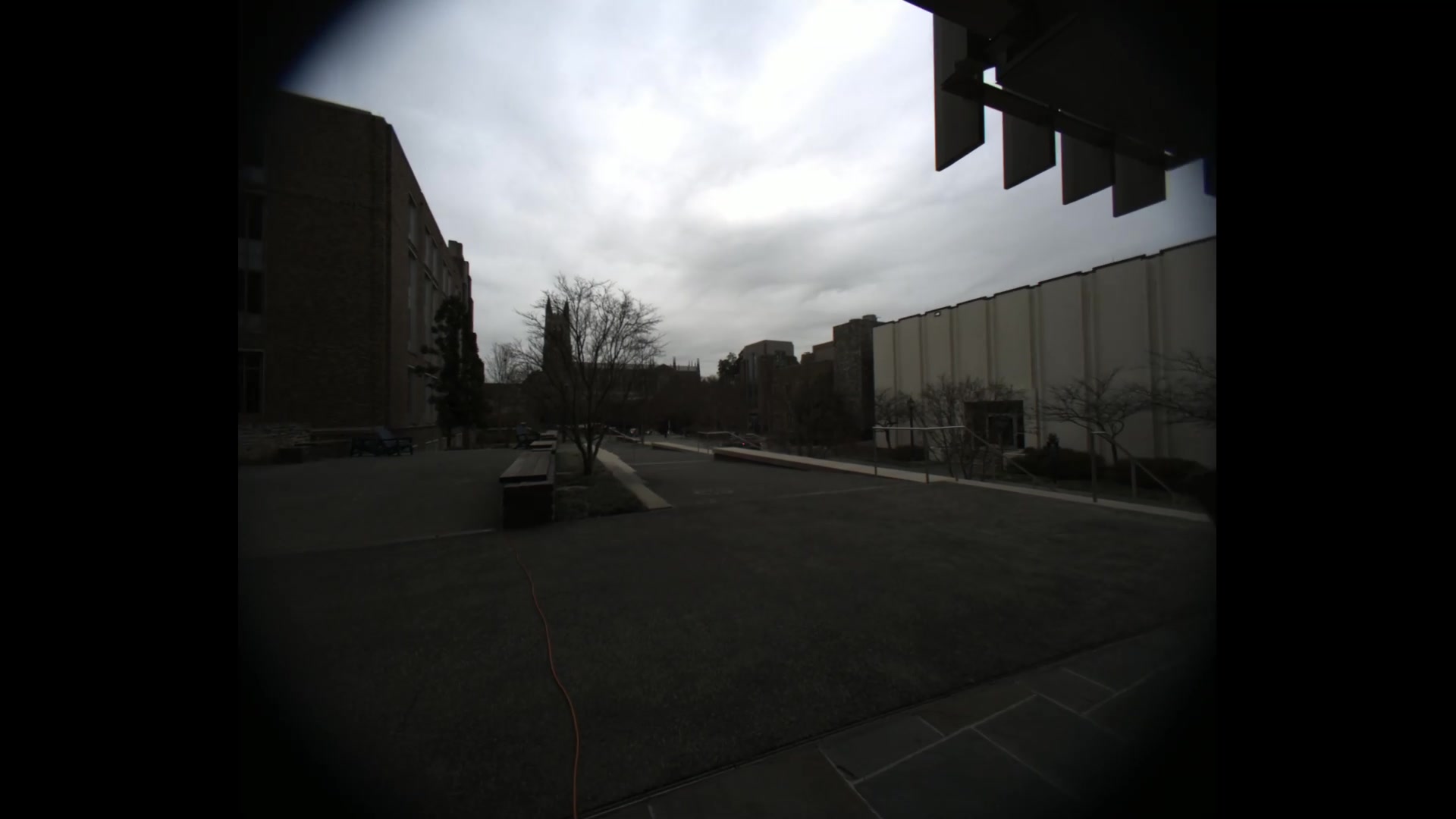} &
        \TriptychLidar{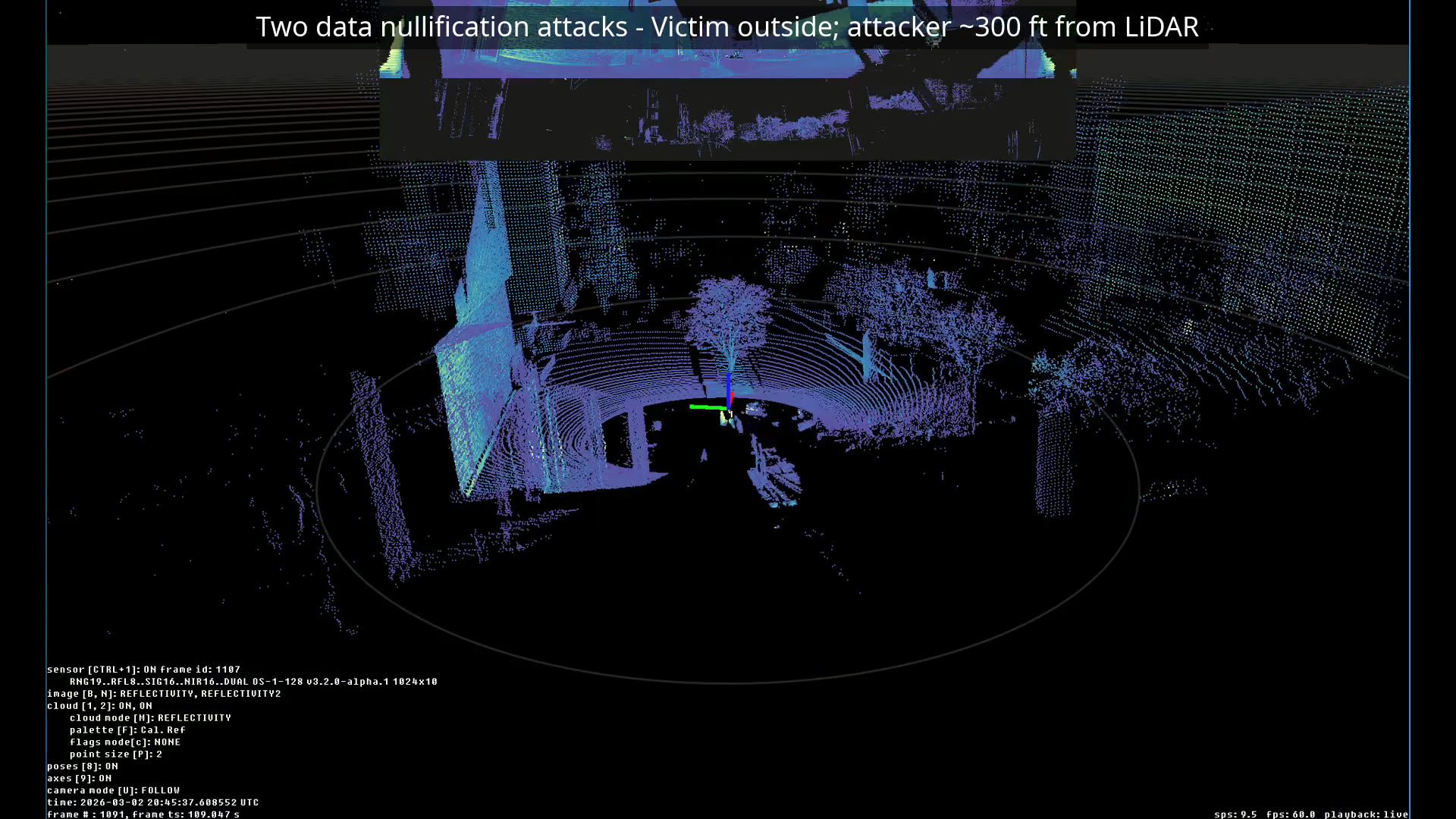} &
        \TriptychLidarSup{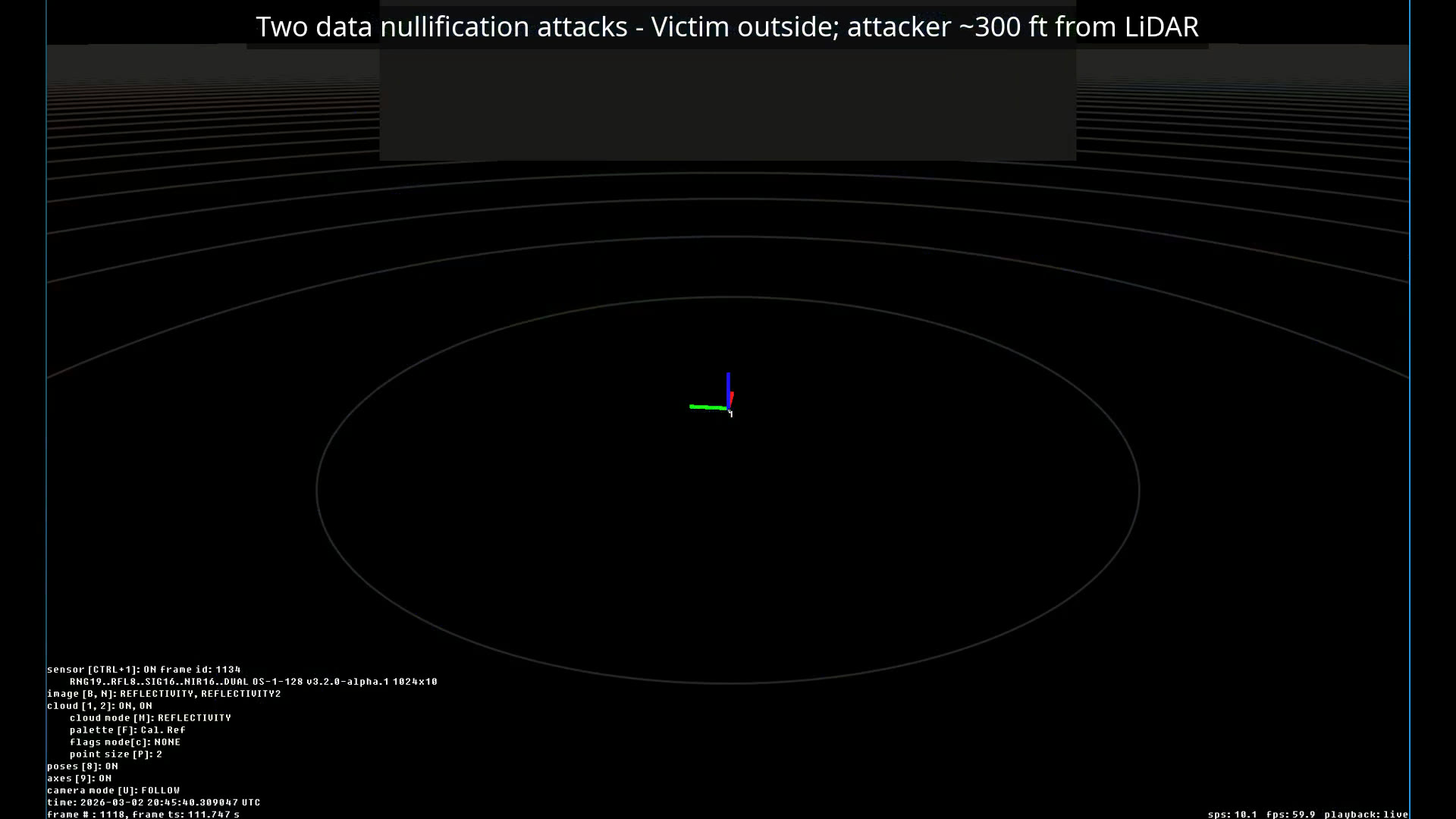} \\
        \bottomrule
    \end{tabular}
    \caption{Additional static-setting examples using the same triptych schema as Fig.~\ref{fig:main-feasibility-triptych}. Dashed outlines mark injected returns (author annotations, not detector outputs).}
    \label{fig:appendix-static-triptych}
\end{figure*}

%% file: figures/figure_appendix_triptych_mobile.tex
\begin{figure*}[t]
    \centering
    \setlength{\tabcolsep}{4pt}
    \begin{tabular}{>{\centering\arraybackslash}m{0.14\textwidth} >{\centering\arraybackslash}m{0.27\textwidth} >{\centering\arraybackslash}m{0.27\textwidth} >{\centering\arraybackslash}m{0.27\textwidth}}
        \toprule
        \textbf{Scenario} & \textbf{Camera} & \textbf{LiDAR Before Attack} & \textbf{LiDAR After Attack} \\
        \midrule
        \textbf{\textit{Suppression}} Drive-by pass 1 &
        \TriptychCamera{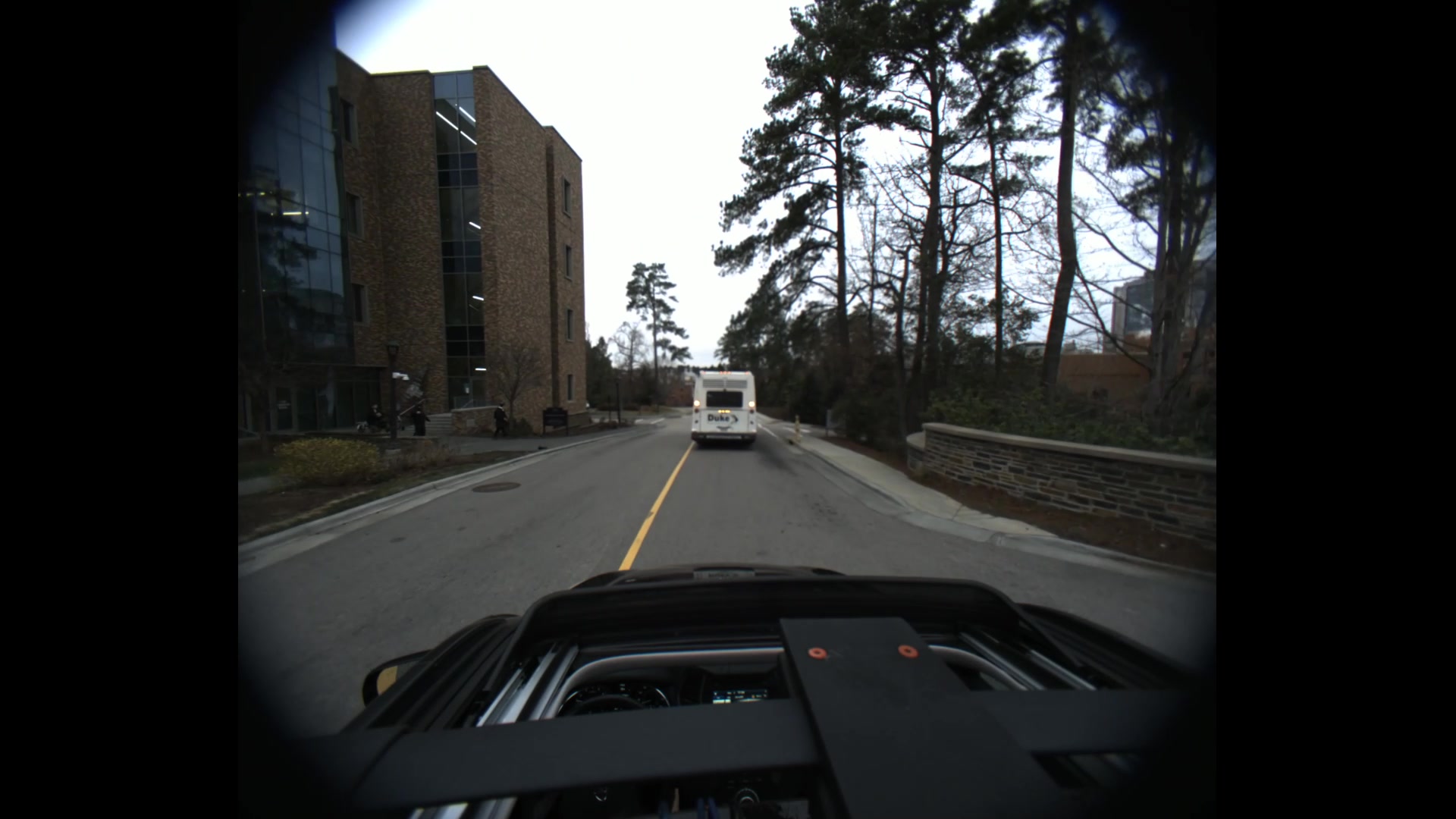} &
        \TriptychLidar{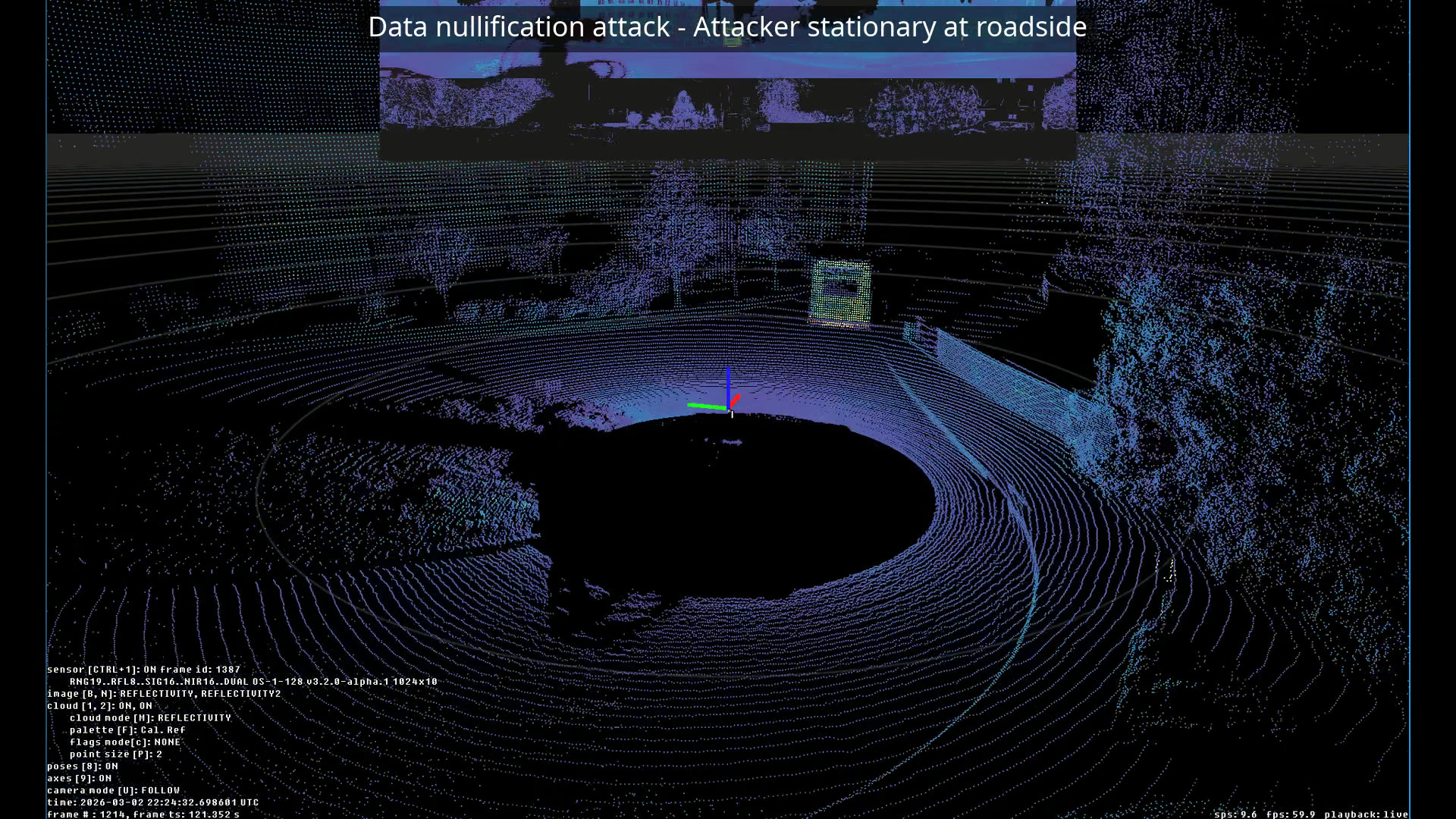} &
        \TriptychLidarSup{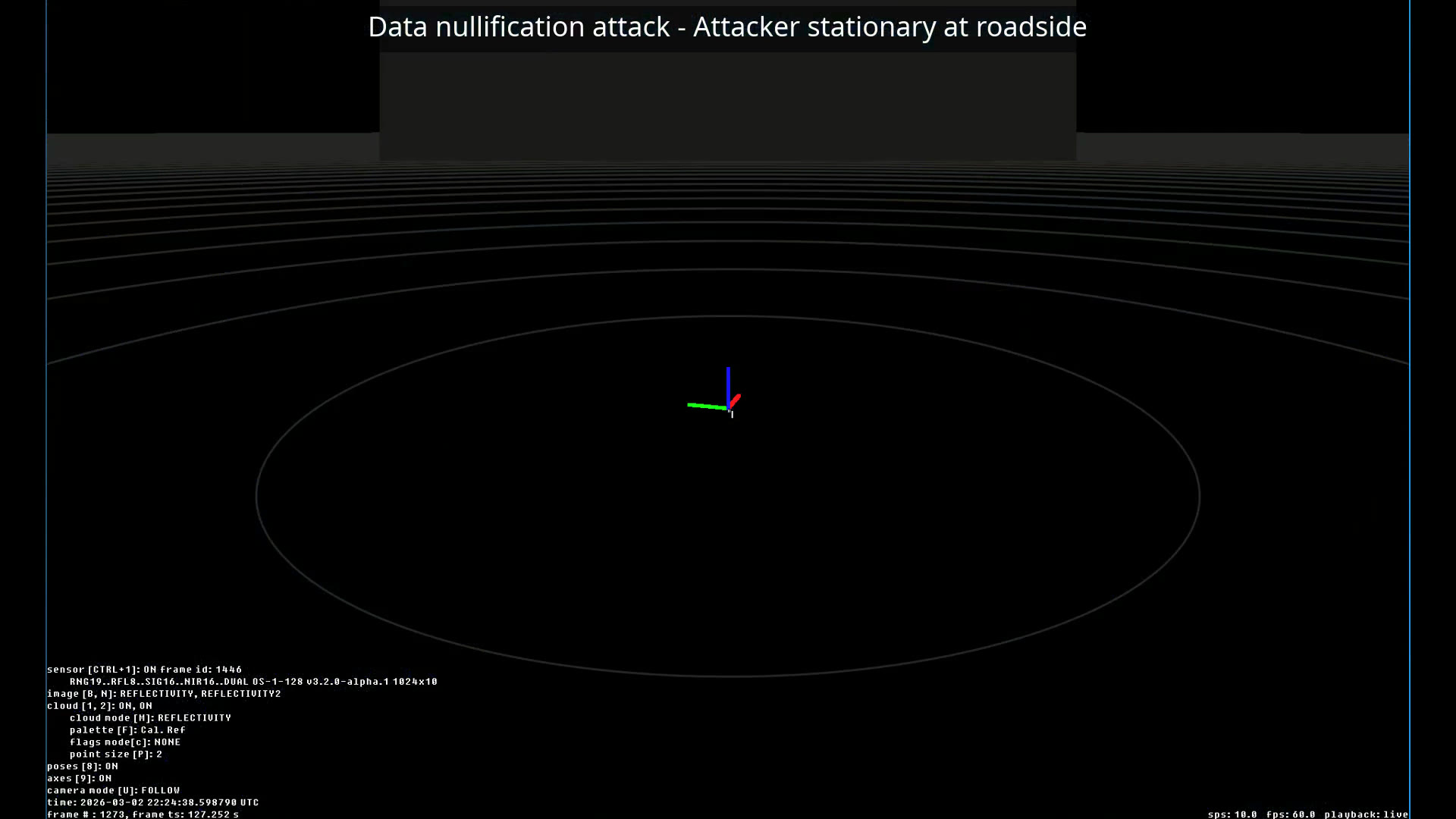} \\

        \textbf{\textit{Injection}} Drive-by pass 2 &
        \TriptychCamera{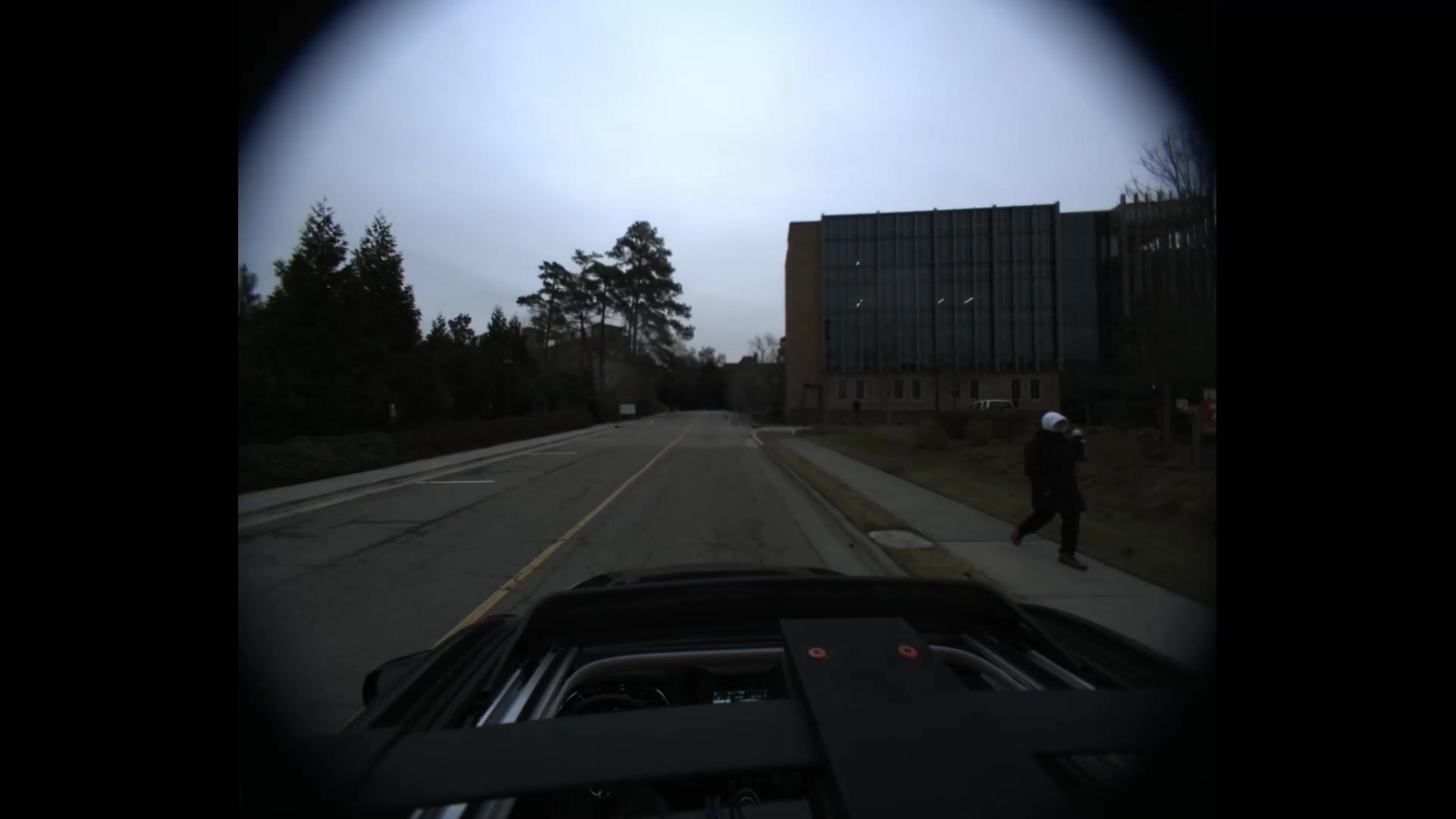} &
        \TriptychLidar{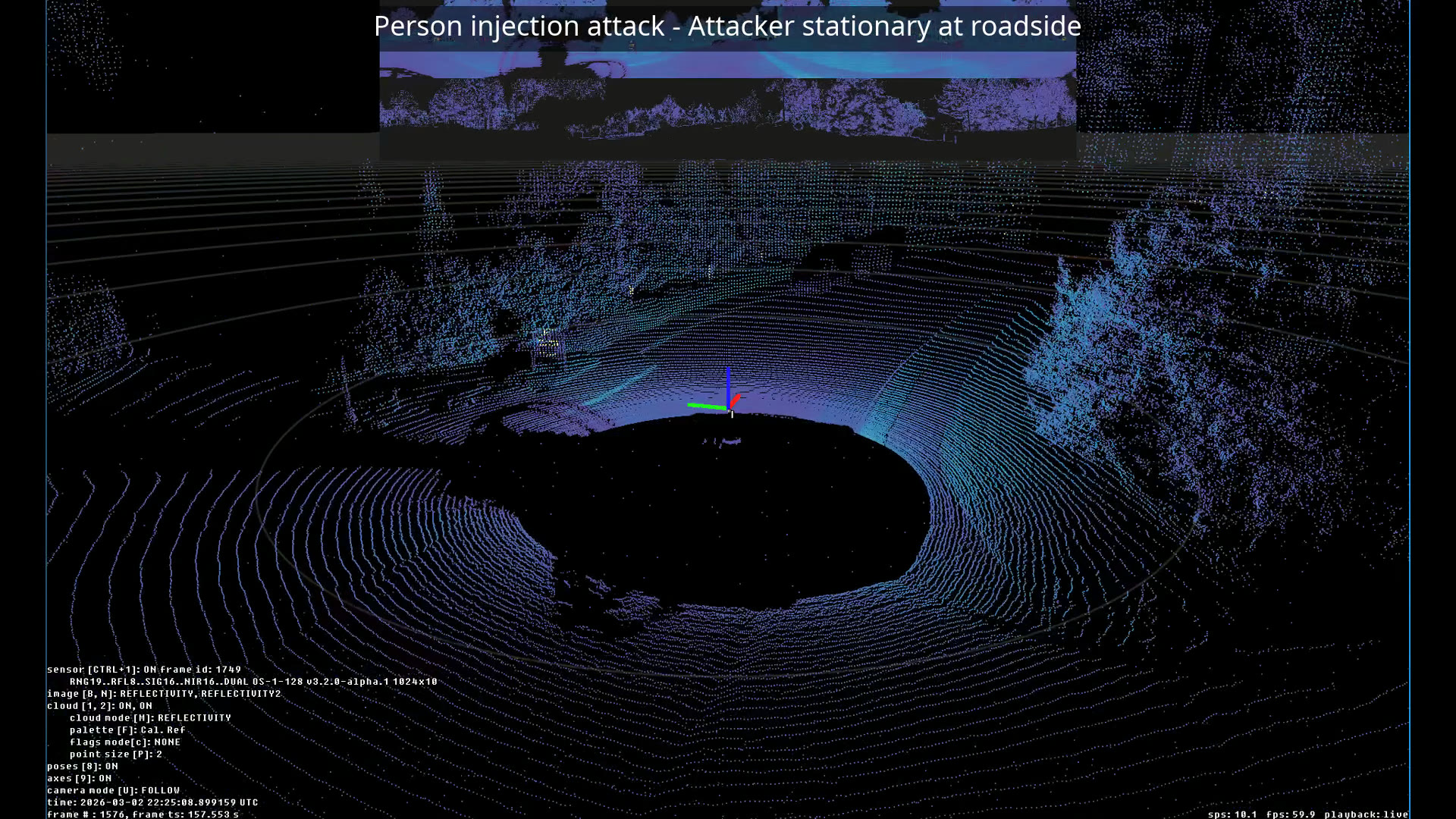} &
        \TriptychLidarInj{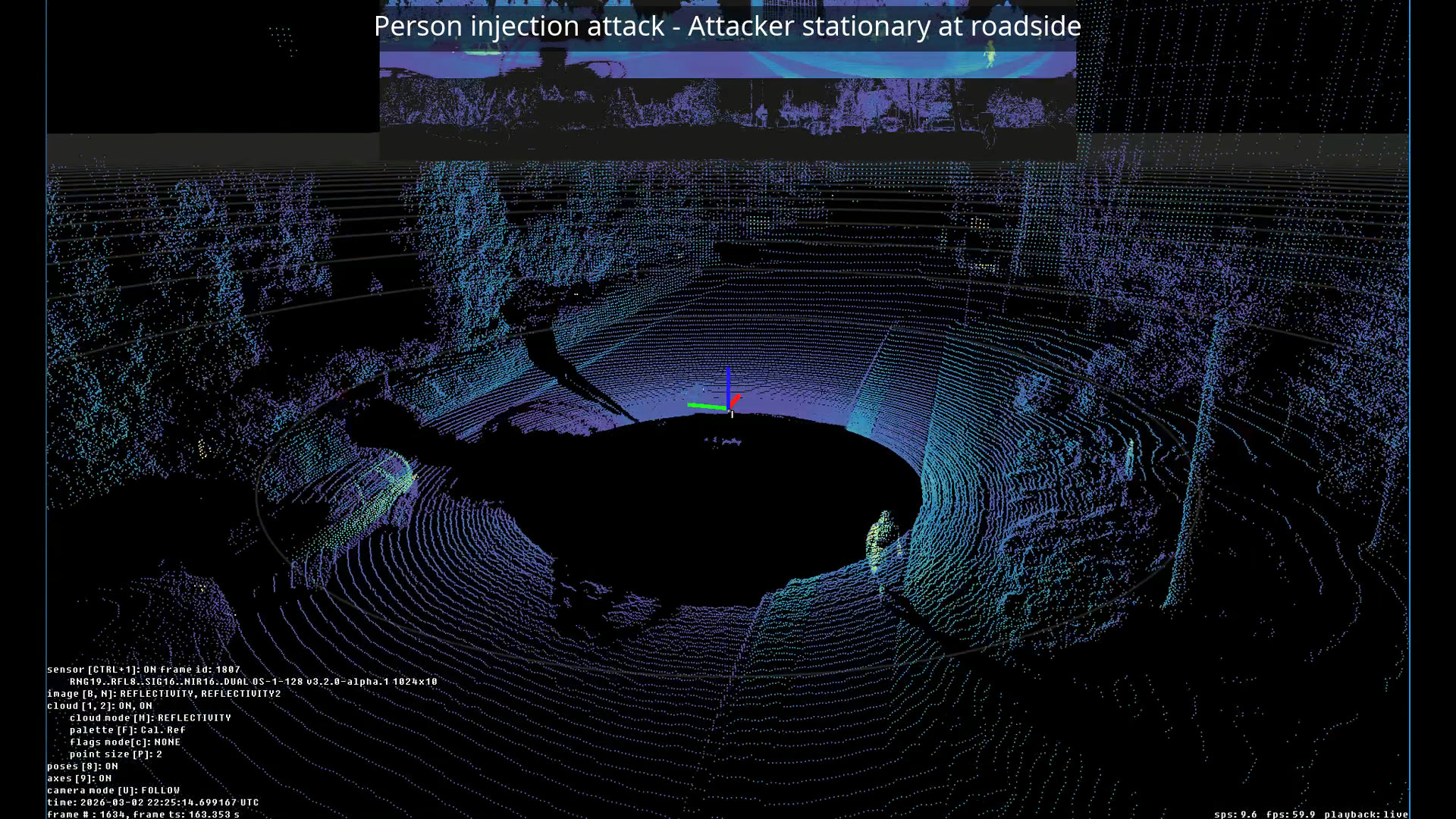}{0.183}{0.453}{0.166}{0.227} \\

        \textbf{\textit{Injection}} Drive-by pass 3 &
        \TriptychCamera{images/triptych_curated/run-outdoor-car-3/camera.jpg} &
        \TriptychLidar{images/triptych_curated/run-outdoor-car-3/lidar_before_unboxed.jpg} &
        \TriptychLidarInj{images/triptych_curated/run-outdoor-car-3/lidar_after_unboxed.jpg}{0.171}{0.464}{0.166}{0.227} \\

        \textbf{\textit{Suppression}} Drive-by pass 4 &
        \TriptychCamera{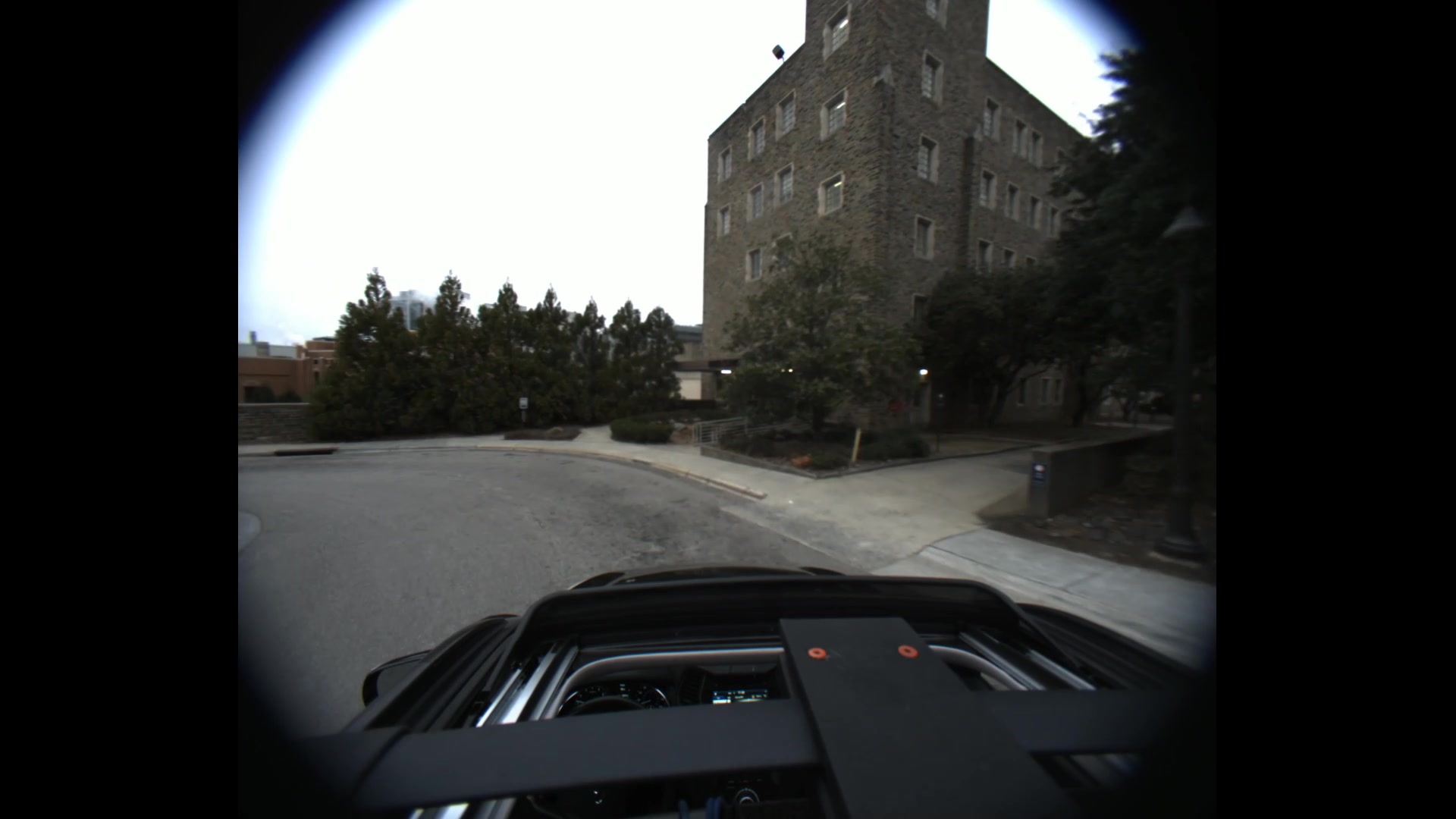} &
        \TriptychLidar{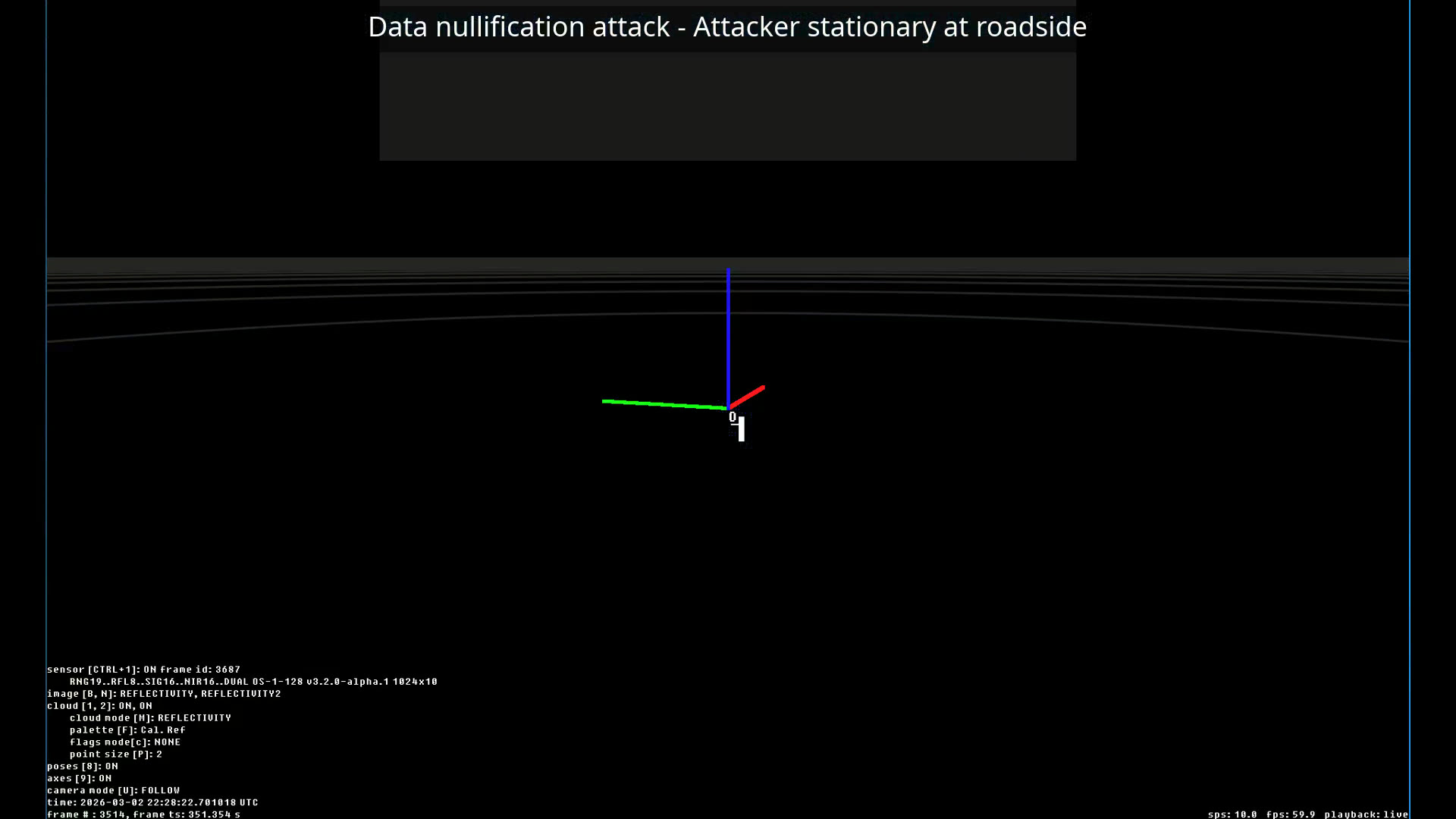} &
        \TriptychLidarSup{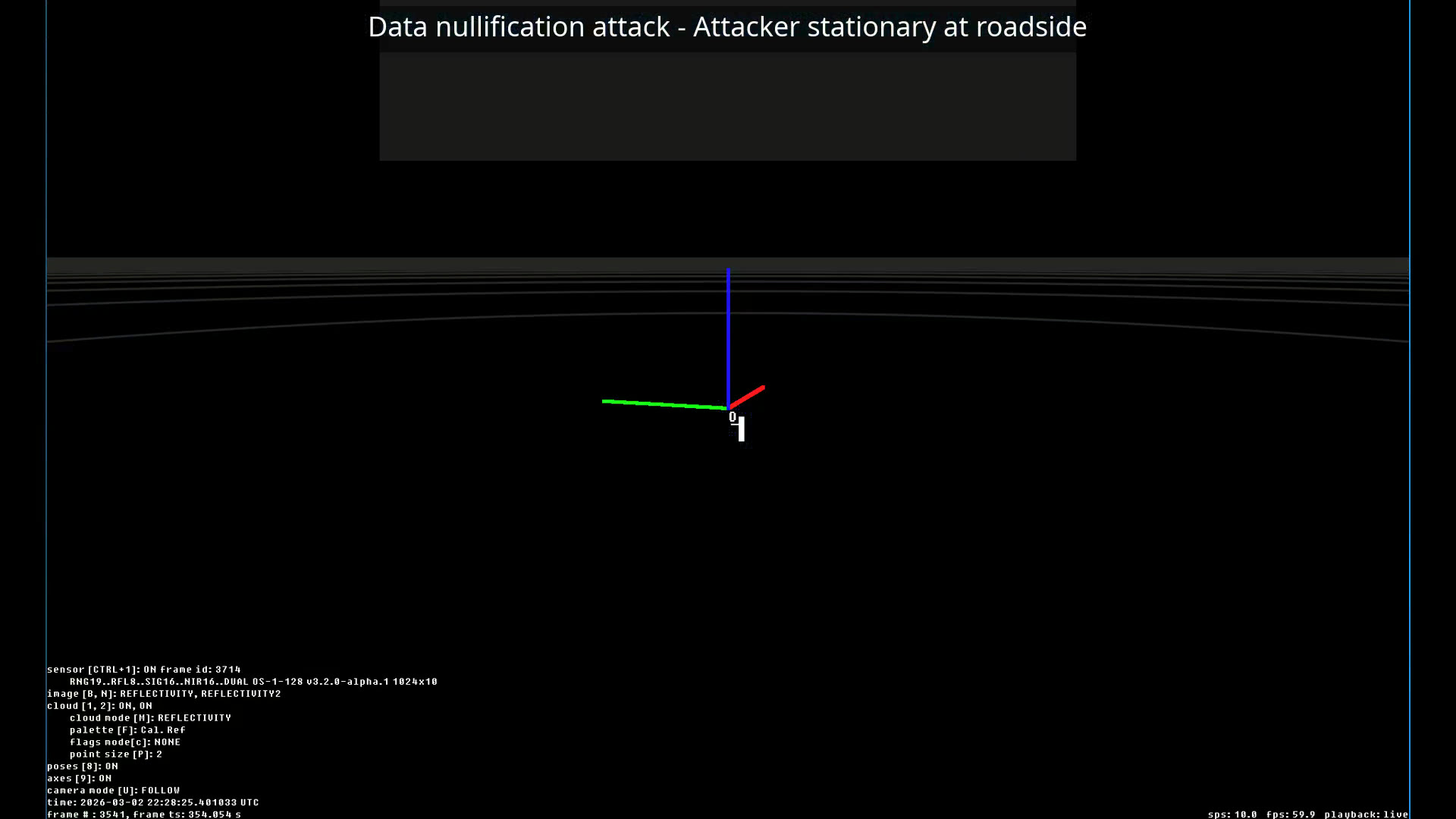} \\
        \bottomrule
    \end{tabular}
    \caption{Additional mobile-setting examples using the same triptych schema as Fig.~\ref{fig:main-feasibility-triptych}. Dashed outlines mark injected returns (author annotations, not detector outputs).}
    \label{fig:appendix-mobile-triptych}
\end{figure*}